%% file: HWSMreview.tex
\documentclass[12pt,a4paper]{article}
\pdfoutput=1
\usepackage[utf8]{inputenc}
\usepackage{amsmath}
\usepackage{amsfonts}
\usepackage{amssymb}
\usepackage{slashed}
\usepackage{epsfig}
\usepackage[T1]{fontenc} 
\usepackage{hyperref}
\usepackage{graphicx}
\usepackage{tikz}
\usepackage{diagbox,pict2e}
\hypersetup{
    colorlinks,
    citecolor=blue,
    filecolor=black,
    linkcolor=red,
    urlcolor=blue,
    linktoc=all
}
\usepackage{cite}

\def\be{\begin{eqnarray}}
\def\ee{\end{eqnarray}}
\def\bea{\begin{eqnarray}}
\def\eea{\end{eqnarray}}

\newcommand{\nn}{\nonumber}

\def\arXiv#1{\href{http://arxiv.org/abs/#1}{arXiv:#1}}
\def\arXiv#1#2{\href{http://arxiv.org/abs/#1}{arXiv:#1}}

\def\doi#1#2{\href{http://doi.org/#1}{#2}}

\author{Karl Landsteiner\\%$^*$\\
{\normalsize\it Instituto de Física Teórica UAM/CSIC,}\\
{\normalsize\it C/ Nicolás Cabrera 13-15, %\\
Campus Cantoblanco, 28049, Spain}\\
{\normalsize E-mail: \tt{karl.landsteiner@csic.es }} \\
\and
Yan Liu\\
{\normalsize\it Center for Gravitational Physics, Department of Space Science,}\\ 
{\normalsize\it Beihang University, Beijing 100191, China}\\
{\normalsize\it Key Laboratory of Space Environment Monitoring and Information Processing,}\\ 
{\normalsize\it Ministry of Industry and Information Technology, Beijing, China}\\
{\normalsize E-mail: \tt{yanliu@buaa.edu.cn}} \\
\and
Ya-Wen Sun\\
{\normalsize\it School of physics \& CAS Center for Excellence in Topological Quantum Computation,}\\
{\normalsize\it University of Chinese Academy of Sciences, Beijing 100049, China}\\
\and
{\normalsize\it Kavli Institute for Theoretical Sciences,}\\ {\normalsize\it  University of Chinese Academy of Sciences,
Beijing 100049, China}\\
{\normalsize E-mail: \tt{yawen.sun@ucas.ac.cn}}
 }

\title{%Holographic Weyl Semimetals\\
\vspace{-2 cm}
\bf Holographic Topological Semimetals}
\date{\vspace{-3ex}}
\begin{document}
\maketitle

 \begin{abstract}
The holographic duality allows  to construct and study models of strongly coupled quantum matter via dual gravitational
theories. In general such models are characterized by the absence of quasiparticles, hydrodynamic behavior and
Planckian dissipation times. 
One particular interesting class of quantum materials are ungapped topological semimetals which have many 
interesting properties from Hall transport to topologically protected edge states.
We review the application of the holographic duality to this type of quantum matter including the construction of holographic
Weyl semimetals, nodal line semimetals, quantum phase transition to trivial states (ungapped and gapped), the holographic
dual of Fermi arcs and how new unexpected transport properties, such as Hall viscosities arise. 
The holographic models promise to lead to new insights into the properties of this type of quantum matter.
\end{abstract}


\newpage

\begingroup
\hypersetup{linkcolor=black}
\tableofcontents
\endgroup

%%%%%%%%%%%%%%%%%%%%%%%%%%%
\section{Introduction}
%%%%%%%%%%%%%%%%%%%%%%%%%%%

Weyl semimetals are a very interesting new form of gapless topological quantum matter \cite{ wsmreview2, Hosur:2013kxa, vishwanath, WSMviewpoint, Witten:2015aoa}.  What makes them special is that the electronic excitations behave in a very unusual way. The electronic quasiparticles can be described by the Weyl equation known from
high energy physics. In the massless limit the Dirac equation can be decomposed into two irreducible parts, called Weyl equations which differ by chirality. In this way the electronics of Weyl semimetals is governed by the laws of relativistic physics. One of the cornerstones of relativistic quantum field theory is that the concept of chirality is sometimes incompatible with quantum theory. Whereas classically the number of left- and right-handed chiral fermions is separately conserved, quantum mechanically this is no longer true. This is the so-called chiral anomaly \cite{abj-bj, Adler:1969gk}. Since the electronics of Weyl semimetals is governed by the chiral Weyl equation the chiral anomaly also has profound consequences on the physics of these materials. Many of the most exotic and interesting phenomena such as the the appearance of surface states (Fermi-arcs) or exotic transport phenomena such as the Hall effect are directly linked to the chiral anomaly. Much of the physics of Weyl semimetals can be understood by analyzing the properties of the one-particle wave function. For example the anomaly can be understood as the effect of non-vanishing Berry flux through the Fermi surface, and chirality manifests itself as monopole like singularity of the Berry connection \cite{Son:2012wh}. These notions are of course bound to the validity of a quasiparticle picture in which interactions play a subordinate role. The question arises then if the salient features of Weyl semimetals do persist in a strong coupling context in which there are no clear quasiparticle excitations. 

The holographic duality (also known as AdS/CFT correspondence and gauge/gravity duality) has been used over the last decade and a half as a tool to investigate precisely this type of questions. No attempt will be made here to give a concise account of the workings of the holographic duality. There are excellent reviews available in the literature \cite{Zaanen:2015oix, book0, review, CasalderreySolana:2011us, Cai:2015cya}. In fact the holographic duality has provided already outstanding results. The modern understanding of hydrodynamics as a derivative
expansion and its validity is put to an all new and sound footing using insights from holography \cite{Baier:2007ix,Rangamani:2009xk}. 
Anomaly induced transport properties \cite{Kharzeev:2013ffa,Landsteiner:2016led}  such as the chiral magnetic and chiral vortical effects and their relations to anomalies can be most easily understood with the means of holographic duality \cite{Erdmenger:2008rm, Banerjee:2008th}. The reader is reminded of the conceptual difficulties in the interpretation of the formula for chiral magnetic effect. This is in sharp contrast with the clarity of its theory in holography. It is by now well-known that the chiral magnetic effect vanishes in equilibrium but it is probably not universally acknowledged that this has been calculated first in holographic models \cite{Gynther:2010ed}.   Another example is the direct connection between the temperature dependence of the chiral vortical effect and the gravitational contribution to the chiral  anomaly \cite{Landsteiner:2011cp, Landsteiner:2011iq}. Therefore holographic duality is not only interesting because of its inherently strongly coupled nature but also because it gives valuable insight via the holographic perspective on conceptually difficult problems such as anomaly induced transport. This provides more than sufficient theoretical
motivation for studying holographic models of Weyl semimetals. It is also noteworthy that hydrodynamics and thus strongly coupling behavior has been reported for the Weyl semimetal WP$_2$ \cite{WP2}. It is possible that holographic models
can serve as models for such type of materials.
 As we will review here, also in the case of Weyl semimetals holography is able to provide new perspectives, leading to new directions of research and even allowing the discovery of new, possibly unexpected transport phenomena.

The review is organized as follows. In section \ref{sec:holography} we give a flash review of the holographic duality.
Then we introduce the holographic model of a (time reversal symmetry breaking) Weyl semimetal in section \ref{sec:holoWSM}. The most
important results stemming from working with this model are also reviewed. These include the existence of a quantum phase transition between the Weyl semimetal and a topological trivial state, calculation of the Hall conductivity, the calculation of topological invariants via fermionic holographic spectral functions, finite temperature and viscosities, in particular the appearance of Hall viscosity at the critical point of the quantum phase transition, the calculation of the axial Hall conductivity, the effects of disorder and the properties of quantum chaos across the quantum phase transition.

All these results are obtained in models that show a transition between the Weyl semimetal and a trivial semimetal. In section \ref{sec:WSMinsulator} a new model is introduced in which the transition is between the Weyl semimetal and an insulating state. In section \ref{sec:nlsm} a generalization to holographic nodal line semimetals is discussed. 
We briefly point to alternative approaches of applications of the holographic duality to the physics of Weyl semimetals in
section \ref{sec:alt}. In the final section \ref{sec:sumout} we first briefly summarize and then give an outlook on possible interesting future directions of research. 

%%%%%%%%%%%%%%%%%%%%%%%%%%%%%%%%%%%%
\section{A short review on the holographic duality}
\label{sec:holography}
%%%%%%%%%%%%%%%%%%%%%%%%%%%%%%%%%%%%

We now give a flash review on the holographic duality \cite{Zaanen:2015oix, book0, review, CasalderreySolana:2011us}. The origin of the holographic duality lies in string theory. In its original form it states that a certain type of string theory (type IIB) on the space AdS$_5$ $\times$ S$^5$ is dual to $N=4$ supersymmetric gauge theory with gauge group $SU(N_c)$ in four dimensions \cite{Maldacena:1997re}. String theory needs ten dimensions and that’s why there is the compact five sphere. The isometry group of this five dimensional sphere is SO(6). In the dual field theory this is the internal global symmetry of the $N=4$ gauge theory. The metric of the five dimensional anti de-Sitter (AdS) space is
\begin{equation}
\label{eq:ads}
ds^2 = \frac{r^2}{L^2}\left( -dt^2 + d\mathbf{x}^2 \right) + \frac{L^2}{r^2} dr^2\,, 
\end{equation}
where $1/L^2$ is a measure for the curvature of the AdS space.

The $N=4$ supersymmetric gauge theory with gauge group $SU(N_c)$ in four dimensions is characterized by two physical parameters the Yang-Mills coupling $g_{YM}$ and the rank $N_c$ of $SU(N_c)$. On the dual 
string theory side, there are two parameters, the fundamental length scale $l_s$ and the string coupling $g_s$ (the amplitude for a string to split in two). 
The AdS$_5$ geometry has curvature $R=-20/L^2$ where is  $L$ is an AdS radius scale. 
%determining the curvature of the AdS$_5$ as $R=-20/L^2$. 
These parameters are related in the AdS/CFT correspondence as the following way
\begin{eqnarray} \label{eq:holodictionary}
g^2_{YM} N_c &\propto& \frac{L^4}{l_s^4}\,,\\
1/N_c &\propto& g_s\,.
\end{eqnarray}
From the above relations, we can see that the AdS/CFT correspondence is a strong weak duality. From (\ref{eq:holodictionary}), for weak curvature on the string theory side the AdS radius $L$ is large which indicates a 
large 't-Hooft coupling constant on the field theory. In this parameter regime we can neglect the stringy effects and use type IIB supergravity to approximate the string theory. If we further take the rank $N_c$ of the gauge group to be very large, i.e. the large $N_c$ limit, for the string theory $g_s$ is very small, we can ignore the quantum loop effects and end up with the theory of classcal supergravity! 

From the above analysis, we found that the classical (super)gravity on AdS$_{d+1}$ space is the infinite coupling and infinite rank limit of a gauge theory in $d$ dimensions, This is known as the AdS/CFT correspondence in its most useful form for applications to quantum many body physics. Here we have allowed ourselves to be already a bit more general. Once we have understood the original example based on the maximally supersymmetric four dimensional field theory, we conjecture that every gravitational theory with some additional suitably chosen matter fields on AdS$_{d+1}$ is dual to a certain $d$ dimensional quantum field theory. We take this point of view in the applications of the AdS/CFT correspondence to quantum many body systems. The additional matter fields chosen on the gravity side is according to a particular symmetric property of the underlying quantum field theoretical system that one is interested in. 

The dual field theory lives in the four dimensional spacetime parametrized by $(t,\mathbf{x})$, where $\mathbf{x}$ denotes the vector of spatial coordinates $(x,y,z)$. It is sometimes said that the dual field theory lives on the the boundary of AdS space when taking the limit $r\rightarrow \infty$. But this is not really true, all of the bulk has a field theory interpretation. The best way of thinking about the “holographic” direction is as an energy scale. The high energy limit of the theory is given by $r\rightarrow \infty$ and vice versa the infrared limit is $r\rightarrow 0$. This allows a direct geometric interpretation of the renormalization group flow of a holographic theory from the ultraviolet (UV) to the infrared (IR).

The other important ingredient of the holographic dictionary, the rules that allow us to extract field theory information from gravitational physics is the identification between fields in AdS and operators and couplings of the dual field theory. If we consider the solution to a (second-order) field equation in anti-de Sitter 
space it allows an expansion for large $r$ of the form
\begin{equation}\label{eq:asymptoticexpansio}
 \Phi = \frac{1}{r^{\Delta_-}} \left(\Phi_0(x) + \mathcal{O}(\frac{1}{r^{2}})\right) + \frac{1}{r^{\Delta_+}} \left(\Phi_1(x) + \mathcal{O}(\frac{1}{r^{2}})\right)\,.
\end{equation}
We assume there that $\Delta_{\pm}\geq 0$ and $\Delta_- < \Delta_+$. $\Phi_0(x)$ is the boundary value (non-normalizable mode) of the field $\Phi(r,x)$ in anti-de Sitter space and at the same time it is interpreted as a coupling or source for an operator in the dual field theory. 
When we do the path integration over
the fields in AdS, we have to keep the boundary values $\Phi_0(x)$ fixed. What we obtained finally is a functional $Z[J] $ of the form 
\begin{equation}\label{eq:pathin}
 Z[J] = \int_{J(x)=\Phi_0 } [d \Phi]\; \exp(-i S[\Phi])\,, 
\end{equation}
where the source $J(x)$ is the boundary field $\Phi_0(x)$. This source $J(x)$ couples to a (gauge invariant)
operator ${\cal O}(x)$ with conformal dimension $\Delta_+$ in the field theory\footnote{This choice is known as standard quantization. When $\frac{d}{2}\leq \Delta_+\leq\frac{d}{2}+1$, for the dual field theory we could add a double trace deformation $\int d^dx {\cal O}(x)^2$ which is irrelevant close to the fixed point, to generate a flow to a new fixed point. In this case $\Delta_+$ and $\Delta_-$ exchange their roles, i.e. $\Phi_1(x)$ is now interpreted as the source $J(x)$ which couples to an operator ${\cal O}(x)$ of conformal dimension $\Delta_-$. This is known as alternative quantization of the bulk theory \cite{Klebanov:1999tb}. }. 
Performing functional
differentiation of (\ref{eq:pathin}) with respect to the sources, we can obtain the connected correlation functions of the gauge invariant operators ${\cal O}(x)$ in the quantum field theory 
\begin{equation}
\left\langle {\cal O}_1(x_1) \dots  {\cal O}_n(x_n) \right\rangle = \frac{\delta^n \log Z}{\delta J_1(x_1) \dots \delta J_n(x_n)}\,.
\end{equation}
These are of course rather formal expressions. In general one does not know how to do this type of path integral including
the metric degrees of freedom or even the proper string theory dual. In the large $N_c$ and large coupling $g^2_{YM} N_c$ limit, the gravitational theory becomes classical. The path integral (\ref{eq:pathin}) now is dominated by the classical solutions from the equations of motion for fields. The generating functional $\log Z$ can be simplified and computed by the
classical action evaluated on the classical solution. In this case  in the asymptotic
expansion (\ref{eq:asymptoticexpansio}) the coefficient $\Phi_1(x)$ is the vacuum expectation value of the dual operator sourced by $\Phi_0$
\begin{equation}
 \left\langle {\cal O}(x)\right\rangle \propto \Phi_1(x)\,.
\end{equation}
This scheme applies to all fields in AdS, also to the metric itself. The operator that corresponds to the metric is the energy-momentum tensor. In the same way the operator that corresponds to a gauge field in AdS is a current. The essentials of the
holographic dictionary are summarized in table \ref{tab:one}.
\begin{table}[h!]
\begin{center}
\begin{tabular}{|c|c|}
\hline
Field in AdS & Dual Operator \\
\hline\hline
Metric $g_{\mu\nu}$ & Energy-momentum tensor $T^{\mu\nu}$\\
Gauge field $A_\mu$ & Current $J^\mu$ \\
Scalar field $\Phi$ & Scalar operator $\mathcal{O}$\\
\hline
\end{tabular}
\caption{\small The essentials of the holographic dictionary \label{tab:one}.}
\end{center}
\end{table}

One can use this dictionary to generate new solutions that are deformations of the simple AdS space %in equation 
(\ref{eq:ads}) by switching on certain couplings. In practice this means that one demands specific boundary conditions on suitably chosen AdS fields that represent couplings in the dual field theory. Let us now explain how this strategy can be implemented to obtain a holographic version of a Weyl semimetal. 

%%%%%%%%%%%%%%%%%%%%%%
%%%%%%%%%%%%%%%%%%%%%%%
\section{The holographic Weyl semimetal}
\label{sec:holoWSM}
%%%%%%%%%%%%%%%%%%%%%%%
%%%%%%%%%%%%%%%%%%%%%
 
To find the holographic background solution we first must identify what kind of deformations we need to introduce in AdS space to mimic the essential features of a Weyl semimetal. In order to do so we first review quickly a quantum field theoretical model of a Weyl semimetal.

%%%%%%%%%%%%%%%%%%%%%%%%
\subsection{Weyl semimetal from Dirac equation}
%%%%%%%%%%%%%%%%%%%%%%%%

In Weyl semimetal, the physics around the nodal points can be described by a quantum field theoretical model %which has the same local properties around the nodal points as a Weyl semimetal  
which takes the following form of a ``Lorentz breaking'' Dirac
equation \cite{Grushin:2012mt, Grushin:2019uuu}
\begin{equation}
\label{eq:lorentzbreaking}
\left( i \slashed\partial -  e \slashed{V} - \gamma_5 \text{\boldmath$\gamma$} \cdot \mathbf{b} + M \right)\psi =0\,.
\end{equation}
Here $\slashed X = \gamma^\mu X_\mu$, $V_\mu$ is the electromagnetic gauge potential, $\gamma^\mu$ are the Dirac matrices, and $\gamma_5 = i \gamma_0\gamma_1\gamma_2\gamma_3$. We can 
define left- or right-handed spinors via $(1\pm\gamma_5)\psi = \psi_{L,R}$. The axial gauge field $\mathbf{b}$ breaks the time reversal symmetry and is introduced to separate the Weyl points in the momentum space as we will show from the perspectives of energy spectrum. For simplicity we take $\mathbf{b}= b \mathbf{e}_z$. $M$ is the mass of the Dirac field.

\begin{figure}[!thb]
\begin{center}
\includegraphics[scale=0.63,clip=true]{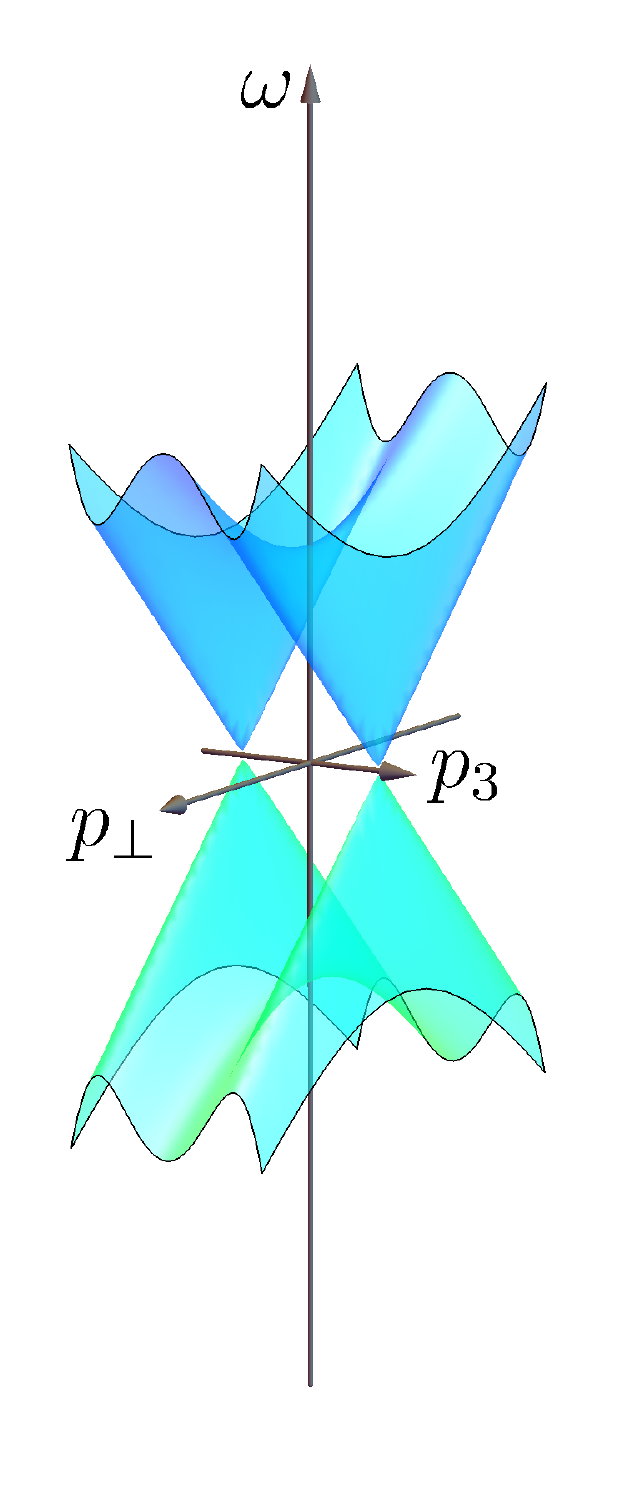}~~~~
\includegraphics[scale=0.63,clip=true]{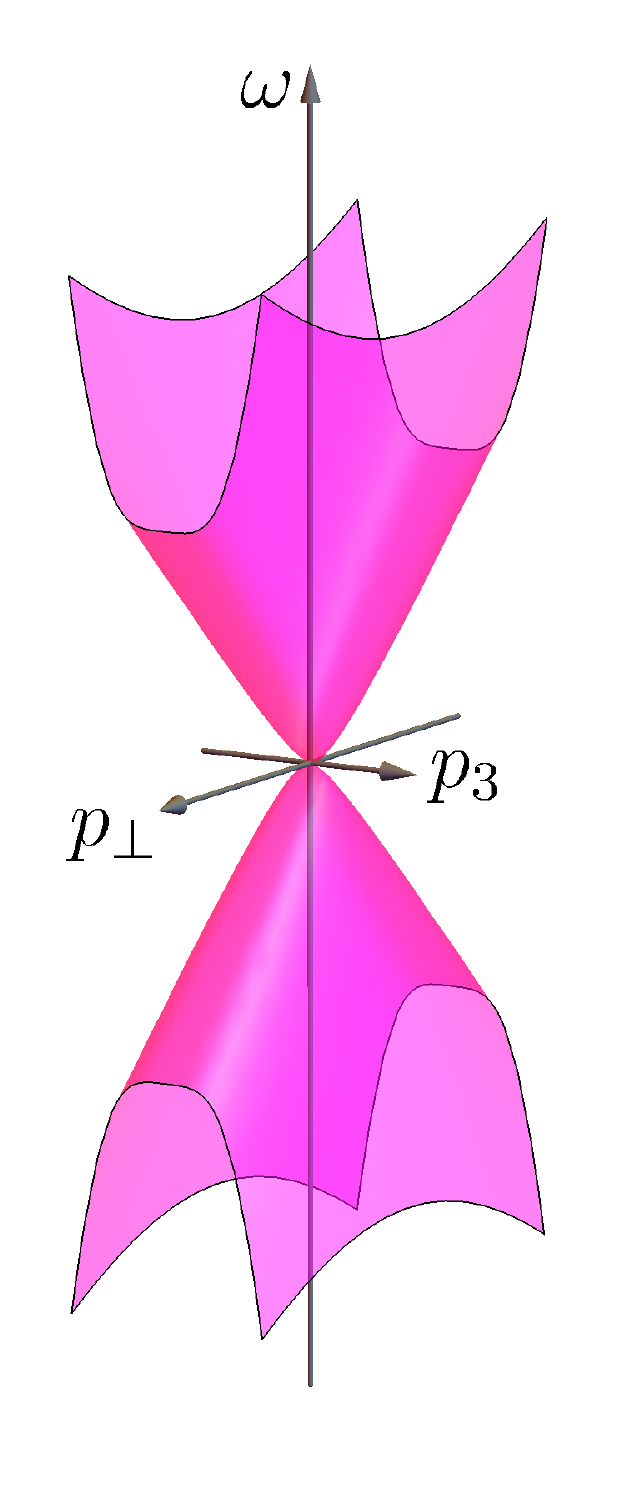}~~~~~
\includegraphics[scale=0.63,clip=true]{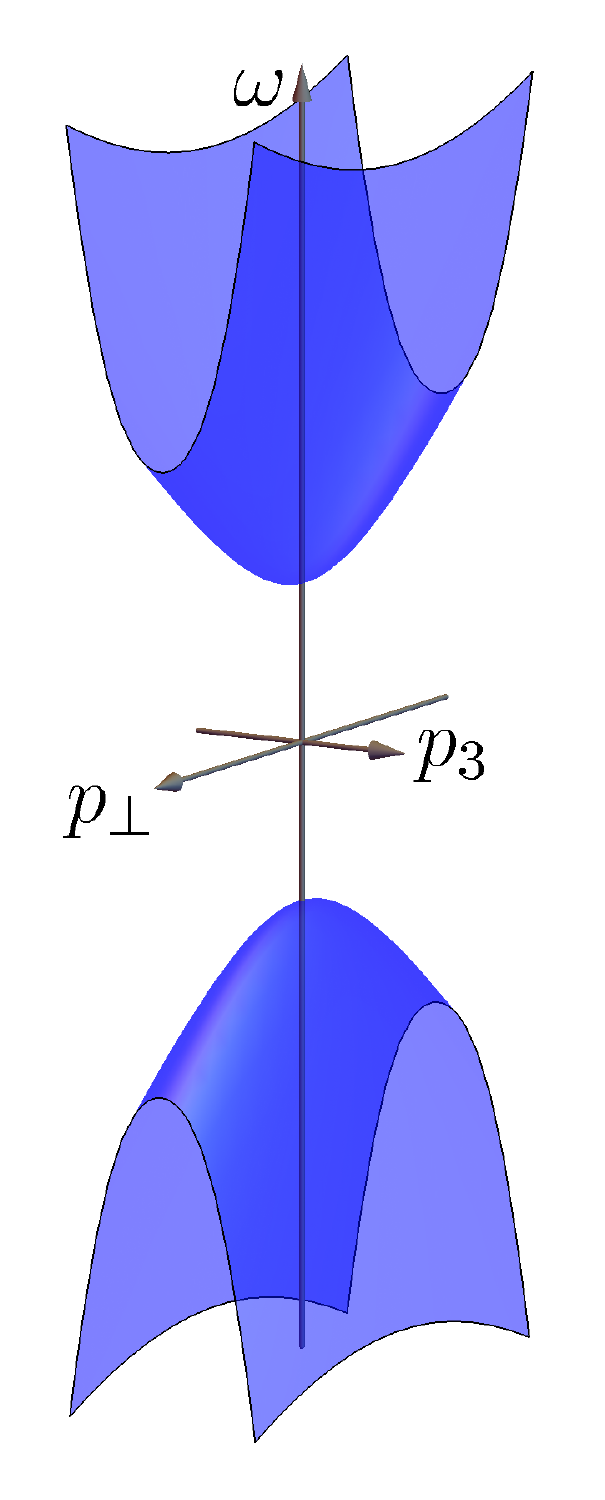}
\end{center}
\vskip -5mm 
\caption{\small Phases of the Dirac equation (\ref{eq:lorentzbreaking}). 
The figure shows the dispersion relation as a function of $p_\perp$ and $p_3$.
The left figure is deep in the Weyl semimetal phase and the middle figure shows the critical point in between the two.
 The right figure is the gapped phase.} 
\label{fig:phases}
\end{figure}

The energy spectrum of (\ref{eq:lorentzbreaking}) is shown %sketched 
in figure \ref{fig:phases}.
When $|b|>|M|$ the spectrum is ungapped. There is a band inversion in the spectrum and at the crossing points the wave function is described by the one of Weyl fermions.
The separation of the Weyl points in momentum space is given by $2\sqrt{b^2-M^2}$ along the direction indicated by
the vector $\mathbf{b}$. At low energies it is described by the  the effective theory with the Lagrangian of the form (\ref{eq:lorentzbreaking}) with $M_\mathrm{eff} =0$ and 
$\mathbf{b}_{\text{eff}} = \sqrt{b^2-M^2}\mathbf{e}_z$.
For $|b|<|M|$ the system is gapped with gap $2 M_\mathrm{eff} = 2(|M|-|b|)$. 
%%%%%%%%%%%%%%%%%%%%%%%%%%%%%%%%%%%%%
%%%%%%%%%%%%%%%%%%%%%%%%%%%%%%%%%%%%%

The axial anomaly
\begin{equation}\label{eq:axialanomaly}
 \partial_\mu J^\mu_5 = \frac{1}{16\pi^2} \varepsilon^{\mu\nu\rho\lambda} F_{\mu\nu} F_{\rho\lambda} + 2  M \bar\psi\gamma_5 \psi\,
\end{equation}
indicates there is an anomalous Hall effect in the Weyl semimetal phase \cite{Haldane:2004zz,Yang,Xu:2011dn,burkovbalents,Zyuzin:2012tv,vazifeh}
\begin{equation}\label{eq:ahe}
 \mathbf{J} = \frac{1}{2\pi^2} \mathbf{b}_\text{eff} \times \mathbf{ E}\,.
\end{equation} 

Thus by tuning $M/b$, from the band structure we see that  there is a quantum phase transition from topologically nontrivial Weyl semimetal phase to a trivial insulating phase. This phase transition is beyond the Landau classification and is an example of a topological phase transition.  In both phases of the system, the same symmetries of the underlying theory are explicitly broken by the the couplings $M,b$.
 Due to the fact that in the topologically nontrivial phase there is a nontrivial Hall effect while in the topological trivial phase there is trivial Hall effect, the Hall conductivity can be taken as the order parameter\footnote{It
is not a traditional order parameter but Hall effect is known to serve  
as signature of topologically non-trivial Fermi surfaces \cite{Haldane:2004zz}.} of this special topological phase transition \cite{haldane}. In more general case, additional massless Dirac fermions might show up, and the topologically trivial phase might be a semimetal instead of a gapped trivial phase. Then this  quantum phase transition goes from a topologically nontrivial semimetal to a trivial semimetal. This will be exactly the case of our holographic model in the next subsection. In section \ref{sec:WSMinsulator} we will improve on this and discuss a holographic model with a phase transition
to a Chern insulator.

The anomalous Hall effect (\ref{eq:ahe}) in the quantum field theory is obtained from a one-loop contribution to the polarization tensor.  However, there are infamous 
regularization ambiguities \cite{Jackiw:1999qq} in the quantum field theory. There are some ways to resolve the ambiguity, e.g. 
by considering anomaly cancellation arising from 
chiral edge states at the boundaries (Fermi arcs) \cite{Goswami:2012db} or by matching to a tight-binding model \cite{Grushin:2012mt, vazifeh}. %Essentially this regularization is gauge invariant in the presence of axial gauge fields which is a source for the axial current. 

What can we learn from this for  building a holographic model? First we see that there are two $U(1)$ symmetries at play. One of them, the axial one, is anomalous and explicitly broken by the mass term in the Dirac equation.
The anomaly gives rise to the quantum Hall effect (\ref{eq:ahe}) as long as $\sqrt{b^2-M^2}>0$. The mass term can be identified as a source for the operator $\bar \psi \psi$. We can take the mass to be the expectation value of a complex classical scalar field that is charged under the axial $U(1)$ symmetry. Because of this an expectation value breaks the axial symmetry already on the classical level. Under a chiral rotation $\psi \rightarrow i \alpha \gamma_5 \psi$ the operator
$\bar \psi \psi$ transforms into $\bar\psi \psi \rightarrow 2 i \alpha \bar \psi \gamma_5 \psi$. Furthermore the parameter $\mathbf{b}$ or more generally $b^\mu$ couples to the axial current $J^\mu_5=\bar \psi \gamma_5 \gamma^\mu \psi$
and can therefore be understood as the background value of an axial gauge field.
These considerations give us the ingredients we need to implement in the holographic model.

%%%%%%%%%%%%%%%%%%%%%%%%
\subsection{Holographic model}
%%%%%%%%%%%%%%%%%%%%%%%%

A holographic action allowing the implementation of the above symmetries and breaking pattern is
\begin{align}
  S=&\int d^5x \sqrt{-g}\bigg[\frac{1}{2\kappa^2}\Big(R+\frac{12}{L^2}\Big)-\frac{1}{4}\mathcal{F}^2-\frac{1}{4}F^2 \nonumber\\& ~~~~ 
+\epsilon^{abcde}A_a\bigg(\frac{\alpha}{3} \Big(F_{bc} F_{de}+3 \mathcal{F}_{bc} \mathcal{F}_{de}\Big)+\zeta  R^{f} _{~gbc}R^{g}_{~fde} \bigg)\nonumber\\
&~~~ -(D_a\Phi)^*(D^a\Phi)-V(\Phi)\bigg]\,,\label{eq:holomodel}
\end{align} where $\kappa^2$ is the Newton constant, $L$ is the AdS radius and $\alpha, \zeta$ are the Chern-Simons coupling constants.\footnote{Note that $\epsilon_{abcde}=\sqrt{-g}\varepsilon_{abcde}$ with $\varepsilon_{0123r}=1.$
Our conventions for indexes are as follows: latin indexes from the beginning of the alphabet $\{a,b,\dots\}$ are five dimensional ones, greek indexes are four dimensional ones and latin indexes from the middle of the alphabet $\{i,j,m,n\}$ are purely spatial indexes.} In field theory, we have conserved electromagnetic current and non-conserved axial current. As shown in table \ref{tab:one}, the conserved currents in the field theory are dual to gauge fields in AdS space. 
The electromagnetic $U(1)$ current is dual to the bulk gauge field $V_a$ in AdS with field strength $\mathcal{F}=dV$. The axial $U(1)$ current is dual to the gauge field $A_a$ in AdS with field strength $F = dA$. Since the axial symmetry is anomalous in the field theory and in the bulk the anomaly is characterized by the Chern-Simons part of the action (\ref{eq:holomodel}) with coupling constants $\alpha$ and $\zeta$. The gauge invariant regularization corresponds to this choice of Chern-Simons term with which the electromagnetic $U(1)$ symmetry remains non-anomalous. 
The anomaly arises in a gauge variation of the axial gauge field $\delta A_a = \partial_a \theta$ as a boundary term
\begin{equation}\label{eq:holoanomaly}
\mathcal{A} = \int d^4x \sqrt{-g} \theta \left( \frac{\alpha}{3}\epsilon^{\mu\nu\rho\lambda}
\Big(F_{\mu\nu} F_{\rho\lambda}+3 \mathcal{F}_{\mu\nu} \mathcal{F}_{\rho\lambda}\Big) + \zeta R^{\alpha} _{~\beta\mu\nu}R^{^\beta}_{~\alpha\rho\lambda}  
\right)\,.
\end{equation}
We have included here the usual axial anomaly due to electromagnetic field, the purely axial anomaly due to axial gauge fields and the gravitational\footnote{A contribution due to the extrisic curvature vanishes on asymptotic boundary of AdS \cite{Landsteiner:2011iq}.} contribution to the axial anomaly. All three will play a role in the physics of holographic Weyl semimetals. The factor of $3$ in the axial anomaly reflects the symmetry factor that arises in the corresponding triangle diagram in quantum field theory.
We will introduce the mass deformation via a boundary value of the scalar field $\Phi$ \cite{Jimenez-Alba:2015awa}. Since the dual axial symmetry is explicitly broken, in the bulk this scalar field is introduced to be only axially charged. The covariant derivative is $D_a\Phi = (\partial_a - i q A_a )\Phi$. The scalar field potential is $m^2 |\Phi|^2 + \frac{\lambda}{2} |\Phi|^4$.
The AdS bulk mass $m^2L^2= -3 $ is chosen such that the dual operator
has conformal dimension three and its source has conformal dimension one. 
The electromagnetic and axial currents\footnote{These are the consistent currents. The vector current $J^\mu$ is conserved while the conservation of the the
axial current $J_5^\mu$ is broken explicitly by the scalar field and spontaneously by the anomaly \cite{Jimenez-Alba:2015awa}.
The covariant currents can be defined by dropping the Chern-Simons terms. }  are defined as
\begin{align}
\label{eq:consVcur}
J^\mu &= \lim_{r\rightarrow\infty}\sqrt{-g}\Big(\mathcal{F}^{\mu r}+4\alpha\epsilon^{r \mu\beta\rho\sigma} A_{\beta} \mathcal{F}_{\rho\sigma}  \Big) \,,\\
\label{eq:consAcur}
 J^\mu_5 &= \lim_{r\rightarrow\infty}\sqrt{-g}\Big(F^{\mu r}+\frac{4\alpha}{3}\epsilon^{r \mu\beta\rho\sigma} A_{\beta}F_{\rho\sigma}  \Big)\,.
\end{align}
The model was first studied in the probe limit in \cite{Landsteiner:2015lsa} and then in the backreacted case in 
\cite{Landsteiner:2015pdh}. 

We are looking for asymptotically AdS solutions.  The mass parameter and the time-reversal symmetry breaking parameter in the field theory are
introduced through the conformal boundary conditions 
\begin{equation}\label{eq:bcs}
 \lim_{r\rightarrow \infty}\,r\Phi = M~,~~~\lim_{r\rightarrow \infty}A_z = b\,.
\end{equation}
We take the following ansatz for the zero temperature solution
\begin{eqnarray}\label{eq:ansatz}
ds^2&=&u(-dt^2+dx^2+dy^2)+\frac{dr^2}{u}+h dz^2\,,~~~A=A_z dz\,,~~~\Phi=\phi\,\,,
\end{eqnarray}
where $u,h,A_z,\phi$ are functions of $r$. 
In this case $M/b$ is the only tunable parameter of the system due to the conformal symmetry.
We set $2\kappa^2=L=1$.
\vspace{.1cm}\\
\noindent{\it Critical solution.} 
The following Lifshitz solution is an exact solution of the system  
\begin{eqnarray}\label{nh-cs}
ds^2&=&u_0r^2(-dt^2+dx^2+dy^2)+\frac{dr^2}{u_0r^2}+h_0 r^{2\beta}dz^2\,,\nonumber\\
A_z&=&r^\beta,~~~\phi=\phi_0\,.
\end{eqnarray}
It exists an anisotropic Lifshitz symmetry $(t,x,y,r^{-1})\to s(t,x,y,r^{-1})$ and $z\to s^\beta z$.
The irrelevant deformations can be introduced to flow it to UV with the boundary conditions (\ref{eq:bcs}). 
%The scaling symmetry $z\to s z$ can be used to set the coefficient in $A_z$  to be 1. 
The four constants $\{u_0, h_0, \beta, \phi_0\}$ in (\ref{nh-cs}) are determined 
by the values of $\lambda$, $m$ and $q$.

It turns out that the following irrelevant perturbations around the Lifshitz fix point can flow the geometry to asymptotic AdS
$
u =u_0r^2\big(1+ \delta u\,r^\alpha\big),~~
h= h_0 r^{\beta}\big(1+ \delta h\, r^\alpha\big), ~~
A_z =r^\beta \big(1+ \delta a\, r^\alpha\big),~~
\phi=\phi_0\big(1+ \delta \phi\, r^\alpha\big)
$.
%Due to the Lifshitz scaling symmetry, only the sign of $\delta \phi$ is a free parameter and others are fully determined by $\delta \phi=\pm 1$. Numerics shows that only $\delta\phi=-1$ corresponds to asymptotic AdS space at the UV.  
Only the sign of $\delta \phi$ is a free parameter and the geometry can flow to AdS with $\delta\phi=-1$. 
In the case $q=1,\lambda=1/10$, we have $(u_0, h_0, \beta, \phi_0,\alpha)\simeq (1.468,0.344, 0.407, 0.947,1.315)$ and 
$(\delta u, \delta h, \delta a)\simeq (0.369, -2.797, 0.137)\delta \phi$.  We obtain the critical value $M/b\simeq 0.744$, which corresponds to the transition point. 
\vspace{.1cm}\\
{\noindent{\it Topological nontrivial phase.} }
At leading order the second type of solution in the IR is 
\begin{align}\label{nearhor-nt}
u=r^2,~~
h=r^2,~~
A_z=a_1+\frac{\pi a_1^2\phi_1^2}{16 r} e^{-\frac{2 a_1 q}{r}},~~\nonumber\\ \phi=\sqrt{\pi}\phi_1\Big(\frac{a_1 q}{2r}\Big)^{3/2} e^{-\frac{a_1 q}{r}}\,; 
\end{align}
%$\lambda$ appears only at higher order terms which become important when $M/b$ is close to the critical value. 
$a_1$ can be set to a numerically convenient value. Later on we rescale to $b=1$. 

Starting from the near horizon solution (\ref{nearhor-nt}), the equations can be numerically integrated 
towards the UV. $\phi_1$ can be taken as the shooting parameter to obtain an 
AdS$_5$ to AdS$_5$ domain wall.  For the values $q=1,\lambda=1/10$ this type of solution exists only for $M/b<0.744$. 
\vspace{.1cm}\\
{\noindent{\it Topological trivial phase.}  }
The third type of solution at leading order in IR is  
\begin{equation}\label{nearhor-tt}
u=\big(1+\frac{3}{8\lambda}\big)r^2,~~
h=r^2,~~
A_z=a_1 r^{\beta_1},~~\phi=\sqrt{\frac{3}{\lambda}}+\phi_1 r^{\beta_2}\,,
\end{equation}
where $(\beta_1,\beta_2)=(\sqrt{1+\frac{48q^2}{3+8\lambda}}-1, 2\sqrt{\frac{3+20\lambda}{3+8\lambda}}-2).$
For our choice of $\lambda$ and $q$  $(\beta_1,\beta_2)=(\sqrt{\frac{259}{19}}-1, \frac{10}{\sqrt{19}}-2)$. 
 $a_1$ can be set to be 1. $\phi_1$ can be taken as the shooting parameter to obtain the AdS$_5$ to AdS$_5$ domain wall.  
 For the values $q=1,\lambda=1/10$ this type of solution only exist for $M/b>0.744$.

\begin{figure}
\begin{center}
\includegraphics[width=0.5\textwidth]{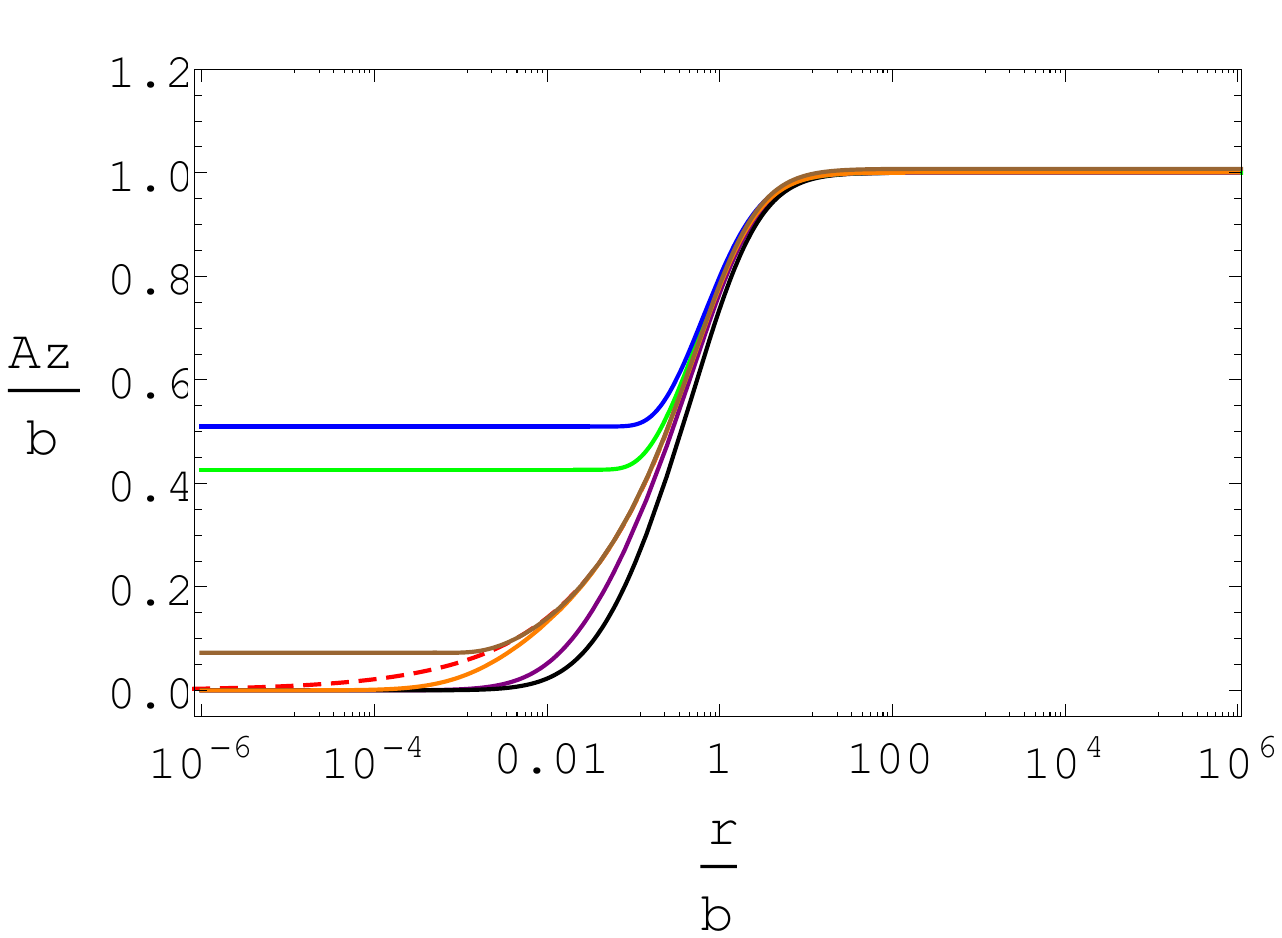}
\includegraphics[width=0.46\textwidth]{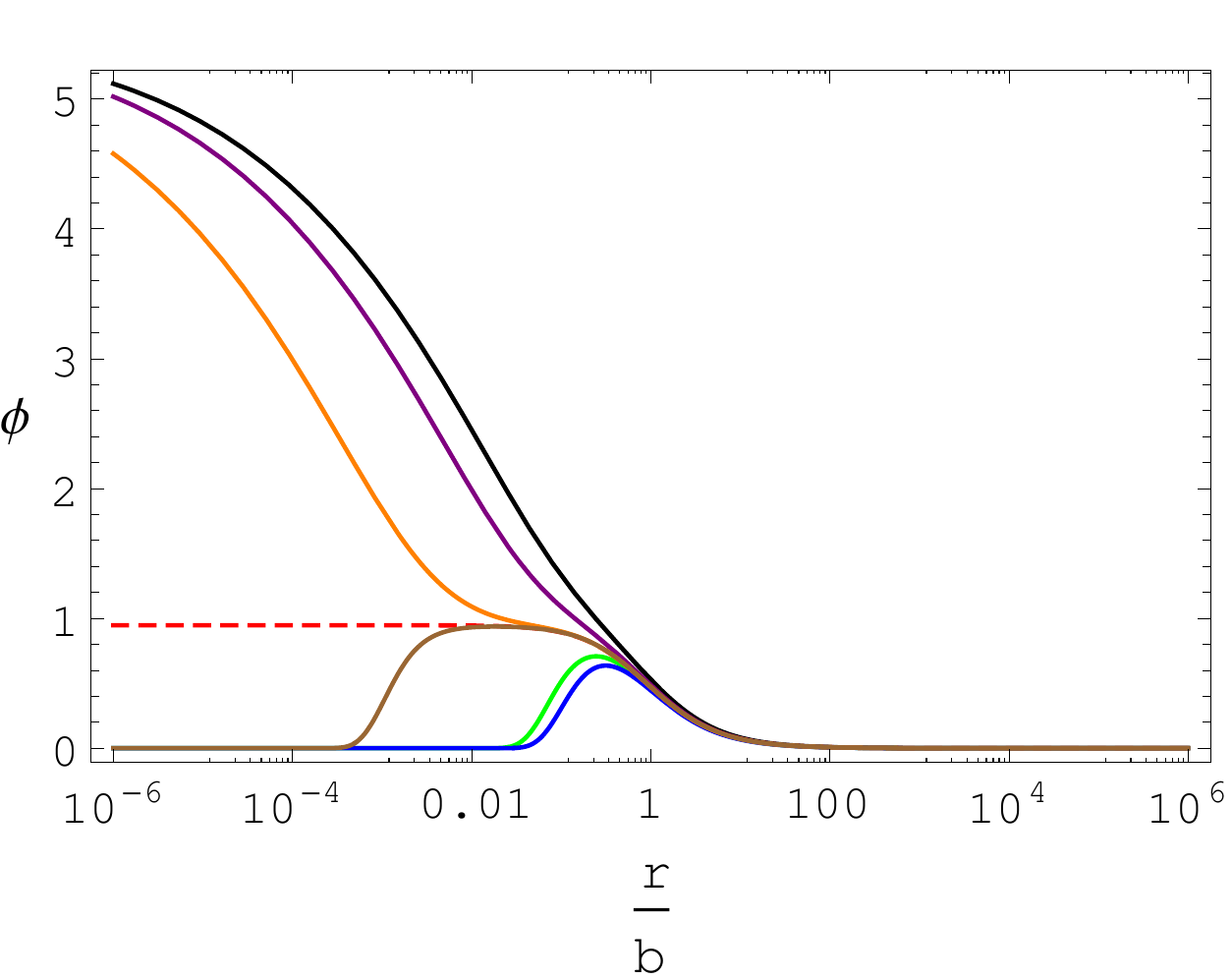}
\end{center}
\vskip -5mm 
\caption{\small The bulk profile of background $A_z$ and $\phi$ for $M/b=0.695$ (blue), $0.719$ (green), $0.743$ (brown), 
$0.744$ (red-dashed), $0.745$ (orange), $0.778$ (purple), $0.856$ (black). Figure from \cite{Landsteiner:2015pdh}.}
\label{fig:profile}
\end{figure}

Figure \ref{fig:profile} shows the profiles of the scalar field $\phi$ and the gauge field $A_z$. %for all three phases at  different values of $M/b$.  
Only one of the above three types of solutions exists at a given value of $M/b$. The horizon value of $A_z$ varies continuously between the
two phases while the horizon value of $\phi$ jumps discontinuously. Close to the phase transition point, the deep IR geometry  (\ref{nearhor-nt}) or (\ref{nearhor-tt}) quickly flows to the critical Lifshitz solution in the intermediate IR region. 

The free energy density can be computed by adding standard holographic counterterms and is behaved 
continuously and smoothly at the critical value \cite{Landsteiner:2015pdh}. Note that the free energy does not depend on the Chern-Simons coupling constant. It does not probe the topological nature of the quantum phase transition,  in contrast to the anomalous Hall conductivity.
%%%%%%%%%%%%%%%%%%%%%%%%%%%%%%%%%%%%%%

%%%%%%%%%%%%%%%%%%%%%%%%%%%%%
\subsection{Anomalous Hall conductivity}
\label{subsecHall}
%%%%%%%%%%%%%%%%%%%%%%%%%%%%%

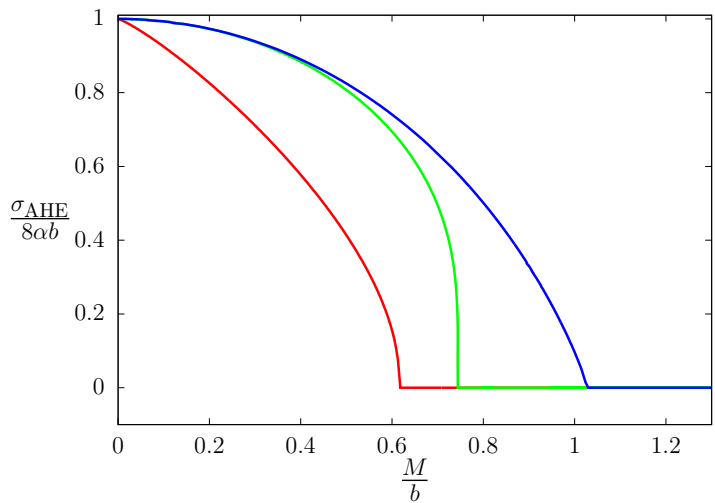
\begin{figure}[!htb]
\begin{center}
\resizebox{0.6\textwidth}{!}{\input{conductivity-differentparameters1.tex}}
\end{center}
\vskip -5mm
\caption{\small The zero temperature anomalous Hall conductivity is given by the value of the axial gauge field on the degeneratre horizon at $r=0$. It is plotted here for different model parameters, specifically $m^2  =-2$, $\lambda=1/10$ (red), $m^2  =-3$, $\lambda=1/10$ (green), $m^2  =-3$, $\lambda=1$ (blue). It can be observed how the critical value for the $M/b$ parameter where the conductivity goes to zero changes in the different cases. Figure from \cite{Copetti:2016ewq}.}
\label{fig:ahe}
\end{figure}

The essential hall-mark of the topological character of the Weyl semimetal state is the presence of anomalous Hall conductivity. In the case we are interested in it is anomalous Hall conductivity for the vector type current. A fast way of calculating it in holography is as follows. First one observes that the quantity
\begin{equation}
 j^\mu(r) = \mathcal{F}^{\mu r}+4\alpha\epsilon^{r \mu\beta\rho\sigma} A_{\beta} \mathcal{F}_{\rho\sigma}  %\Big)
\end{equation}
fulfills a radial conservation equations as a consequence of the holographic equations of motion in the bulk space-time 
\begin{equation}
\frac{d}{dr} j^\mu(r) = \partial^\mu  X \,.
\end{equation}
The precise form of $X$ is not important since from now on we integrate over space and take the zero frequency limit such
that the right hand side of this conservation equation vanishes. It follows then that in this situation the holographic expectation value of the current is given by the value of $j^\mu(r)$ at the horizon 
\begin{equation}
J^\mu = j^\mu(r_h)\,.
\end{equation}
Since due to the Bianchi identity the electric field is constant along
the AdS bulk direction $r$, the current at zero temperature is given by the Hall current \cite{Copetti:2016ewq}
\begin{equation}\label{eq:ahc}
J^x = 8 \alpha A_z(0) E_y\,.
\end{equation} 
The anomalous Hall conductivity is completely determined by the horizon value of the axial gauge field \cite{Landsteiner:2015lsa,Landsteiner:2015pdh}. 
In particular in holography it is only non-vanishing in the topological phase but vanishes at the quantum critical point and in the non-topological phase. 
This is exactly the same behavior the weak coupling Dirac like model shows. The anomalous Hall conductivity (\ref{eq:ahc})
is shown in figure \ref{fig:ahe} for different values of model parameters. 
%%%%%%%%%%%%%%%%%%%%%%%%%%%%%%%%%

%%%%%%%%%%%%%%%%%%%%%%%%%%%%%
\subsection{Universality of the quantum phase transition}
%%%%%%%%%%%%%%%%%%%%%%%%%%%%%

\begin{figure}[!htp]
\begin{center}
\resizebox{0.56\textwidth}{!}{\input{lambda-map1.tex}}
\end{center}
\vskip -5mm
\caption{\small The critical value of $M/b$ as a function of the quartic scalar self coupling $\lambda$ for different values of the axial charge $q=1$, $q=1.5$ and $q=2$. As can be seen the value of $M/b$ diverges at some finite values fo $\lambda$. This means that in these cases the scalar self interaction suppresses the phase transition to the trivial phase. Figure from \cite{Copetti:2016ewq}.}
\label{fig:universality}
\end{figure}
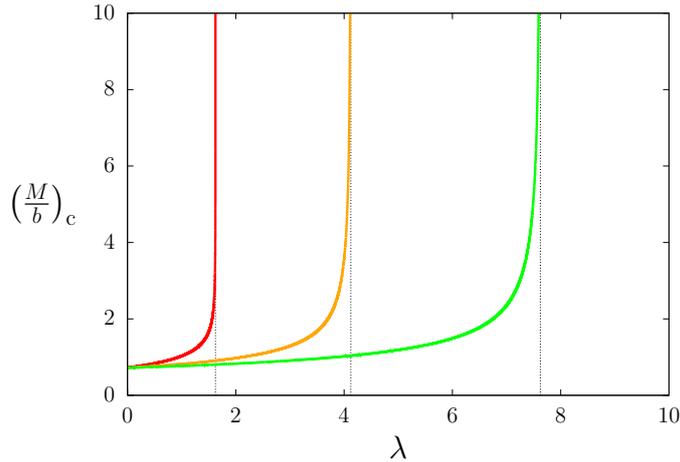

The precise value of $M/b$ at which the topological quantum phase transition from the Weyl semimetal to the trivial theory arises depends on the model parameters. Using the holographic duality one can investigate the critical values of $M/b$ as a function of the quartic scalar self coupling $\lambda$ for various values of the axial charge $q$ of the scalar field \cite{Copetti:2016ewq}. 
The result is shown in figure \ref{fig:universality}.
Interestingly for a given charge value there is a maximum value of $\lambda$ beyond which the phase transition does not occur at any finite value of $M/b$. This has an interesting interpretation in terms of the holographic duality. The spacetime curvature can be taken as a measure of the degrees of freedom. In the cases in which the phase transition cannot take place anymore it turns out that the holographic number of degrees of freedom in the infrared in the trivial phase would be larger than in the critical phase. Intuitively one expects that more degrees of freedom are gapped out in the IR than at the critical point. This intuition would be violated if the trivial phase could still be reached for high values of the self coupling. Fortunately direct inspection shows that this is not the case. 

\subsection{Holographic Fermi arcs}
One of the key signatures of Weyl semimetals is the presence of topologically protected surface states, the so-called Fermi-arcs. A simple and efficient field theory model of Fermi arcs follows from the thinking about the low energy description in terms of the deformed Dirac operator (\ref{eq:lorentzbreaking}). The effective separation of the Weyl cones in the Brillouin zone enters the Dirac equation like a gauge field with the key difference that it couples with different signs to left- and
right-handed Weyl fermions. If this parameter varies spatially it can induce an axial magnetic field $\mathbf{b}_5 =  \mathbf{\nabla} \times \mathbf{b}$. A standard argument shows now that such a spatial dependence is inevitable. Inside the Weyl semimetal we describe the system by a Dirac equation with axial gauge field $\mathbf{ A}_5 = \mathbf{b}$. But outside the material there are no
low energy states available for the electrons. This is equivalent to describing the outside by a Dirac equation with a very
large mass and vanishing axial gauge field. In turn this means that on the edge of the material there is necessarily a strongly localized axial magnetic field. Now one can invoke an index theorem that states that the number of zero modes in
the axial magnetic field is given by the degeneracy of the Lowest Landau level $|\mathbf{b}_5 |/(2\pi)$. In a usual magnetic field there
would be zero-modes of both chiralities but in the axial magnetic field the zero modes from both the right-handed and left-handed fermions have the same chirality. If one populates these zero modes by turning on a chemical potential they will lead to an edge current of the form
\begin{equation}\label{eq:ame}
{\mathbf J}_\mathrm{edge} = \frac{\mu}{2\pi^2}{\mathbf{b}}_5\,.
\end{equation}
This can also be viewed as an instance of an anomaly induced transport phenomenon, the axial magnetic effect \cite{Landsteiner:2016led}.

\begin{figure}[!thb]
\begin{center}
\includegraphics[width=0.67\textwidth]{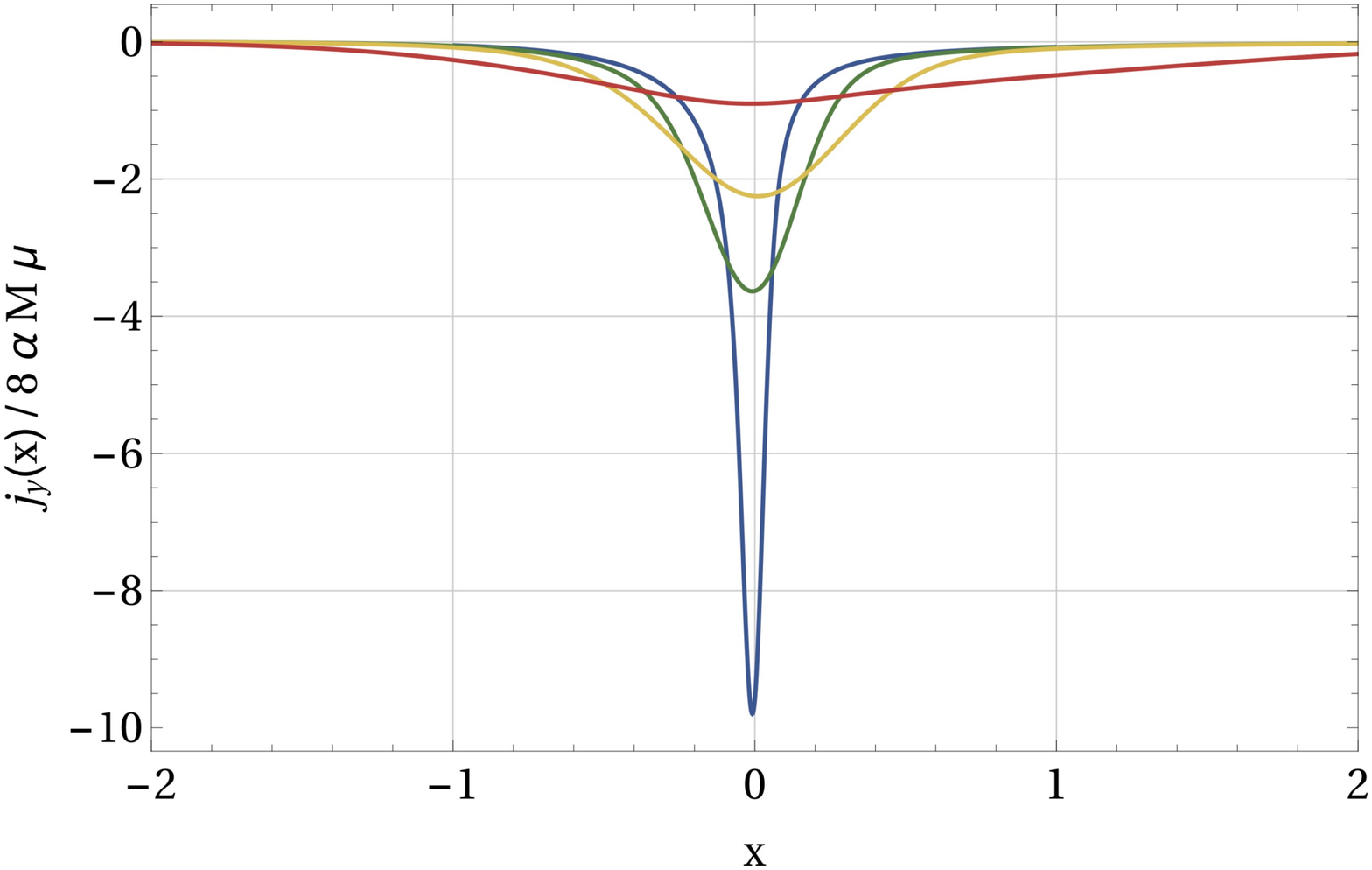}
\end{center}
\vskip -5mm 
\caption{\small Current distribution along the $x$-direction for different choices of interpolating functions with widths $l=0.1, 0.5,1,4$. As expected the current in localized in the region in which the effective low energy axial gauge field shows non-vanishing curl. This is the direct holographic counterpart of the Fermi-arcs associated to the surface of a Weyl semimetals. Figure with permission reproduced from \cite{Ammon:2016mwa}, \textcopyright 2017 by the American Physical Society.} 
\label{fig:edgecurrent}
\end{figure}

At strong coupling or more generally in the absence of quasiparticle excitations Fermi-arcs per se can not be expected to be seen. But the topologically protected edge currents should still exist\footnote{It might be however that
Fermi arcs exist also in the spectral functions of probe fermions. Indeed as we will review in the next section
probe fermions do carry the signatures of the non-trivial topology of momentum space.}. This is exactly what \cite{Ammon:2016mwa} investigated.
The authors numerically constructed solutions with spatial dependent boundary conditions of the form
\begin{equation}
A_z(r,x) \Big{|}_{r\rightarrow\infty} = 
\left\{
\begin{array}{cc}
b_L & ~\mathrm{for }\; ~~~x<-l\\
p(x) &
~~~~~\mathrm{for }\; -l\geq x\leq l \\
b_R & \mathrm{for }\; ~~~x>l
\end{array}
\right.
\end{equation}
where $p(x)$ is a suitably chosen smooth interpolating function. The scalar field was kept fixed and the chemical potential
is introduced as
\begin{equation}
\lim_{r\rightarrow \infty} r\Phi(r) = M ~~,~~ \lim_{r\rightarrow \infty}V_t = \mu\,,
\end{equation}
with the understanding that $V_t=0$ at the horizon. They found that indeed a current flows on the interface between the two
asumptotic regions and the total current in the $y$ direction is given by
\begin{equation}
J_y = 8 \alpha \mu (b_{\mathrm{eff},L} - b_{\mathrm{eff},L}) = 8\alpha \mu \left(\sigma_{\text{AHE},L} -  \sigma_{\text{AHE},R}\right)\,.
\end{equation}
This is exactly what one can expect since in the effective low energy theory $\int dx B_5 = b_{\mathrm{eff},L} - 
b_{\mathrm{eff},R}$. Moreover, the current distribution is concentrated on the interface as can be seen from figure \ref{fig:edgecurrent}.

%%%%%%%%%%%%%%%%%%%%%%%%%%%
\subsection{Fermionic probes and the topological invariant}
\label{secti}
%%%%%%%%%%%%%%%%%%%%%%%%%%%

Various bulk calculations have shown that the dual system should be a Weyl semimetal having Weyl cones with an effective momentum separation in the direction of $\mathbf{b}$. A direct observation of the Weyl cones needs to employ the holographic fermionic probes whose spectral function at zero temperature would show two poles separated in the momentum space in the $\mathbf{b}$ direction. Besides a direct observation of the two Weyl cones, an important further evidence is to compute the topological invariant for the dual strongly coupled topological semimetal states, which would also require a calculation of the Green's function of probe fermions in the bulk. In the following we will first introduce how topological invariants for strongly coupled systems could be calculated directly from the bulk fermionic Green's function and then show how fermionic spectral functions could be calculated for the holographic Weyl semimetal background and obtain the corresponding topological invariant accordingly.

Mathematically topological invariants are properties that are invariant under homeomorphisms. %, which could be numbers, e.g. the genus of a closed surface, or could also be groups, e.g. the fundamental group. 
%In the same way, 
Similarly in physics topological invariants can be defined for topological matter, which are invariant under adiabatic deformations of the Hamiltonian that protect the topology of the underlying system. 

For weakly coupled topological systems, in momentum space  we can define the topological invariants from the Bloch states, i.e. the eigenstates of the weakly coupled Hamiltonians. The Berry phase \cite{berry}, which is the phase accumulated along a closed loop $\gamma$ in the momentum space for the Bloch states $|n_{\bf k}\rangle$, is defined as 
$ \phi=\oint_\gamma \mathcal{A}_{\bf k}\cdot d {\bf k}$
where the Berry connection is determined by the eigenstates  
$|n_{\bf k}\rangle$ of the momentum space Hamiltonian as $
\mathcal{A}_{\bf k}=i \sum_{j}\langle n_{\bf k}|\partial_{\bf k}|n_{\bf k}\rangle$ with $j$ runs over all occupied bands. 
The Berry phase with value $0$ or $\pi$ is one simple example of a topological invariant.  

There is another way to compute the Berry phase. Using the Berry curvature $
\Omega_i=\epsilon_{ijl } \big( \partial_{k_j}\mathcal{A}_{k_l}-\partial_{k_l}\mathcal{A}_{k_j}\big)
$ associated to the the Berry connection and choosing a surface $S$ whose boundary is the closed loop $\gamma$, we have 
$
\phi=\int_S {\bf  \Omega}\cdot d{\bf S}\,.
$ 

An equivalent way to calculate the topological invariant is to use the Green's function 
\be
N(k_z)=\frac{1}{24\pi^2} \int dk_0 dk_x d k_y \text{Tr} \Big[\epsilon^{\mu\nu\rho z}G\partial_{\mu}G^{-1}G\partial_{\nu}G^{-1}G\partial_{\rho}G^{-1}\Big]\,,
\ee
where $\mu,\nu,\rho \in k_0,k_x,k_y$ and $k_0=i\omega$ is the Matsubara frequency. For free systems, the Green's function takes the form $G(i\omega, k)=1/(i\omega-h(k))$ where $h(k)$ is the Hamiltonian matrix $H=\sum_k c_k^\dagger h(k)c_k$. We can also use this formula to compute topological invariant for interacting systems, however, for strongly interacting systems it is difficult to compute practically since it involves an integration in the $i\omega$ direction.

Refs. \cite{Wang:2012ig,wang-prx,interaction1} shows that the  Green's function at zero frequency $G(0,{\bf k})$ contains all the topological information of the system. An effective topological Hamiltonian can be defined 
\be \label{eq:topH}
\mathcal{H}_t({\bf k})=-G^{-1}(0,{\bf k})
\ee
and the eigenvectors can be obtained from this effective topological Hamiltonian.  As long as there is no pole at nonzero $\omega$ in $G(i\omega,{\bf k})$, the topological invariants derived from the effective Hamiltonian $\mathcal{H}_t({\bf k})$ 
%as if the system is a weakly coupled theory described by $\mathcal{H}_t({\bf k})$ 
would be the same as those defined in the original system. Thus topological invariants can be defined using negative valued eigenvectors of $\mathcal{H}_t({\bf k})$, i.e. effective occupied states $n_{\bf k}$ with $\mathcal{H}_t({\bf k})|n_{\bf k}\rangle=-E_t |n_{\bf k}\rangle$ and $E_t>0$.

%When there is no pole in the imaginary frequency axis in the Green's function, the topological invariant can be computed from the weakly coupled formula using the effective topological Hamiltonian method. 

Having shown that topological invariants for strongly coupled fermionic systems could be directly calculated from the zero frequency Green's function, we now show how fermionic Green's function could be calculated for the holographic Weyl semimetal. In holography, probe fermions in the bulk have been studied for various backgrounds at finite density in 
\cite{Liu:2009dm, Cubrovic:2009ye} 
to obtain the dual fermion spectral functions. Here for the holographic Weyl semimetal, to probe the dual fermion spectrum we add a probe fermion on the background geometry (\ref{eq:ansatz}) and calculate the dual Green's functions from the holographic dictionary \cite{Liu:2018djq}.  One important difference here is that we work in five dimensions, and now a bulk four component spinor corresponds to a two component spinor of the dual field theory in four dimensions \cite{Iqbal:2009fd}. Therefore in the bulk we use two spinors $\Psi_1$ and $\Psi_2$ with opposite sign masses and axial charges and choose one with standard quantization while the other with alternative quantization to correspond to four component spinor with two opposite chiralities.

From the point of view of the dual field theory 
these probe fermions correspond to composite operators of a scalar field
with the fundamental fermions. A priori it is these fundamental fermions that carry the non-trivial topology.
As we will show now this topology is still present in the strongly coupled bound state that are the probe fermions.

The action of probe fermions is as follows, 
\bea\label{eq:probeDirac1}
S&=&S_1+S_2+S_\text{int}\,,\\
S_1&=&\int d^5x \sqrt{-g} i\bar{\Psi}_1\big(\Gamma^a D_a -m_f-i A_a \Gamma^a \big)\Psi_1\,,\nonumber \\
S_2&=&\int d^5x \sqrt{-g} i\bar{\Psi}_2\big(\Gamma^a D_a +m_f +i A_a \Gamma^a \big)\Psi_2\,,\nonumber \\
S_\text{int}&=&-\int d^5x \sqrt{-g}\big( i\eta_1\Phi\bar{\Psi}_1 \Psi_2+i \eta_1^*\Phi^*\bar{\Psi}_2 \Psi_1\big)\,,\nonumber
\eea  
where $
D_a=\partial_a-\frac{i}{4}\omega_{\underline{m}\underline{n},a}\Gamma^{\underline{m}\underline{n}}\,. 
$
The coupling constant in front of $A_z$ is opposite for the two spinors. Here $\Gamma^a=e_{\underline m}^{~a} \Gamma^{\underline m} $ with $\Gamma^{\underline m}$ the $\Gamma$-matrices in five dimensional Minkowski spacetime. 

From this form of bulk action for probe fermions, we can obtain the retarded Green's function from the boundary values of the two bulk fermionic fields at different momenta  \cite{Liu:2018djq}. In the simplest $M/b\to 0$ limit, it could easily been shown that two poles exist separately in the $z$ direction in the momentum space. At small $M/b$ limit, this could also be obtained with some semi-analytic method.

%When there is no pole in the imaginary frequency axis in the Green's function, the topological invariant can be computed from the weakly coupled formula using the effective topological Hamiltonian method. %Once we have obtained the topological Hamiltonian, the procedure would be the same as the weakly coupled case. 

As a simple example we first show how this procedure works for the pure AdS case, which of course would give a trivial topological invariant. In this case, in fact the system is degenerate at zero frequency, i.e. the two Weyl nodes coincide to form a Dirac node.  The fermionic retarded Green's functions for one chirality for $\omega>k$ has already been obtained in \cite{Iqbal:2009fd}. 
The topological Hamiltonian $\mathcal{H}_t$ is defined as $-G^{-1}(0,\bf{k})$ from (\ref{eq:topH}).  The two eigenvectors are $
| n_1\rangle=n_1^0\big(k_z+k,k_x+i k_y,0,0\big)^T$ and $ 
| n_2\rangle=n_2^0\big(0,0,k_z-k,k_x+i k_y\big)^T$
where $n_{l}^0=1/\sqrt{2k(k-(-1)^l k_z)}$ with $l\in \{1,2\}$. In fact these two eigenvectors are the same as the ones in the free massless Dirac Hamiltonian. $|n_1\rangle$ has positive chirality and is the eigenvector of the positive chirality Hamiltonian while $|n_2\rangle$ has negative chirality and is the eigenvector of the negative chirality Hamiltonian. 

The topological invariants can be calculated as follows. Around the Dirac node, we can define a sphere ${\bf S}$ to enclose it. 
% We first define a sphere ${\bf S}: k=k_0$ enclosing the Dirac node $k=0$ where $k_0$ is a constant.
On this sphere the system is gapped. The topological invariant  can be computed from 
$ C_l=\frac{1}{2\pi}\oint_S {\bf \Omega}_l \cdot d{\bf S}\,, $ where $
\Omega^i=\epsilon^{ijk }\mathcal{F}_{ij}$ with $(i\,,j\,,k)\in\{k_x\,,k_y\,,k_z\}$
 and $\mathcal{F}$ is the Berry curvature. Note that the topological number is an integer number and it stays as a constant when we deform the shape and exact shape and radius of the sphere without passing through a Dirac node. 
We can parameterize the sphere as ${\bf S}=k_0(\sin\theta\cos\phi,\,\sin\theta\sin\phi,\,\cos\theta)$ and we have 
${\bf \Omega}_l=(-1)^{l}{\bf e_\rho}/2k_0^2$. We obtain $C_1=-1$ for $| n_1\rangle$ 
and $C_2=1$ for $| n_2\rangle$. Then the total topological invariant is zero. This is due to the fact that the zero density state dual to pure AdS$_5$ is a Dirac semimetal. %only consists massless Dirac excitations. 

Now we continue to calculate the topological invariants for the holographic Weyl semimetal. In Weyl semimetals, we can define the topological invariant as the integration of Berry curvature on a closed surface ${\bf S}$ which encloses one of the Weyl nodes in the momentum space. This result will be insensitive to the shape and size of the closed surface. 
%When there is only one Weyl node inside the closed surface, the result does not depend on the exact shape and size of ${\bf S}$. 
From semi-analytic calculations, we obtained that when $M/b$ is very small,  the topological invariants are $\pm 1$ which are precisely the same as the results from weakly coupled WSM model. For larger $M/b$ numerics has to be involved. 
The topological invariant for finite temperature case has been studied in \cite{Song:2019asj}. The total topological invariants  are zero due to the Nielsen-Ninomiya theorem \cite{Nielsen}. 

In addition to the anomalous Hall conductivity and edge states, the nontrivial topological invariants serve as further nontrivial  
evidence that the holographic Weyl semimetal models are strongly coupled topologically nontrivial semimetals.

%%%%%%%%%%%%%%%%%%%%%%%%%%%
\subsection{Finite temperature, conductivities and viscosities}
%%%%%%%%%%%%%%%%%%%%%%%%%%%

In the previous subsections, the studies are mainly for the zero temperature case. Now we will turn to finite temperature physics and the interesting transport physics. 

We use the following ansatz to study the finite temperature solutions %with a regular horizon at a finite value of $r$ the following ansatz is used 
\cite{Landsteiner:2015pdh}
\begin{eqnarray}
\label{ansatzforfiniteT}
ds^2&=&-udt^2+\frac{dr^2}{u}+f(dx^2+dy^2)+h dz^2\,,\nonumber\\
A&=&A_z dz\,,~~~\Phi=\phi\,,
\end{eqnarray}
where all the fields $u, f, h, A_z, \phi$ are functions of $r$.  At the regular horizon $r=r_0$, $u$ has a simple zero whereas all these functions are analytic. This geometry is a black hole with horizon located at $r=r_0$ and Hawking temperature $4 \pi T =  u'(r_0)$. 
According to the holographic dictionary, the Hawking temperature of this black hole geometry corresponds to the physical temperature of the dual field theory. 
By using the scaling symmetries of the system and the constraints from the equations of motion near the horizon, there are only two independent dimensionless parameters, which can be parametrized by $M/b$ and $T/b$ in the UV. 

\begin{figure}[t]
\begin{center}
\begin{tikzpicture}
\draw [<->, thick] (0,4) -- (0,0) -- (6,0);
\draw[fill=white!20!white,draw=white] 
   (0.04,3.4) -- (0.04,0.06) -- (5.84,0.06)--(5.84,3.4)--(0.04,3.4);
\draw[fill=lightgray!15!white, draw=gray, dashed, thick,domain=1.2:4.8] plot (\x, {(4.6/1.8)*sqrt(abs(\x-3)))});
\draw node at (-0.2,-0.2){ $0$} node at (3,-0.4)  { $(M/b)_c$} node at (5.8,-0.4)  { $M/b$}
node at (-0.3,3.7)  { $T$};  
\draw node at (1.2,1.45){ Weyl}
node at (1.2,0.95){ semimetal}; 
\draw[black] node at (3,2.9){ quantum} node at (3,2.4){ critical}; 
\draw node at (4.8,1.45) { topologically}
node at (4.8,0.95) { trivial semimetal};
\draw [fill] (3,0) circle [radius=2pt];
\end{tikzpicture}
\caption{\small  The cartoon picture for the holographic Weyl semimetal at different temperature as a function of $M/b$.  %At $T=0$ there is a quantum phase transition at the critical value $(M/b)_c$.  At finite temperature the dashed line is a smooth crossover and in the quantum critical regime the physics is governed by the quantum critical behaviour. 
Figure from \cite{Landsteiner:2015pdh}.} 
\label{fig-cartoon}
\end{center}
\end{figure}
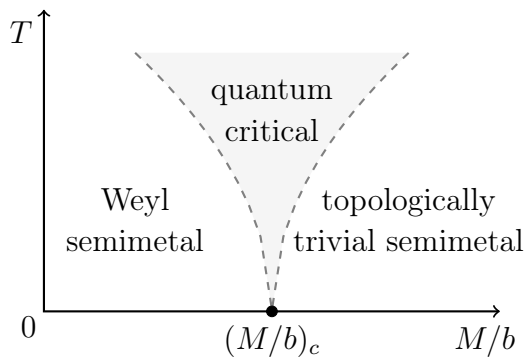

A cartoon illustration for the phases is shown in figure \ref{fig-cartoon} \cite{Landsteiner:2016stv}. 
At zero temperature the model undergoes the already discussed topological quantum phase transition between a topological semimetal
state and a trivial semimetal state. At the critical phase transition point %$M/b \simeq 0.744$, 
there is an emergent Lifshitz symmetry at zero temperature. %and it governs the quantum critical physics at finite temperature \cite{subir}. 
At finite temperature there is a quantum critical regime whose physics is governed by the Lifshitz symmetry. Meanwhile, 
this quantum phase transition becomes a smooth crossover behavior.  %The far IR physics is covered by a horizon at some finite value of  theholographic coordinate $r=r_0$. 

%%%%%%%%%%%%%%%%%%%%%%%%%%%%%%%%%%%%%%%%%%
\begin{figure}
\begin{center}
\includegraphics[width=0.6\textwidth]{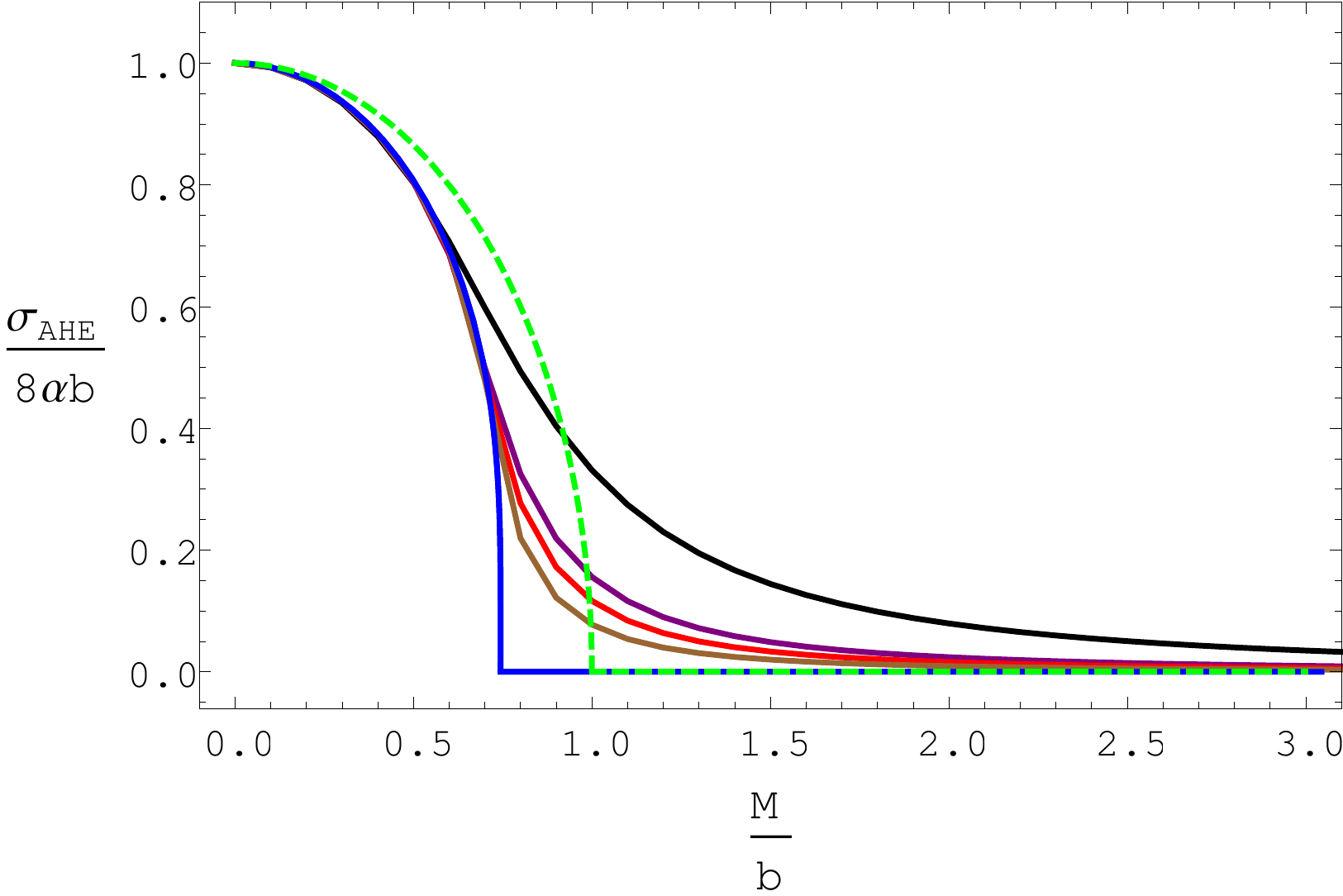}
\end{center}
\vskip -5mm 
\caption{\small Anomalous Hall conductivity as a function of $M/b$ at different temperatures. The solid lines are obtained from the holographic Weyl semimetal. For zero temperature a
sharp but continuous phase transition occurs at a critical value of $M/b$ (blue), which becomes a smooth crossover at finite temperature. The curves are for
$T/b=0.1$ (black), $0.05$ (purple), $0.04$ (red), $0.03$ (brown). The dashed (green) line is for the weak coupling model. Figure from \cite{Landsteiner:2015pdh}. 
}
\label{fig:ahe0}
\end{figure}

%%%%%%%%%%%%%%%%%%%%%%%%%%%%%%%%%%%%%%%%%%
Conductivities can be computed with Kubo formula via retarded correlation functions 
\begin{equation}\label{eq:sigma}
 \sigma_{mn} = \lim_{\omega\rightarrow 0}\frac{1}{i\omega} \langle J_m J_n \rangle (\omega,\mathbf{ k} =0)\,.
\end{equation}
According to the holographic dictionary, we can obtain the retarded Green's functions by studying the gauge field fluctuations around the background with infalling boundary conditions at the horizon. We obtain 
\begin{equation}
\sigma_\text{AHE}=8\alpha A_z(r_0) \,,~~~\sigma_{xx}=\sigma_{yy}= \sqrt{h (r_0)}\,. 
\end{equation}
From the fact that $r_0=0$ and $h(0)=0$ at zero temperature one concludes that the diagonal conductivities vanish. 
The anomalous Hall effect (figure \ref{fig:ahe0}) is completely determined by the horizon value of the axial gauge field. 
%%%%%%%%%%%%%%%%%%%%%%%%%%%%%%%%%%%%%
%%%%%%%%%%%%%%%%%%%%%%%%%%%%%%%%%%%%%

The longitudinal electric conductivity can be computed by studying the fluctuation $\delta V_z=v_z e^{-i\omega t}$ in the bulk.  At zero temperature we obtain $\sigma_{zz} =0$ and for finite temperature $ \sigma_{zz} = \left.\frac{f}{\sqrt{h}}\right|_{r=r_0}. $

\begin{figure}
\begin{center}
\includegraphics[width=0.6\textwidth]{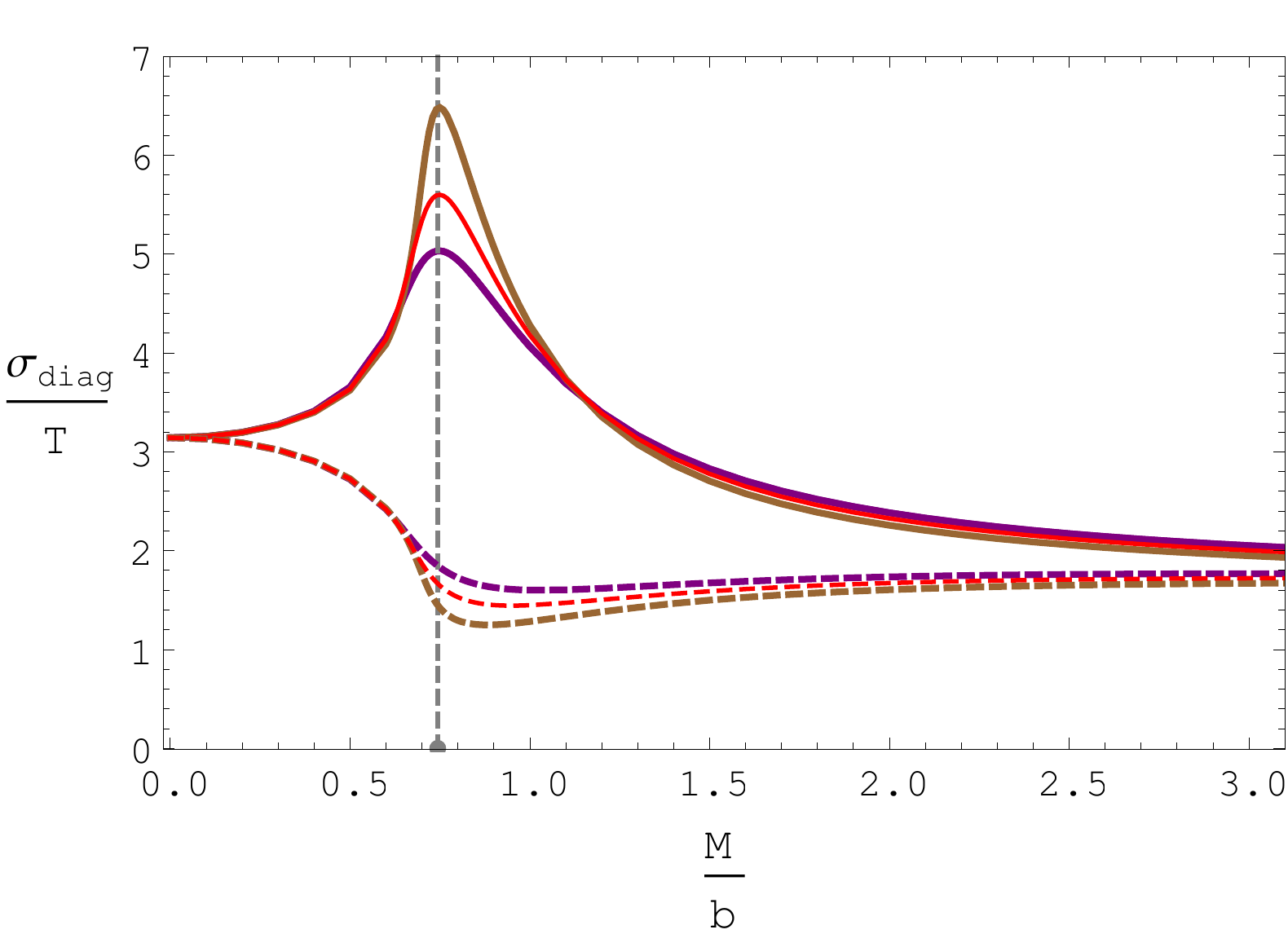}
\end{center}
\vskip -5mm 
\caption{\small The diagonal components of electric conductivities as a functional of $M/b$ for different temperatures. The solid lines are for $\sigma_{xx}=\sigma_{yy}$ and the dashed lines are for $\sigma_{zz}$ from holographic Weyl semimetal with $T/b= 0.05$ (purple), $0.04$ (red), $0.03$ (brown). The dashed gray line is the critical value of $M/b$ at the topological quantum phase transition. Figure from \cite{Landsteiner:2015pdh}. 
}
\label{fig:con}
\end{figure}

The diagonal components of electric conductivities at finite temperature as a function of $M/b$ is shown in figure \ref{fig:con}. We can see that at the critical value there is a peak (minimum) for the transverse (longitudianl) diagonal conductivities. %have a peak roughly at the critical value whereas the longitudinal one develops a minimum. The height of the peak
Both the height of the peak and depth of the minimum grow with temperature. At zero $M$ we have $\sigma_{xx,yy,zz}=\pi T$ and for large $M$ the conductivities $\sigma_{xx,yy,zz}= c \pi T$ with a temperature independent $c$ smaller than $1$. This is due to that in the trivial phase some but not all degrees of
freedom are gapped out. The quantum phase transition is between a topological semimetal and a trivial semimetal. 

%%%%%%%%%%%%%%%%%%%%%%%%%%%%%%%%%%%%%

Now let us explain the behavior of viscosities in this system. 
It is known that in an axisymmetric system which has time reversal symmetry breaking by vector ${\mathbf{b}}$ there are 
seven\footnote{These seven components includ three shear viscosities, two odd viscosities and two bulk viscosities. We focus on four of them and will not consider the other two bulk viscosities and one shear viscosity which are from the spin zero sector.} %components of $xx+yy$ and $zz$.} 
independent viscosities \cite{LandauLifschitz} in which 
two of them are independent odd viscosity tensor components. 
The viscosities can be computed from the Kubo formula
\be\eta_{ij,kl}=\lim_{\omega\to 0}\frac{1}{\omega} \text{Im}\big[G^R_{ij,kl}(\omega,0)\big]\,,
\ee
where the retarded Green's function of the energy momentum tensor
\be
G^R_{ij,kl}(\omega, 0)=-\int dt d^3x e^{i\omega t} \theta(t) \langle[T_{ij}(t,{\mathbf x}), T_{kl}(0, 0)]\rangle\,.
\ee
Since we chose ${\mathbf{b}} = b \hat e_z$,  the two shear viscosities \cite{ASZ, Hoyos:2014pba, Haehl:2015pja} are obtained from the symmetric part of the retarded Green's function under the exchange of $(ij)\leftrightarrow(kl)$
\be
\eta_\parallel=\eta_{xz,xz}=\eta_{yz,yz}\,,~~~~~~
\eta_\perp=\eta_{xy,xy}=\eta_{T,T}
\ee 
and the two odd or Hall components of viscosity are related to the antisymmetric part by
\be\label{eq:defoddvisoc}
\eta_{H_\parallel}=-\eta_{xz,yz}=\eta_{yz,xz}\,,~~~~\eta_{H_\perp}=\eta_{xy,T}=-\eta_{T,xy}
\ee
where the index $T$ denotes the component $xx-yy$. 
Note that $\eta_{H_\perp}$ is the odd or Hall viscosity in the plane orthogonal to $\mathbf{b}$ while 
$\eta_{H_\parallel}$ is specific to axisymmetric three dimensional systems.\footnote{There exists odd viscosity $\eta_{H_\parallel}$ 
by considering 
the coupling of elastic gauge fields to the electron gas in Weyl semimetals \cite{Cortijo:2016yph}. 
%which can be thought of as a property of the phonon gas arising through the specific electron-phonon Chern-Simons interactions. 
It was shown in \cite{Cortijo:2016yph} that this effective odd viscosity is related to the Hall conductivity of the electron gas and arises from the electronic point of view as an axial Hall conductivity. Here in holography the Hall viscosity should be viewed as an intrinsic property of the strongly coupled electron fluid.}

\begin{figure}[h!]
\begin{center}
\includegraphics[width=0.47\textwidth]{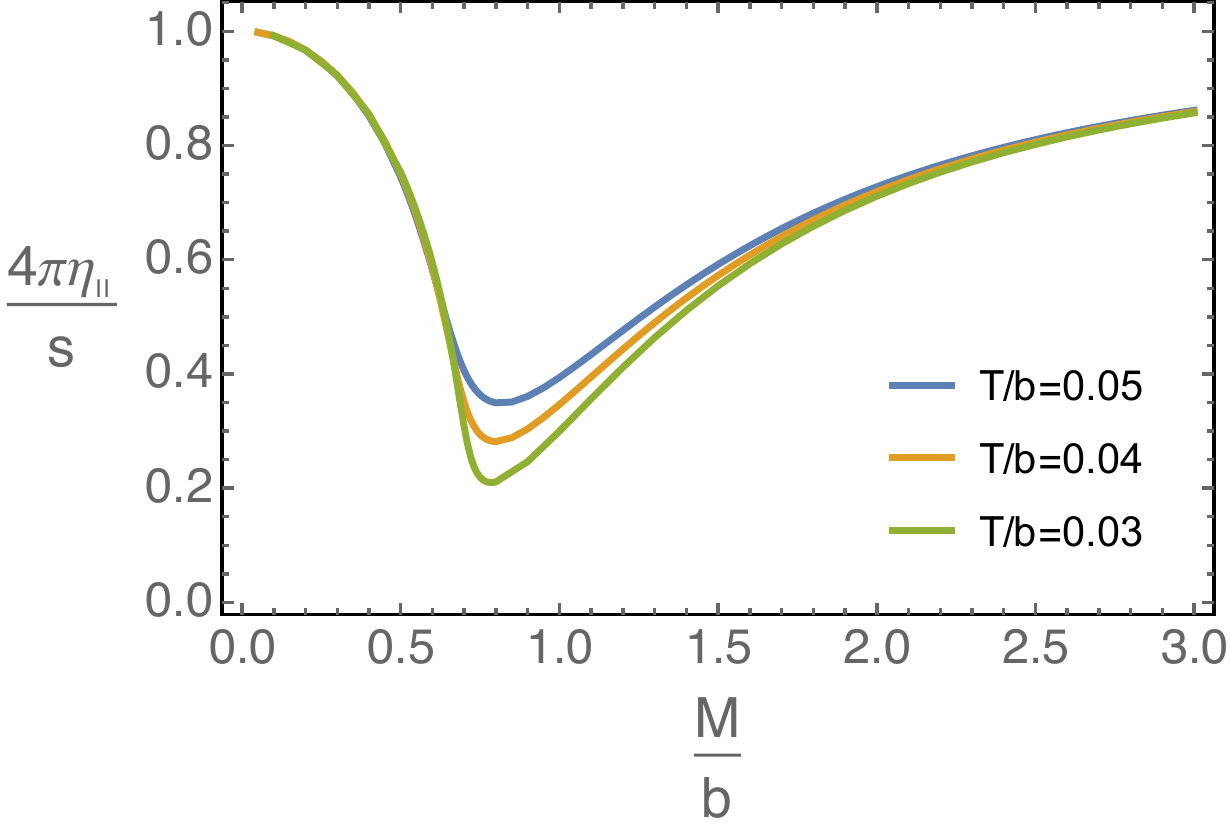}
\vskip -5mm 
\caption{\small The longitudinal shear viscosity over entropy density $4\pi \frac{\eta_\parallel}{s} $ as a function of $M/b$ at different temperatures. Figure from \cite{Landsteiner:2016stv}.}
\label{fig:shearparallel}
\end{center}
\end{figure}

In holography the viscosities can be computed via switching on the following perturbations
$\delta g_{iz} = h_{iz} (r) e^{-i\omega t} $ , $ \delta A_i = a_i(r) e^{-i\omega t}$ for $i\in\{x,y\}$.  
For the other components of viscosities can be computed by considering the perturbations $\delta g_{xx} - \delta g_{yy} = 2 h_T(r) e^{- i \omega t}$ , $\delta g_{xy} = h_{xy}(r)  e^{- i \omega t}$. 
From the holographic dictionary we obtain the following viscosity coefficients,
\begin{align}\label{eq:shearperp}
\text{dissipative~viscosity:}~~~&\eta_\parallel = \eta_{xz,xz} = \eta_{yz,yz} = \frac{f^2}{\sqrt{h}}\bigg{|}_{r=r_0}\\
&\eta_\perp = \eta_{xy,xy} = \eta_{T,T} = f\sqrt{h} \Big{|}_{r=r_0}\\
\label{eq:oddvisperp}
\text{dissipationless~odd~viscosity:}~~~& \eta_{H_\parallel} = \eta_{yz,xz} = -\eta_{xz,yz}  = 4 \zeta \frac{q^2 A_z \phi^2 f^2}{h} \bigg{|}_{r=r_0}\,\\
& \eta_{H_\perp} = \eta_{xy, T} =- \eta_{T,xy}= 8 \zeta q^2 \phi^2 f A_z \Big{|}_{r=r_0}\,.
\end{align}
The dissipative viscosity is a form of shear viscosity and it is interesting to express it normalized to the entropy density $\frac{\eta_\parallel}{s}  = \frac{f}{4 \pi h}|_{r=r_0}$. As can be seen from figure \ref{fig:shearparallel} the shear viscosity drops significantly below the standard result of KSS bound \cite{Kovtun:2004de}. In view of the various results of 
violation of the KSS bound in anisotropic theories \cite{Rebhan:2011vd, {Jain:2014vka}} this is not  
unexpected. Still it is very interesting to note that the shear viscosity reaches a minimum in the quantum critical region of $M/b \approx 0.744$.  
In contrast the transverse viscosity obeys the KSS bound is exactly  $\eta_\perp/s = 1/4\pi$.

\begin{figure}[h]
\begin{center}
\includegraphics[width=0.49\textwidth]{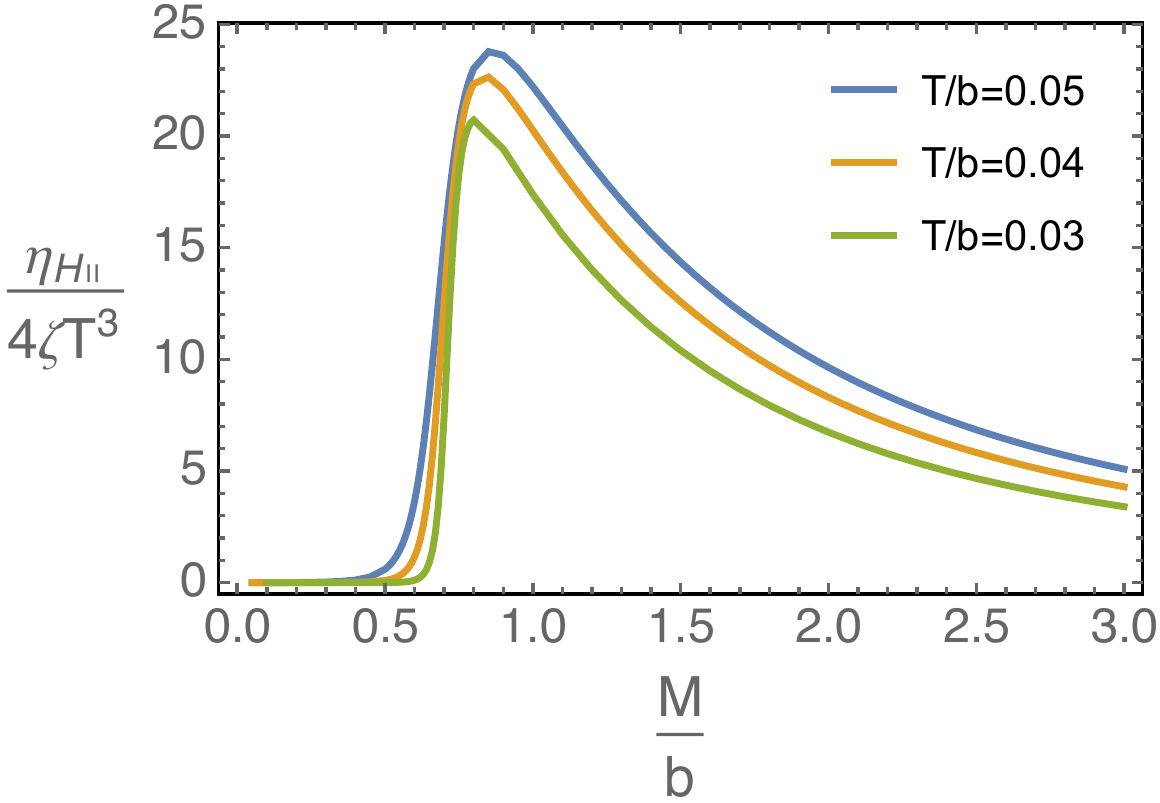}
\includegraphics[width=0.49\textwidth]{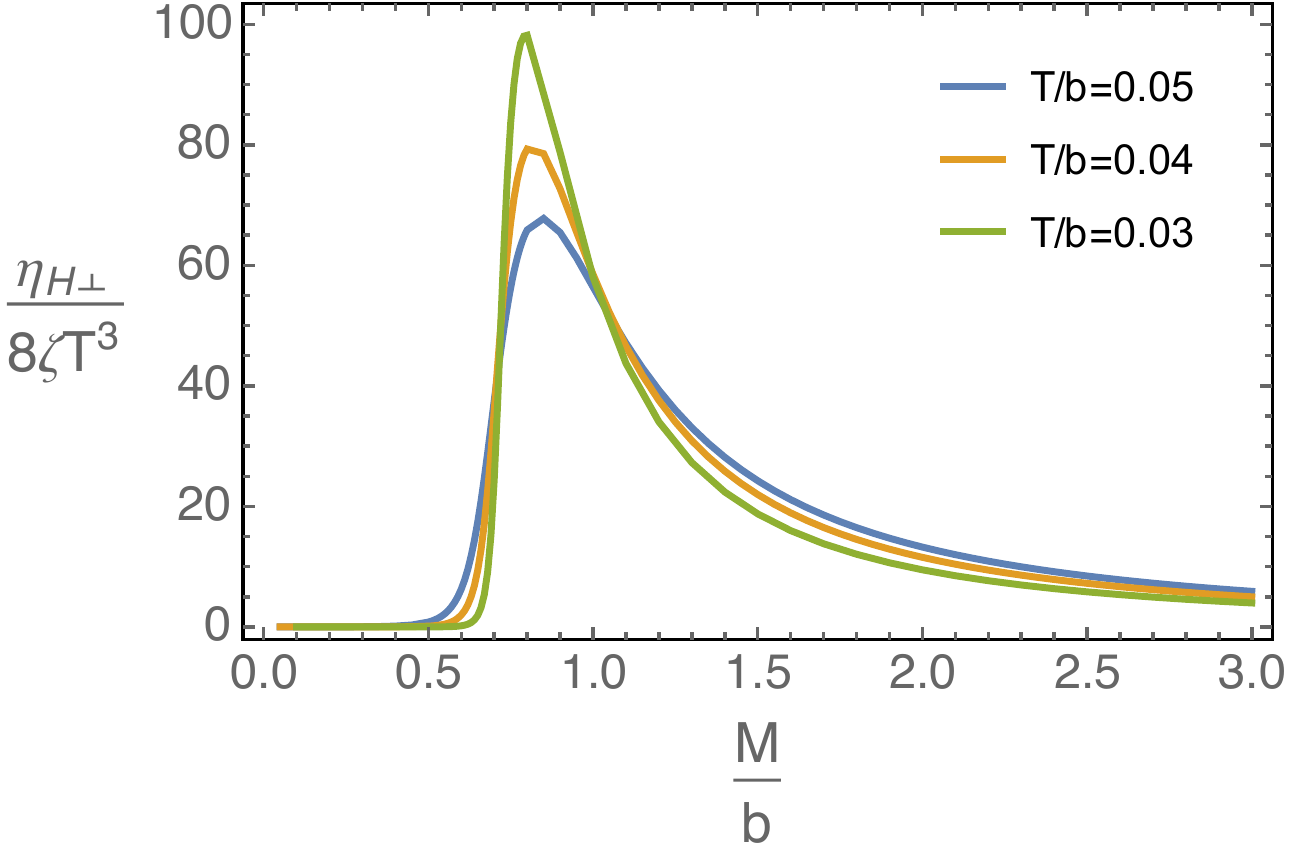}
\vskip -5mm 
\caption{\small Odd viscosity $\eta_{H_\parallel}$ and $\eta_{H_\perp}$ as a function of $M/b$ at different temperatures. 
Figure from \cite{Landsteiner:2016stv}.}
\label{fig:oddvisxzyz}
\end{center}
\end{figure}

A particular interesting fact about odd viscosities is that they are directly proportional to the mixed axial-gravitational anomalous constant which is the gravitational contribution to the axial anomaly $\zeta$ in (\ref{eq:holoanomaly}). Therefore at least in this holographic model they are
a new example of an anomaly induced transport coefficient.  
Fig \ref{fig:oddvisxzyz} shows the odd viscosities $\eta_{H_\parallel}$ and  $\eta_{H_\perp}$ as a function of $M/b$ at small but finite temperatures. In the topologically nontrivial phase the odd viscosity is highly suppressed. It rises steeply when $M/b$ enters into the quantum critical region, peaks around the critical value of $M/b$ and then falls off slowly when $M/b$ increases.
In the limit $M/b\to\infty$ 
the odd viscosity vanishes. 

The appearance of odd viscosity in the quantum critical region %of the phase transition between the Weyl semimetal and the trivial semimetal 
can be considered to be a prediction from holography. Its relation to the gravitational anomaly
suggests that this is a universal property. Indeed recently anomalous Hall viscosity has also been obtained in a weakly
coupled quantum field theory model of the quantum critical point in \cite{Copetti:2019rfp}. The relation to anomalies in the weakly
coupled theory is far from clear. We expect that further investigation of the holographic model and its RG flow to the
critical point can give valuable insight into the origin of this type of odd viscosity.

Note that from the analytic results on the viscosities and conductivities we obtain the non-trivial relation
\be
\frac{\eta_\parallel}{\eta_\perp} = \frac{2\eta_{H_\parallel}}{\eta_{H_\perp}}=\frac{\sigma_\parallel}{\sigma_\perp}=\frac{f}{h}\bigg{|}_{r=r_0} 
\ee
where $\sigma_\parallel=\sigma_{zz}=\frac{f}{\sqrt{h}}\big{|}_{r=r_0}, \sigma_\perp=\sigma_{xx}=\sigma_{yy}=\sqrt{h}\big{|}_{r=r_0}.$

\begin{figure}[h]
\begin{center}
\includegraphics[width=0.43\textwidth]{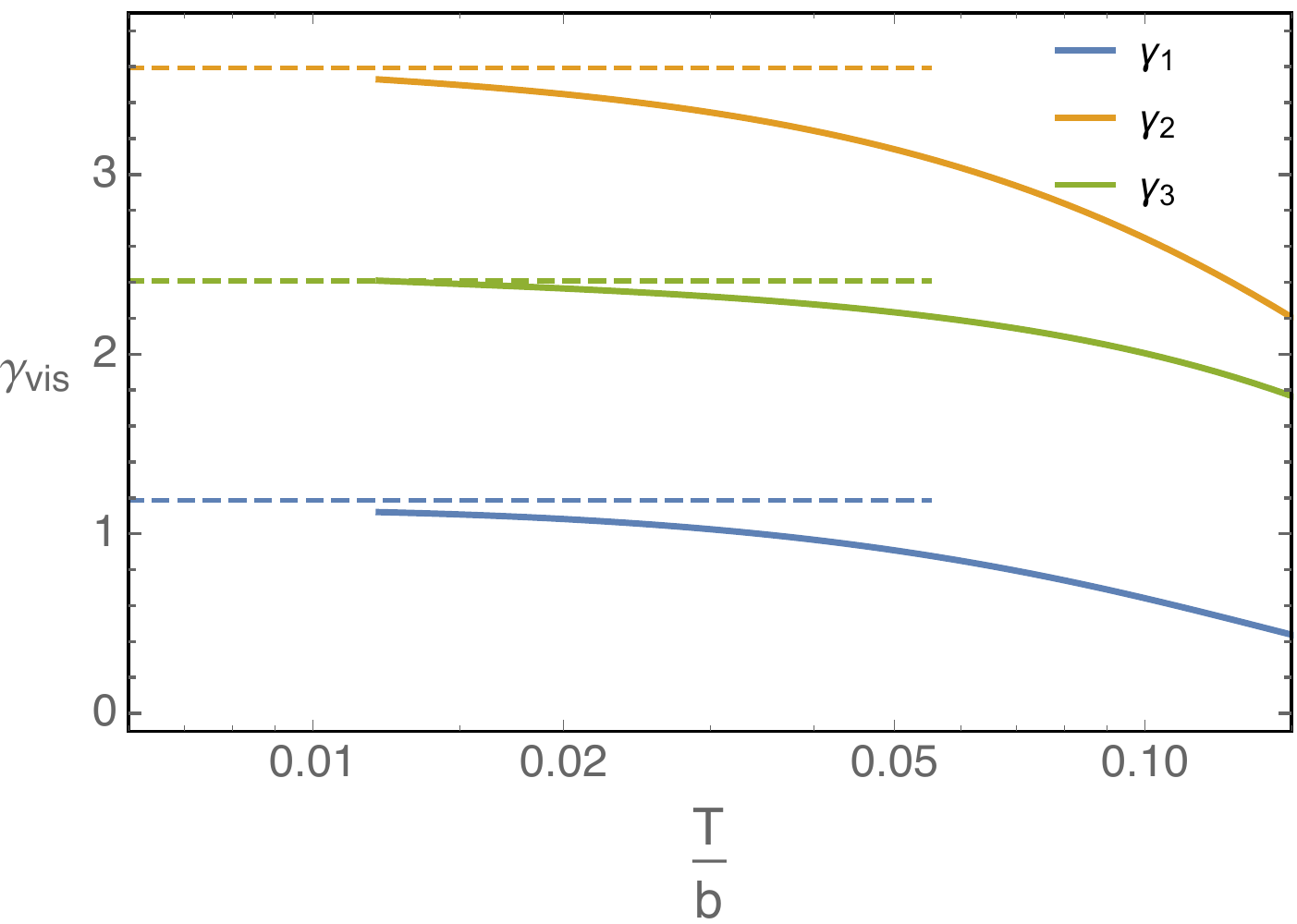}
\includegraphics[width=0.45\textwidth]{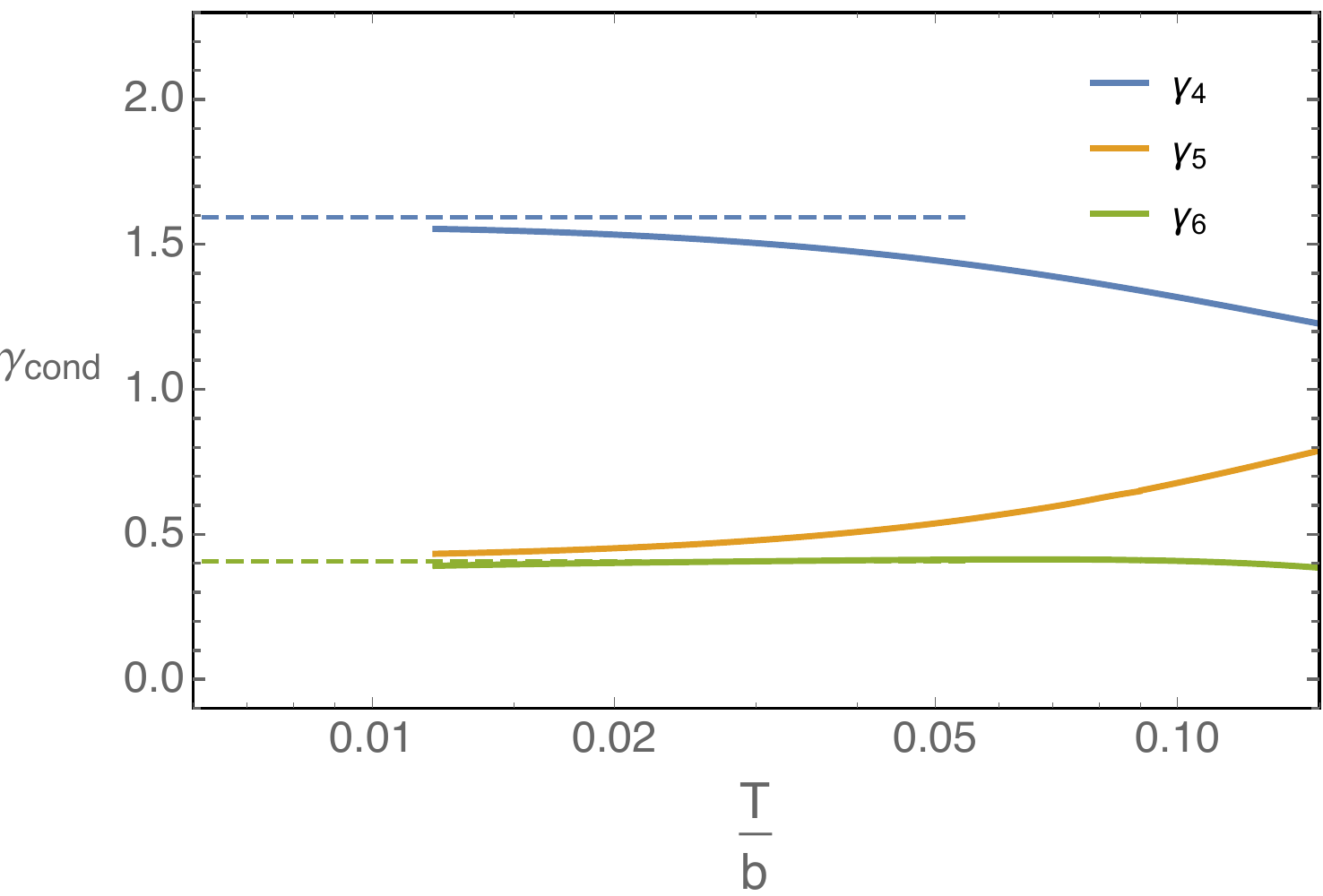}
\caption{\small The temperature scaling exponents $\gamma_i$ with $i\in\{1,\dots,6\}$  for viscosities $\eta_\parallel$, $\eta_{H_\parallel}$ and $\eta_{H_\perp}$ ({\em left}) and for conductivities $\sigma_\parallel$, $\sigma_\perp$ and $\sigma_\text{AHE}$ ({\em right}) in the quantum critical regime. The dashed lines represents the analytic values of the scaling exponents from the scaling analysis. Figure from \cite{Landsteiner:2016stv}.}
\label{fig:exponent}
\end{center}
\end{figure}

Furthermore,  in the quantum critical regime there exists interesting temperature scaling behaviour of conductivities and viscosities. At $T=0$, there is an emergent Lifshitz symmetry in the IR at the quantum phase transition point. The IR physics is invariant under $(t,x,y, r^{-1}) \to l (t,x,y, r^{-1}), z\to l^\beta z$ %$\beta\simeq 0.407$ 
and  $f\to l^{-2} f, h\to l^{-2\beta} h, A_z\to l^{-\beta} A_z, \phi\to \phi$ where $\beta$ is the anisotropic scaling exponent \cite{Landsteiner:2015pdh}.  At very low temperature, since $T\to l^{-1} T$ the temperature scaling dependence of the viscosities and conductivities near the critical region can be obtained from the scaling arguments.  At the critical regime, %when $M/b\to 0.744$, 
we have $\eta_\parallel/s\propto T^{\gamma_1}, \eta_{H_\parallel}\propto T^{\gamma_2},  \eta_{H_\perp}\propto T^{\gamma_3} $ with $(\gamma_1,\gamma_2, \gamma_3)=(2-2\beta, 4-\beta, 2+\beta)$ and $\sigma_\parallel\propto T^{\gamma_4}, \sigma_\perp\propto T^{\gamma_5},  \sigma_\text{AHE}\propto T^{\gamma_6} $ with $(\gamma_4,\gamma_5, \gamma_6)=( 2-\beta, \beta, \beta)$ for low temperatures. Figure \ref{fig:exponent} shows the temperature scaling exponents $\gamma_i$ with $i\in\{1,\dots,6\}$ of the numerical results at low temperatures at the critical value of $M/b$. 
At sufficient low temperature these scaling exponents approaches the analytic values from scaling analysics. Furthermore, the  scaling behaviors explain the peak/dip behaviors of the conductivities/viscosities of holographic Weyl semimetal in the quantum critical regime. 

The chiral vortical conductivity for the holographic Weyl semimetal has been calculated in \cite{Ji:2019pxx}. We can first compute the chiral vortical conductivity and then perform a suitable renormalization via the anomalous Hall conductivity and temperature squared. It was shown that at sufficiently low temperature this renormalized ratio stays as universal constants in both the Weyl semimetal phase and the quantum critical region. Furthermore, in the critical region the renormalized ratio is fully determined by the emergent Lifshitz scaling exponent at the critical point \cite{Ji:2019pxx}.

%%%%%%%%%%%%%%%%%%%%%%%%%%%
\subsection{Axial Hall conductivity}
%%%%%%%%%%%%%%%%%%%%%%%%%%%
Formally in quantum field theory the axial current can be coupled to an axial gauge field just as the electric current couples to the electric gauge
field. There is however a big difference in the possible dynamics of these fields. The dynamics of the true gauge field
is given by Maxwell's equations. For a gauge field that couples to an anomalous current, such as the axial current this is
mathematically inconsistent. A simple way of seeing this is to note that Maxwells equations imply that the divergence
of the current vanishes
\begin{equation}
\partial_\mu J^\mu = \partial_\mu \partial_\nu F^{\mu\nu} =0\,.
\end{equation}
In nature on a fundamental level anomalous currents are not coupled to gauge fields.

Nevertheless such fields can arise as effective fields in condensed matter systems. It has been shown in \cite{Cortijo:2016yph, Cortijo:2016wnf, Pikulin:2016wfj,Grushin:axial}
that axial electric and magnetic fields can be induced by applying strain on Weyl semimetals. 
These are effective low energy couplings in the theory and certainly do not obey Maxwell's equations and
consequently do not  jeopardize the consistency of the theory.  
Therefore it seems a legitimate physics question to ask if there is a purely axial analogue of the anomalous Hall effect
in Weyl semimetals. This again is a question that can be nicely addressed in holographic models and leads to some important
insights. 

Since the anomalous Hall effect is a direct consequence of the anomaly let us have another look to it and see what could
be expected. 
A remarkable fact is that the anomaly (\ref{eq:holoanomaly}) in the axial gauge fields is weaker by a factor of $1/3$ compared to the 
electromagnetic contribution. One useful way to think about this factor is to consider the origin of the anomaly
in a triangle Feynman diagram. In the case of the purely axial anomaly this is a diagram with three identical axial
currents on the vertices. Elementary Feynman rules instruct us therefore to multiply the diagram with a symmetry factor
of $1/3!$. In comparison the electromagnetic contribution comes from a triangle diagram with two electric currents
and one axial current. There are only two identical operators on the vertices and thus the symmetry factor is only $1/2!$
with gives a relative factor of $1/3$. 

The natural expectation is therefore, that the purely axial Hall conductivity, e.g. the transverse axial current induced by an axial electric current is weaker by a factor of $1/3$ compared to the electric Hall conductivity.

While the electric Hall conductivity can be computed in an easy way from the horizon data the calculation of the
axial Hall conductivity is more cumbersome. It is complicated by the fact that the axial symmetry is broken not only by
the anomaly but also by the expectation value of the scalar field (the dual of the mass term in the Dirac equation). 

The effect of this scalar field is that the axial background field $b$ is screened along the holographic direction.
Since the holographic direction encodes the RG flow we can define the analogue of a wave-function renormalization
factor by
\begin{equation}
\sqrt{Z_A} b^{\text{UV}} = b^{\text{IR}} \,.
\end{equation}
This implies that the axial Hall conductivity also suffers from this wave-function renormalization. Taking it into account one arrives at the prediction
\begin{equation}
\sigma^5_{\text{AHE}} = \frac{Z_A}{3} \sigma_{\text{AHE}} \,.
\end{equation}
In other words the axial Hall conductivity is exactly $1/3$ of the electric Hall conductivity once the wave-function 
renormalisation of the axial gauge fields in IR is taken into account. 
This was investigated in \cite{Copetti:2016ewq} and  indeed found to be correct. 
Moreover since the prediction that the axial Hall conductivity is $1/3$ of the electric Hall conductivity is a 
fundamental property of the theory it should hold for all states. Again this can be checked by using the finite temperature
backgrounds and indeed it was found that the relation holds exactly and independent of the temperature.

\subsection{Disorder}
Disorder is an integral component of any real condensed matter system. It is therefore not only interesting
but also mandatory to study the effects of disorder even in semi-realistic models. In
the case of the holographic Weyl semimetal this has been initiated in \cite{Ammon:2018wzb}. The authors studied the effect
of disorder in form of random Gaussian noise in the boundary value of the axial gauge field
\begin{equation}
\lim_{r\rightarrow \infty} A_z(r) = b_0 + 2 \gamma \sum_{i=1}^{N-1} \sqrt{S(k_i)}\sqrt{\Delta k} \cos(k_i x +\delta_i)
\end{equation}
with equally distributed momenta $k_i = i k_0/N$, and random phases $\delta_i$.
The analysis was restricted to the so-called decoupling limit of holography in which the backreaction of
the bulk matter fields on the AdS metric is neglected. Nevertheless the authors find rather interesting
signatures of disorder on the quantum phase transition.  
In general the quantum phase transition is smeared due to the disorder. They also show the appearance of
rare regions and indications of log-oscillatory structures in the Hall conductivity. 

\subsection{AC conductivities}
\label{sec:accon}
The electrical AC conductivity in a holographic Weyl semimetal model was investigated in \cite{Grignani:2016wyz}. A particularly
interesting effect was pointed out in relation to the quantum critical behavior near the phase transition.
On general grounds one expects the (zero temperature) optical conductivity to scale linearily with the frequency
for low enough frequencies
\begin{equation}
\sigma(\omega) = c \omega\,.
\end{equation}
This is simply enforced by the scaling symmetry of the Weyl fermions. The constant $c$ is proportional to the number
of active Weyl fermions. It was pointed out in \cite{Grignani:2016wyz} that this can get modified for higher frequencies near the
quantum phase transition. The frequency dependent optical conductivity enters then the quantum critical region whose
scaling properties are determined by the scaling exponents of the Lifshitz critical point at the phase transition.
The expected change in scaling is
\begin{equation}
\sigma_\parallel(\omega) \propto \omega^{2-\beta} \,,~~~~ \sigma_\perp \propto \omega^\beta \,. \label{eq:ACsigmas}
\end{equation}
Most interestingly such a sudden change in frequency dependence of the optical conductivity at low temperature
of the Weyl semimetal TaAs was experimentally observed in \cite{expAC}. The authors of \cite{Grignani:2016wyz} suggest that this might
be explained by assuming that that one enters the quantum critical region in TaAs at  a frequency around $30$meV.
They find that a fit to the data gives a scaling exponent of $\beta=0.14$ for the transverse conductivity in the
Lifshitz quantum critical region. It should be noted that there are also other candidate explanations, such as activating
of additional Weyl points at higher frequencies. Nevertheless the predicted change in scaling of the optical conductivity
once the frequency is high enough to enter the quantum critical regime seems a robust prediction and is in principle
accessible by experiments. It would be very interesting to see then if the scaling exponents of longitudinal and
transverse optical conductivities can be fitted to weak coupling models or to predictions from holographic models.

%%%%%%%%%%%%%%%%%%%%%%%%%%%
\subsection{Butterfly velocity}
\begin{figure}[h]
\begin{center}
\includegraphics[width=0.43\textwidth]{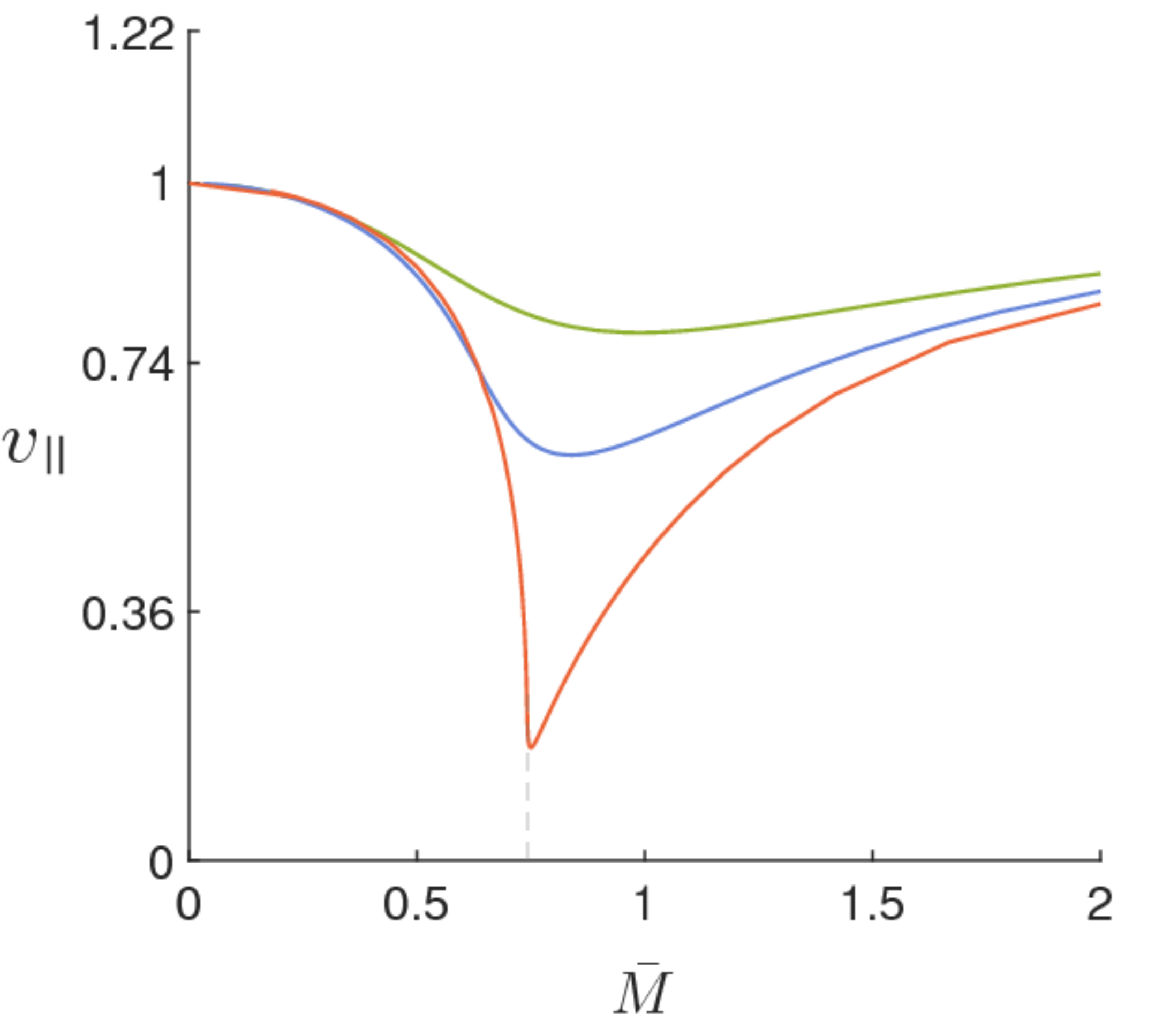}
\includegraphics[width=0.45\textwidth]{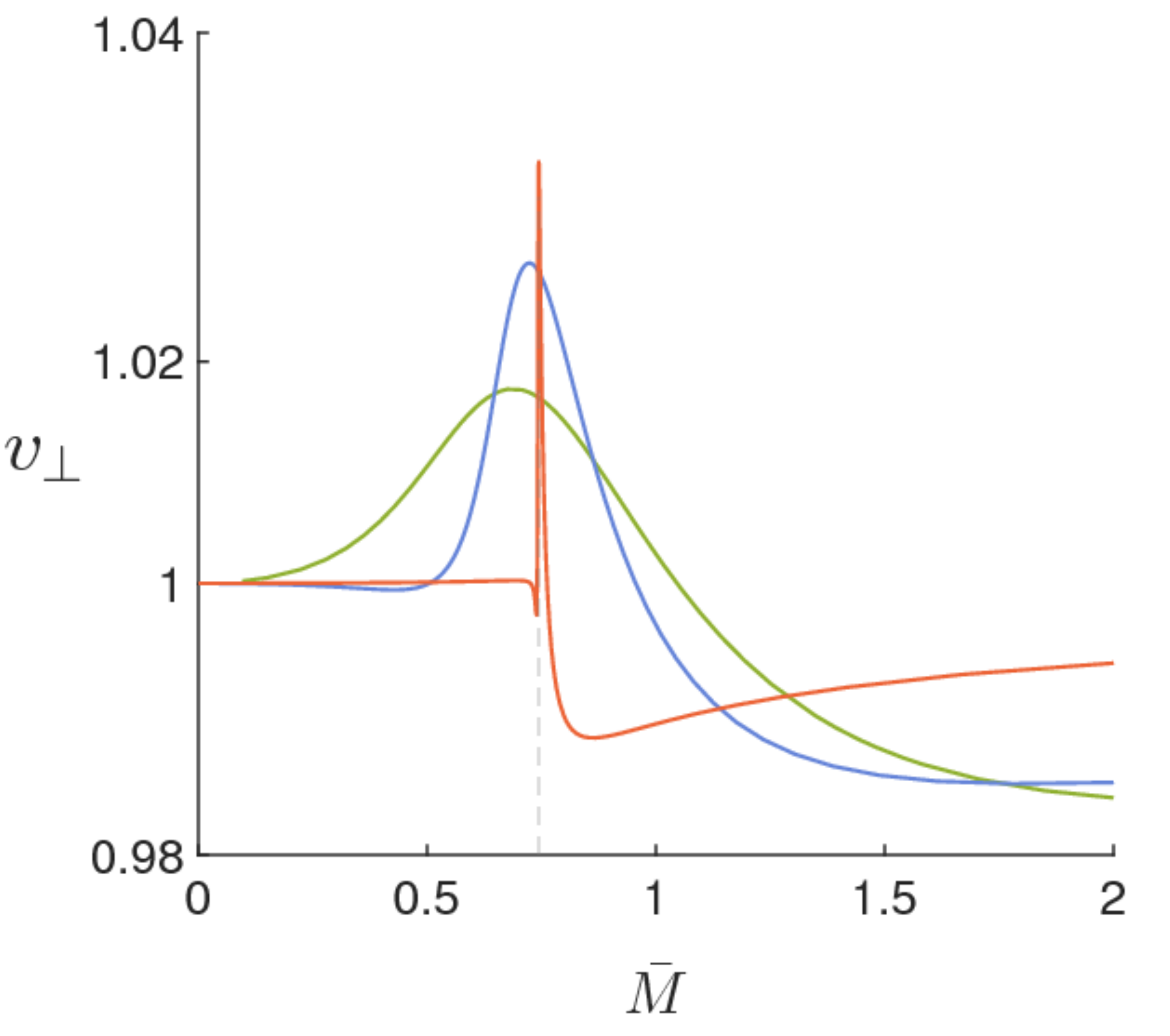}
\caption{\small Butterfly velocities as function of the dimensionless parameter $\bar M = M/b$ for different
values of the Temperature.
On the left panel the parallel butterfly velocity is shown and on the right panel the perpendicular butterfly velocity.
As one can see the results are quite different, whereas the parallel one has a minimum the perpendicular one develps a 
minimum. The lines show different temperatures $T/b = 0.005, 0.05, 0.1$ corresponding to the colors red, blue and green. 
Figure from \cite{Baggioli:2018afg} \href{https://creativecommons.org/licenses/by/4.0/}{CC BY 4.0}. 
}
\label{fig:butterfly}
\end{center}
\end{figure}

%%%%%%%%%%%%%%%%%%%%%%%%%%%
Holography has also contributed in recent years to the understanding of chaos in quantum many body systems.
Quantum chaos can be characterized by the late time behavior of the out of time order correlation function (OTOC)
\begin{equation}
\langle [ {\mathcal V}(t,\mathbf{x}), {\mathcal W}(0,0)]^2\rangle \sim%\approx 
e^{ \lambda_L\left(t-t^* - |\mathbf{x} |/v_B\right)}\,.
\end{equation}
Here $\lambda_L$ is the Lyapunov exponent, $t^*$ the so-called scrambling time and $v_B$ the Butterfly velocity.
The Lyapunov exponent obeys the bound \cite{Maldacena:2015waa}
\begin{equation}
\lambda_L \leq 2\pi T%\frac{2\pi}{\beta}\,,
\end{equation}
where $T$ is the temperature of the system. %$\beta$ is the inverse temperature. 
It is saturated by holographic field theories. For a review on
quantum chaos and holography see \cite{Jahnke:2018off}.
Of particular interest is the behavior of the Butterfly velocity across a quantum phase transition.
This question was addressed for the holographic Weyl semimetal model in \cite{Baggioli:2018afg}. 
There the authors computed the Butterfly velocity in a holographic model with metric
\begin{equation}
ds^2 = - g_{tt}(r) dt^2 + g_{rr} dr^2 + h_\perp(r) d\mathbf{x}_\perp^2 +h_\parallel(r) d\mathbf{x}_\parallel^2\,.
\end{equation}
This is precisely the metric that arises in the holographic Weyl semimetal where we chose $\mathbf{x}_\parallel = z$
and $\mathbf{x}_\perp = (x,y)$.
The result can be seen in figure \ref{fig:butterfly}.

It shows the surprising feature that the parallel component of the Butterfly velocity has a minimum whereas the
perpendicular component shows a maximum at the quantum phase transition. The Butterfly velocity therefore does not
show a universal behavior across the phase transition. This motivates the authors of \cite{Baggioli:2018afg} to introduce a new
quantity, the information screening length $L=1/\mu$. It is defined as follows. First one introduces
$M_\eta = \lambda_l/v_B^\eta$ where $\eta \in \{\perp,\parallel\}$ and defines then
\begin{equation}
\mu^2 = \frac{1}{L^2} = \frac{M_\eta^2}{h_\eta(r_h)} \,.
\end{equation}
This definition manages to get rid of the anisotropy such that a unique $L$ independent of the direction can be defined.
The authors show that $L$ is maximal at the quantum critical point and they conjecture that the information screening length
obeys
\begin{equation}
2 L \leq \frac{1}{D_\perp + \beta D_\parallel} = 2 L_c
\end{equation}
where $\beta$ is the Lifshitz scaling exponent of the anisotropic directions, $D_\perp$ is the number of spatial
dimensions with scaling exponent equal to one and $D_\parallel$ is the number of spatial directions with scaling $\beta$. 
They also point out that in holography the null energy condition restricts the scaling $\beta \leq 1$.

%%%%%%%%%%%%%%%%%%%%%%%%%%%
\section{Weyl semimetal/Chern insulator transition}
\label{sec:WSMinsulator}
%%%%%%%%%%%%%%%%%%%%%%%%%%%
 
The phase diagram for Weyl semimetal is richer and it could go through a phase transition to a normal band insulator \cite{burkov1, Roy:2016rqw} or to a Chern insulator \cite{burkov1,cm-1,Roy:2016amv} etc. However,    the holographic Weyl semimetal model in section \ref{sec:holoWSM} goes through a quantum phase transition to a trivial semimetal phase in which only a part of degrees of freedom has been gapped out. %In condensed matter physics, from weakly coupled theory Weyl semimetals could go through a phase transition to a normal band insulator \cite{burkov1, Roy:2016rqw} or to a Chern insulator \cite{burkov1,cm-1,Roy:2016amv} etc. 
This section reviews a holographic model describing a topologically nontrivial Weyl semimetal goes through a quantum phase transition to an insulating phase where all the degrees of freedom are gapped. 
 
A generation of holographic Weyl semimetal model can be constructed as follows. We use the Stueckelberg trick to replace the complex scalar field $\Phi$ in (\ref{eq:holomodel}) by two real scalar fields $\phi$ and $\theta$ 
and introduce the general dilatonic couplings in front of the kinetic terms \cite{Liu:2018spp}
\bea
S
&=&\int d^5x\sqrt{-g}\bigg[\frac{1}{2\kappa^2}\bigg(R+12-\frac{1}{2}(\partial\phi)^2-V(\phi)\bigg)-\frac{Y(\phi)}{4}\mathcal{F}^2-\frac{Z(\phi)}{4}(F)^2\nn\\
&&~~~~~\label{eq:action}
+
\frac{\alpha}{3}\epsilon^{abcde}A_a\Big(F_{bc}F_{de}+3 \mathcal{F}_{bc} \mathcal{F}_{de}\Big)-\frac{W(\phi)}{2}(A_a-\partial_a\theta)^2
\bigg]\,.
\eea
Up to the anomaly the action (\ref{eq:action}) is invariant under $\theta\to \theta+\chi, ~A_a\to A_a+\partial_a\chi$. One can show that the dual Ward identity for the conserved currents, which should be independent of the coupling strength of the system, is exactly the same as the one obtained from weakly coupled theory. 
The holographic Weyl semimetal model in section \ref{sec:holoWSM} can be recovered via defining $\Phi=\frac{1}{\sqrt{2}}\phi e^{i\theta}$ which is axially charged under the axial gauge field and setting $Y(\phi)=Z(\phi)=1, ~W(\phi)=q^2\phi^2, ~ V(\phi)=\frac{m^2}{2}\phi^2$.\footnote{Studies on constructing insulating phases from generic holography with dilatonic coupling can be found in e.g. \cite{Kiritsis:2015oxa,Charmousis:2010zz}. }

This section will focus on the physics at zero temperature. At $T=0$ we use the same 
ansatz for the background fields as (\ref{eq:ansatz}) in section \ref{sec:holoWSM}.  
To realise a holographic Weyl semimetal/insulator transition, 
the following dilatonic couplings are chosen 
\be\label{eq:modelpara}
Z(\phi)=1 \,,~~~W(\phi)=-q_0 \Big[1-\cosh\big[\sqrt{\frac{2}{3}}\phi\big]\Big]\,,~~~V(\phi)= \frac{9}{2}\Big[1-\cosh\big[\sqrt{\frac{2}{3}}\phi\big]\Big]\,,
\ee
and \be\label{eq:formY}
Y(\phi)=\cosh\Big[\sqrt{\frac{2}{3}}\phi \Big]\,.
\ee
Note there is a Z$_2$ symmetry $\phi\to -\phi$ for (\ref{eq:action}). When $\phi\to 0$, $W(\phi)\simeq\frac{q_0}{3}\phi^2$ and $V(\phi)\simeq -\frac{3}{2}\phi^2.$ Thus $q_0$ plays a similar role as the axial charge. In the following we restrict to $q_0>0$. Near the conformal boundary, $\phi\to 0$, the potential in (\ref{eq:modelpara}) behaves as 
$V(\phi)=\frac{1}{2}m^2 \phi^2+\dots$
with $m^2=-3$. 
As already discussed in section \ref{sec:holoWSM}, close to the  boundary ($r\to\infty$) we have
$
\phi=\frac{M}{r}+\dots, A_z=b+\dots.
$ %Note that $M$ and $b$ play the same roles as the sources of the scalar operator $\bar{\psi}\psi$ and chiral current $\bar{\psi}\gamma^5\gamma^z\psi$. Turning on these two sources, the dual field theory has the same structure as the weakly coupled field theory.

At zero temperature, we first give the near horizon solutions.  Similar to the holographic model in section \ref{sec:holoWSM}, there are again three different kinds of near horizon geometries. Then we turn on irrelevant perturbations to get the full solutions.

\noindent{\em Insulating phase.} The first kind of near horizon geometry is 
\bea
\label{eq:in1}
u=r(1+r)\,,~~
h=r(1+r)\,,~~
A_z=a_1 r^{\frac{1}{4}(\sqrt{1+8q_0}-1)}\,,~~\phi=-\sqrt{\frac{3}{2}}\log \frac{r}{1+r}\,,
\eea
where $a_1$ is a free parameter which will make the geometry flow to AdS$_5$ with different value $M/b$.  At the leading order, the geometry is known as the GPPZ  gapped geometry \cite{Girardello:1999hj}. Here we have a nontrivial $A_z$ such that we will have an anisotropic geometry. %There is a hard gap in the real part of diagonal optical conductivities while the anomalous Hall conductivity is nonzero at zero frequency, therefore this phase corresponds to a Chern insulator phase.

\noindent{\em Weyl semimetal phase.} The second kind of  geometry near the horizon is
\bea\label{eq:wsm1}
u&=&r^2 \,,~~~
h=r^2\,,~~~
A_z=a_0+\frac{\phi_0^2}{4a_0 r}e^{-\frac{2a_0\sqrt{q_0}}{\sqrt{3}r}}\,,~~~
\phi =\frac{\phi_0}{r^{3/2}}e^{-\frac{a_0\sqrt{q_0}}{\sqrt{3}r}}\,.
\eea
At the leading order, the IR geometry is an AdS$_5$ with a constant $A_z$. $a_0$ can be rescaled to $1$. The exponential terms play the role of the irrelevant perturbations. 
This near horizon geometry also shows up as the holographic Weyl semimetal phase in section \ref{sec:holoWSM}. 

\noindent{\em Critical point.} The third kind of near horizon geometry is
\bea\label{eq:cp1}
&&u=u_0 r^2 \big(1+\delta u r^{\alpha_c} \big)\,,~~h=\frac{q_0}{9}r^{2\beta}\big(1+\delta h r^{\alpha_c}\big)\,,\\
&&A_z=r^\beta\big(1+\delta a r^{\alpha_c} \big)\,,~~~\label{eq:cp4}
\phi=\sqrt{\frac{3}{2}} (\log\phi_1) \Big(1+\delta \phi r^{\alpha_c} \Big)\,.
\eea
In the case of $q_0=15$, we have $(u_0,\beta,\phi_1,\alpha_c)\simeq (1.150,0.769,1.797,1.230)$ and
$(\delta u,\delta h,\delta a)\simeq(0.147,-1.043,0.591)\delta \phi$. At the leading order the system has a Lifshitz symmetry $(t, x, y, r^{-1})\to c (t, x, y, r^{-1})$, $z\to c^\beta z$. We can use it to set $\delta\phi=-1$. The irrelevant perturbations can flow the above geometry to AdS$_5$. In the boundary we get $(M/b)_c\simeq0.986$. For arbitrary $q_0>0$, all the relevant perturbations around the above fixed point have complex scaling exponent, indicating that this fixed point is unstable \cite{Hartnoll:2011pp, Donos:2012js} which will be confirmed by studying the free energy.

\begin{figure}[h!]
\begin{center}
\includegraphics[width=0.46\textwidth]{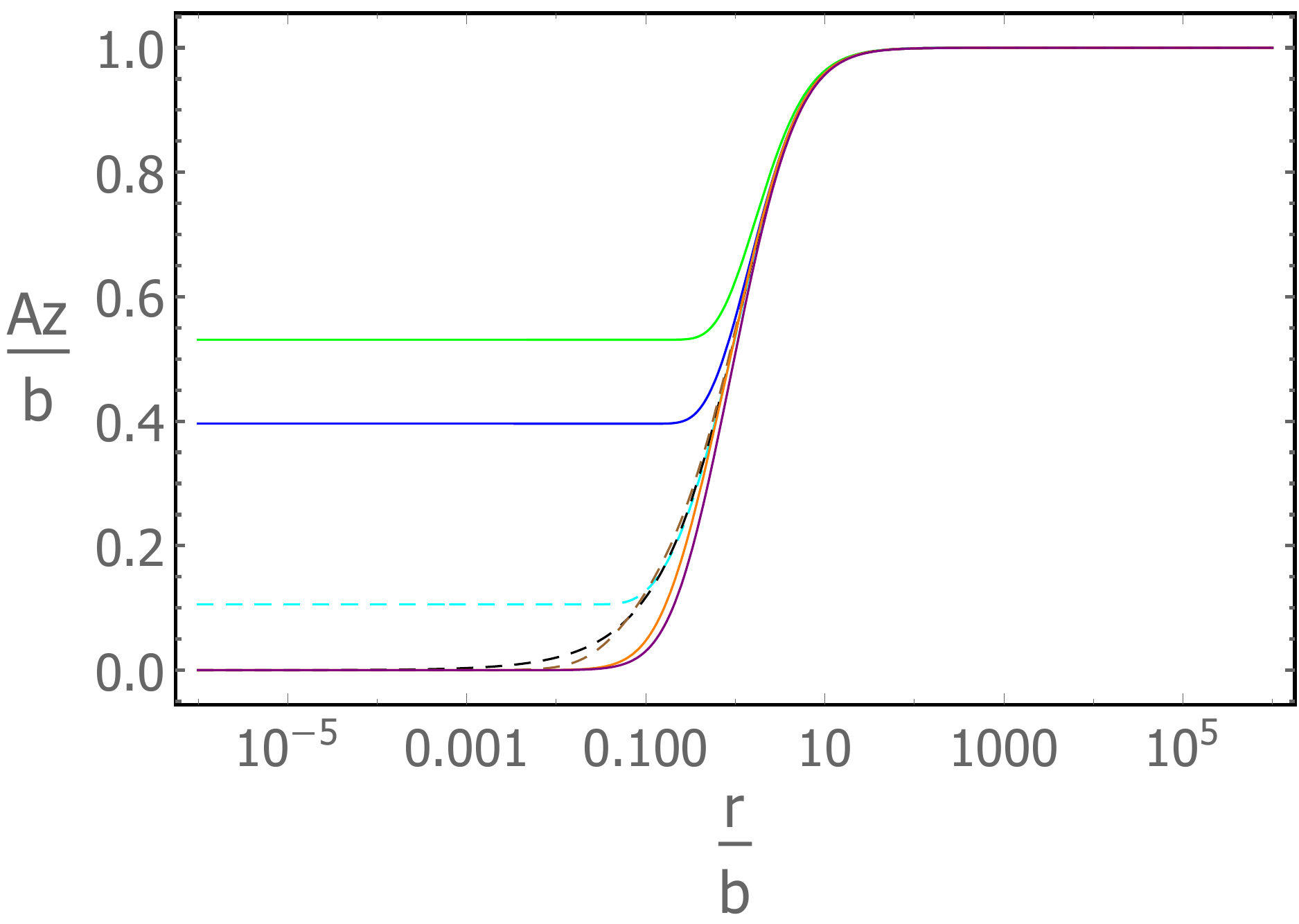}
\includegraphics[width=0.445\textwidth]{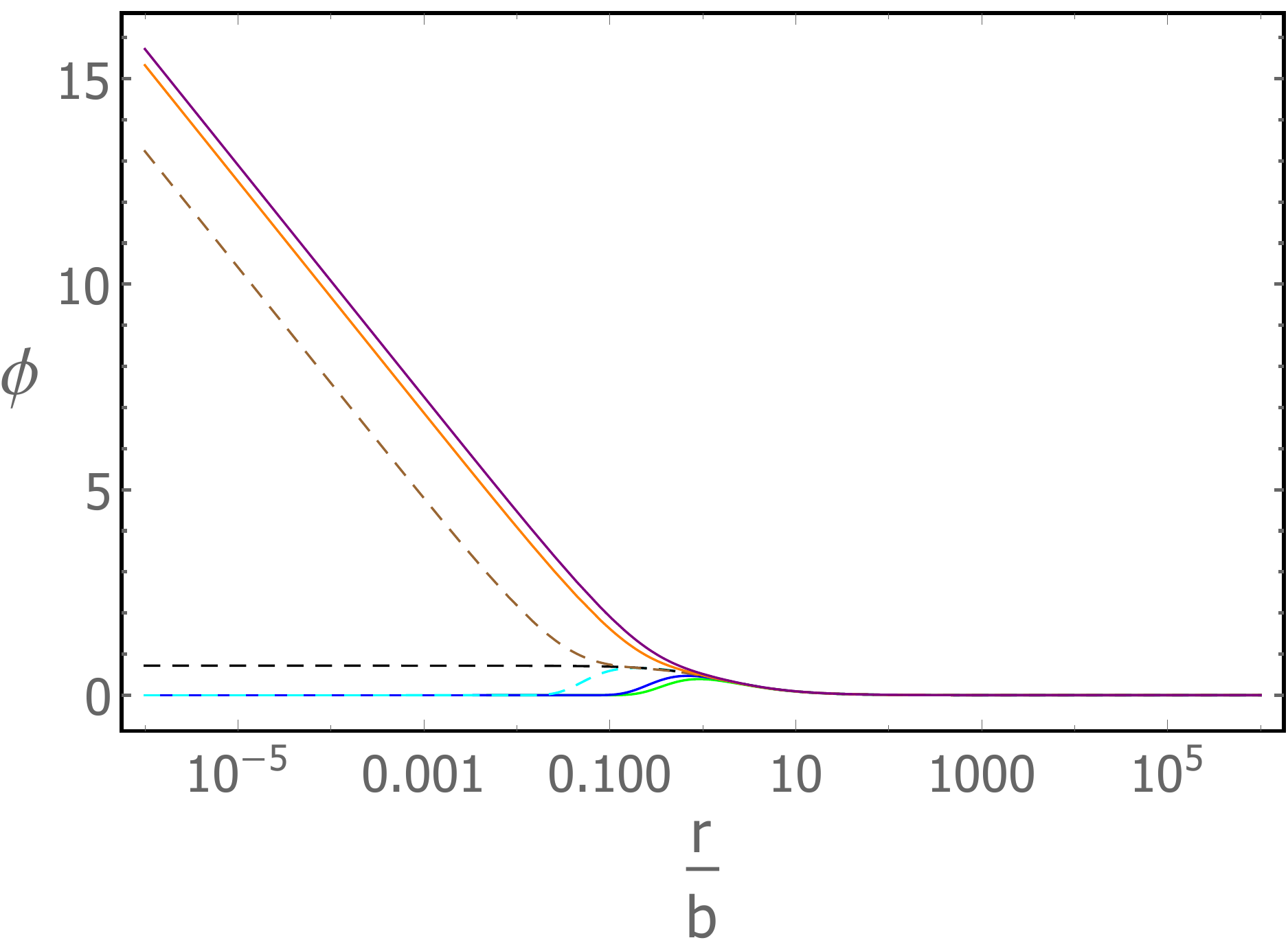}
\end{center}
\vspace{-0.6cm}
\caption{\small{The plots are for the profiles of $A_z$ and $\phi$ at $M/b=0.941$ (green), $0.983$ (blue),
 $0.987$ (dashed cyan), $0.986$ (dashed black), $0.984$  (dashed brown), $0.987$ (orange), $1.019$ (purple). The solid lines are for the stable phase while dashed lines are for the unstable phase. Figure from \cite{Liu:2018spp}.}}
\label{fig:bgsec4}
\end{figure}

The full solutions can be obtained by integrating the above near horizon geometries to the boundary. Different from the holographic semimetals in section \ref{sec:holoWSM}, the near horizon behavior (\ref{eq:wsm1}) flows to AdS$_5$ with a nontrivial $M/b$ that takes from zero to  %$(M/b)_c$, keeps increasing to 
$(M/b)_{t+}$ with $(M/b)_{t+}>(M/b)_c$ and then decreasing to $(M/b)_{c}$. The near horizon geometry (\ref{eq:in1}) flows to AdS$_5$ with $M/b$ whose value is from infinity to $(M/b)_{t-}$ with $(M/b)_{t-}<(M/b)_c$ and then increasing to reach $(M/b)_{c}$ finally. Figure \ref{fig:bgsec4} shows the profiles of the matter fields at different $M/b$. Near the critical $M/b$, the matter fields shows oscillatory behavior (dashed color lines), which can be viewed as a sign of unstable critical solution.%, indicating that the phase transition is not continuous. 

%%%%%%%%%%%%%%%%%%%%%%%%%%%%%%%%%%%%%

\begin{figure}[h!]
\begin{center}
\includegraphics[width=0.7\textwidth]{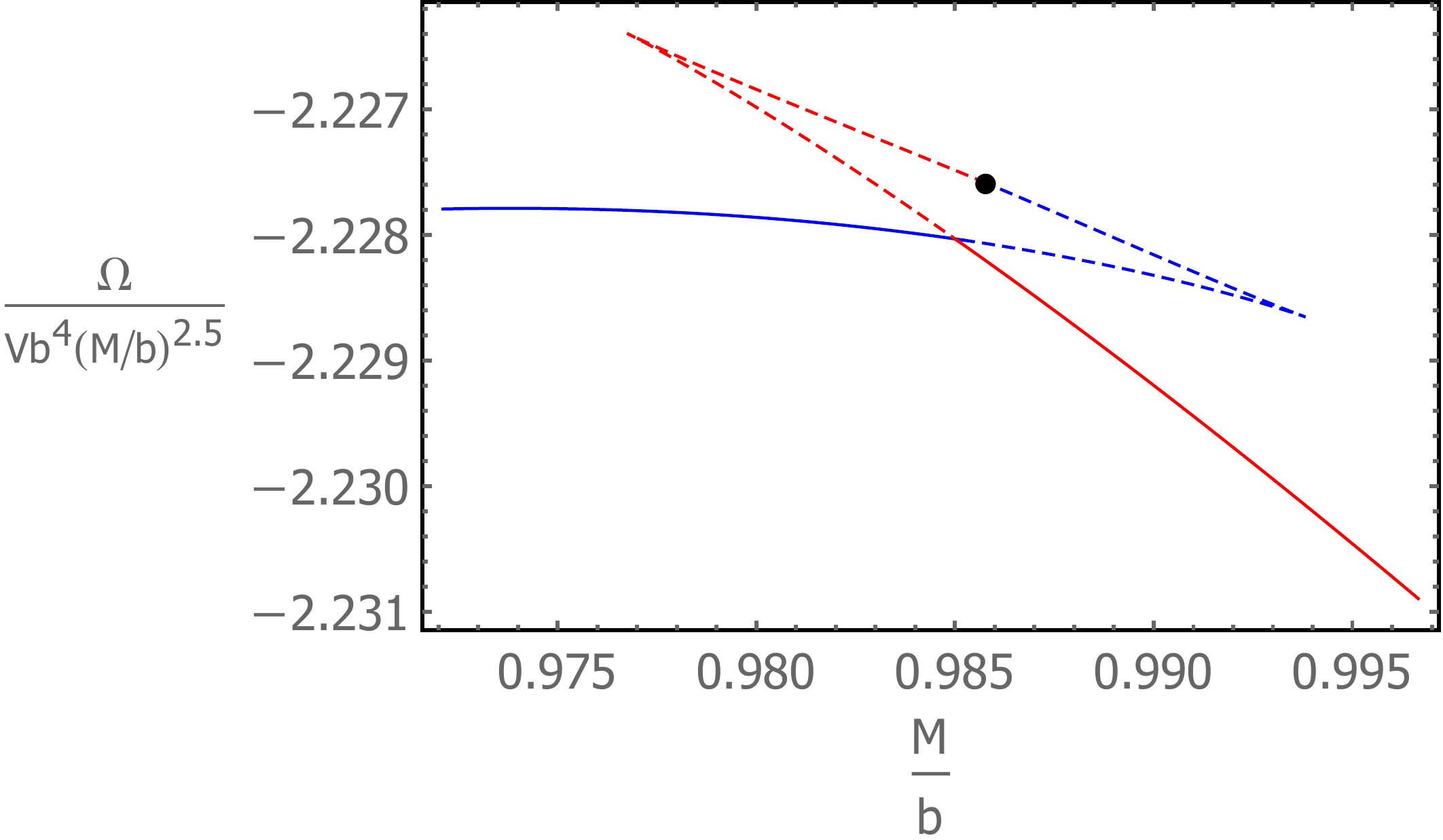}
\end{center}
\vspace{-0.7cm}
\caption{\small This plot shows the free energy density.  Here $q_0=15$. The blue (dashed) lines are for solutions from (\ref{eq:wsm1}) while the red (dashed) line are from (\ref{eq:in1}). The black dot is for the free energy at the unstable critical point. 
%The system undergoes a first order quantum phase transition from the Weyl semimetal phase to an insulating phase. 
Figure from \cite{Liu:2018spp}.}
\label{fig:free}
\end{figure}

With the bulk solution the free energy can be obtained numerically. Near the phase transition the behavior for free energy %as a function of $M/b$ 
 is shown in figure \ref{fig:free}. %The critical point generated by IR geometry (\ref{eq:cp1}, \ref{eq:cp4}) is unstable and t
Different from the holographic model in section \ref{sec:holoWSM},  for this holographic system at zero temperature there is a first order phase transition from the topologically nontrivial Weyl semimetal phase to an insulator phase. The different order of the quantum phase transition may indicate different underlying mechanisms for these two kinds of phase transitions. It can be easily checked that the phase transition is always of first order for any $q_0>0$.

\begin{figure}[h!]
\begin{center}
\includegraphics[width=0.6\textwidth]{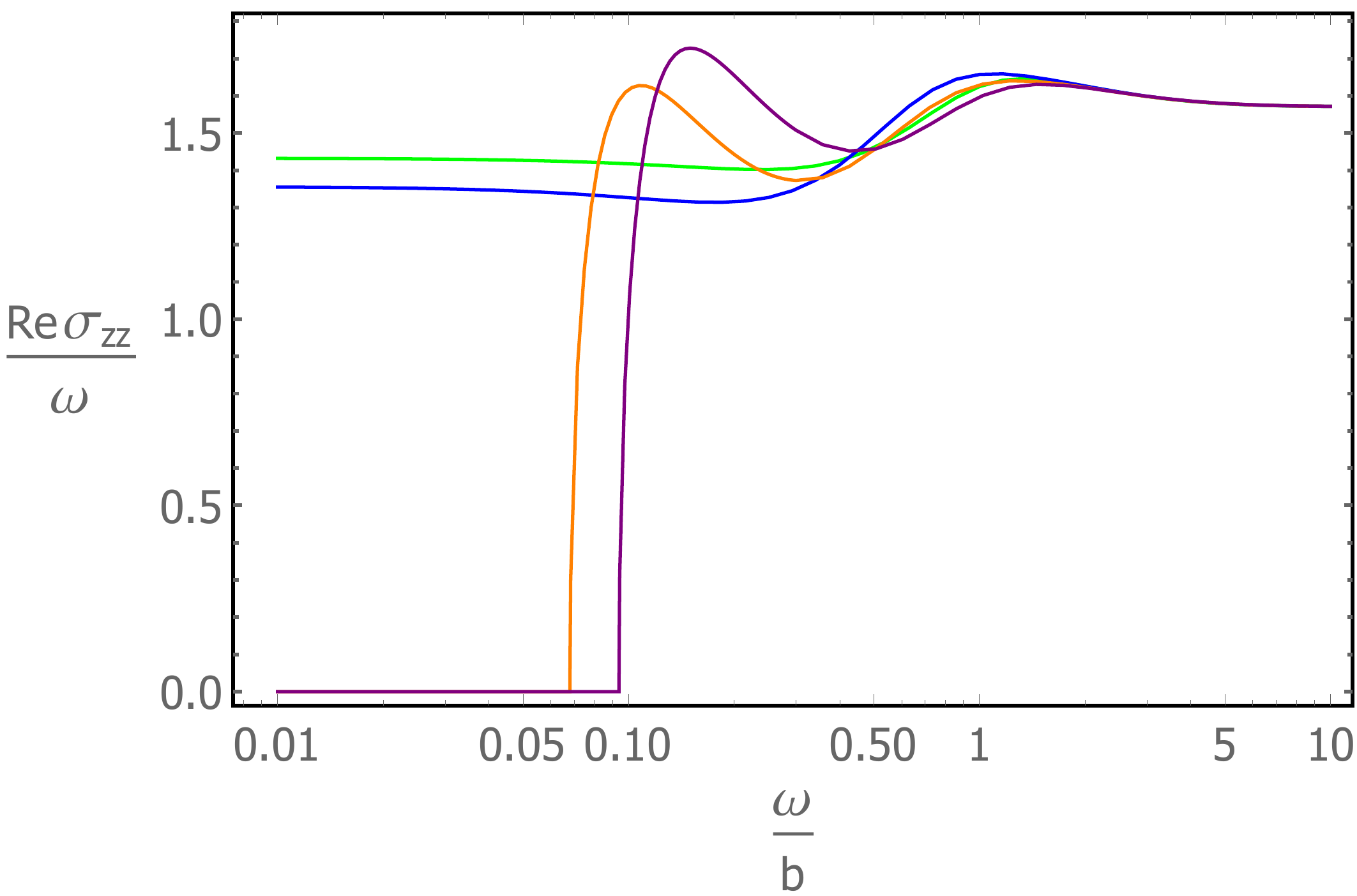}
\end{center}
\vspace{-0.5cm}
\caption{\small The real part of optical longitudinal electric conductivity at $M/b=0.941$ (green), $0.983$ (blue), $0.987$ (orange), $1.019$ (purple). 
Figure from \cite{Liu:2018spp}.}
\label{fig:gap}
\end{figure}

The exact nature of the stable phases can be figured out by studying the conductivities. 
The real part of the optical longitudinal electric conductivity $\sigma_{zz}$ of the holographic system at different $M/b$ is shown in figure \ref{fig:gap}. In the Weyl semimetal phase, $\sigma_{zz}$ is linear in frequency at both small and large frequency regimes, which is quite similar to the discussion in section \ref{sec:accon}. There is a hard gap in $\sigma_{zz}$ in the insulating phase, which indicates that it is indeed an insulating phase. Above the gap there is a continuous gapless spectrum and $\sigma_{zz}$ eventually becomes also linear in frequency at large frequency. The width of the gap depends on $M/b$ in a similar way comparing to the weakly coupled result, i.e. it monotonically increases when $M/b$ increases and for sufficient large $M/b$, $\Delta/b \propto 0.22 (M/b-0.3)$.

\begin{figure}[h!]
\begin{center}
\includegraphics[width=0.48\textwidth]{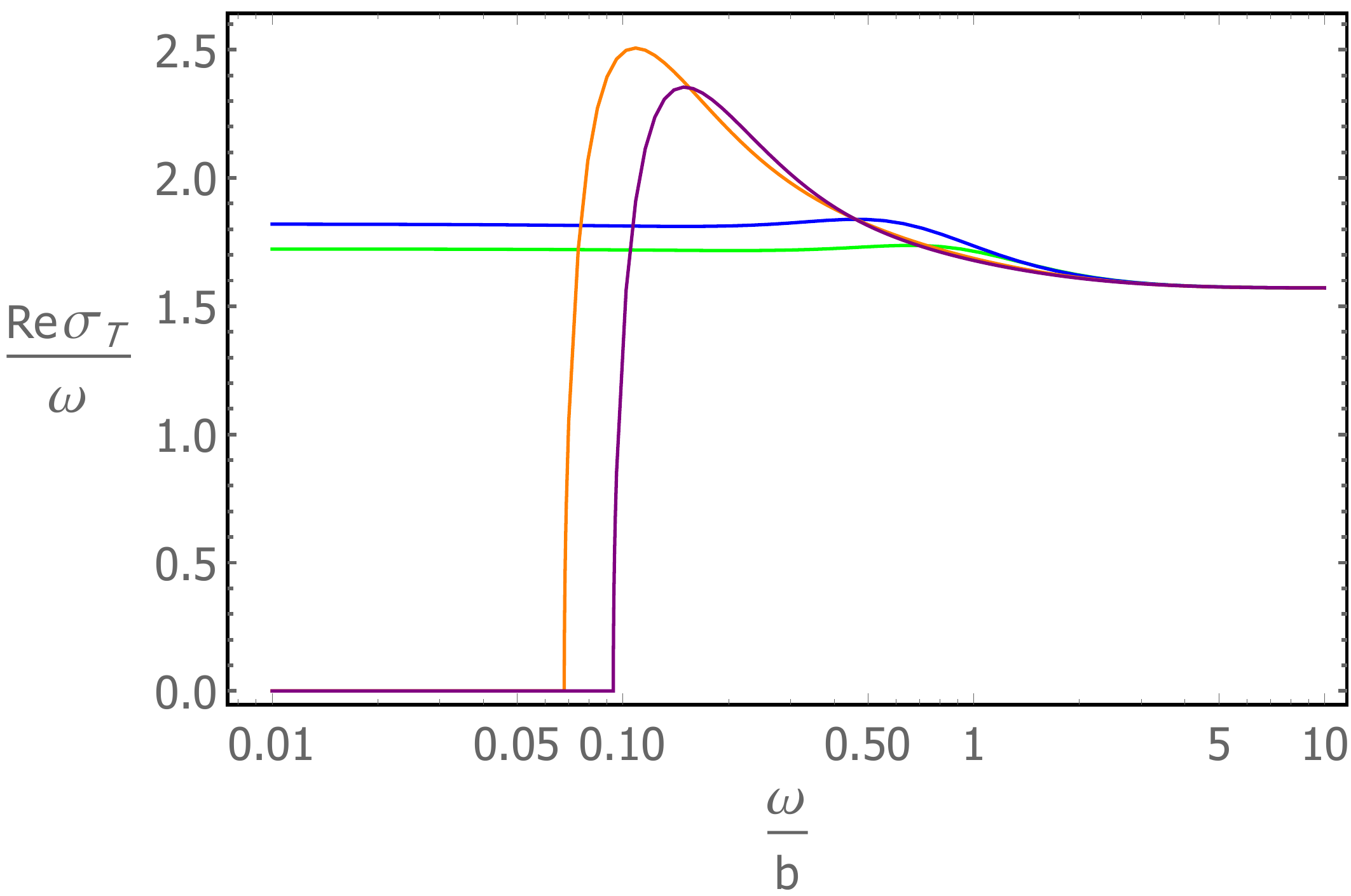}
\includegraphics[width=0.48\textwidth]{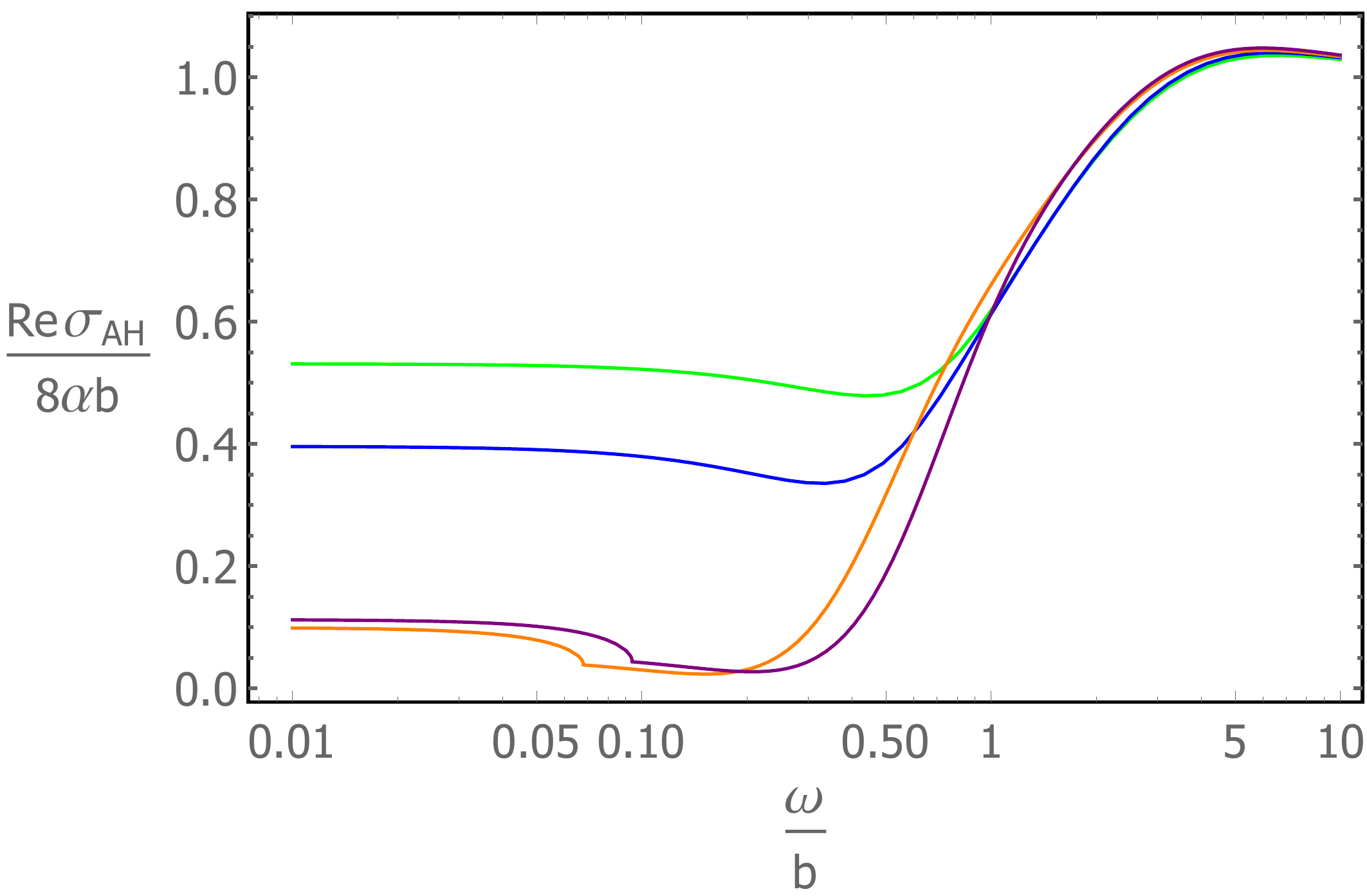}
\end{center}
\vspace{-0.6cm}
\caption{\small {The real part of the transverse optical electric  conductivity (left) and the optical anomalous Hall conductivity (right) at $M/b=0.941$ (green), $0.983$ (blue), $0.987$ (orange), $1.019$ (purple) in the topological and the insulating phases.  Figure from \cite{Liu:2018spp}.}}
\label{fig:cxy}
\end{figure}

The transverse conductivities can be calculated by considering fluctuations
$\delta V_x=v_x(r) e^{-i\omega t},
\delta V_y=v_y(r) e^{-i\omega t}.$ 
Define $v_\pm=v_x\pm i v_y$, from the holographic dictionary we can obtain  the Green's functions $G_\pm$, from which we can compute $G_{xx}$, $G_{yy}$ and $G_{xy}$. We have $\sigma_{xy}\pm i\sigma_{xx}=\pm \frac{G_\pm}{\omega}$, i.e.
\be\label{eq:trancond}
 \sigma_T=\sigma_{xx}=\sigma_{yy}=\frac{G_++G_-}{2i \omega}\,,~~~\sigma_\text{AH}=8\alpha b-\sigma_{xy}=8\alpha b-\frac{G_+-G_-}{2\omega}\,.
 \ee
Figure \ref{fig:cxy} shows the full frequency dependence of transverse conductivities.
Similar to the longitudinal conductivities, there is a gapless spectrum for $\text{Re}[\sigma_{xx}(\omega)]$ and $\text{Re}[\sigma_{yy}(\omega)]$ in the Weyl semimetal phase, while there exists a continuous gapless spectrum above a hard gap  
$\Delta/b$ in the insulating phase. 
The difference comparing to the longitudinal one is that if we increase $M/b$ in the Weyl semimetal phase  
$\text{Re}[\sigma_T]/\omega$  
increases at small $\omega$. 
%This reason for this difference is that the emergent Lifshitz symmetry in the critical point leads to $\text{Re}[\sigma_{zz}(\omega)]\propto \omega^{2-\beta}$ while both $\text{Re}[\sigma_{T}(\omega)]$ and $\text{Re}[\sigma_\text{AH}(\omega)]$ are proportional to $\omega^{\beta}$ when $M/b$ is approaching the (unstable) critical value \cite{Landsteiner:2016stv,Grignani:2016wyz}. 
The right plot in figure \ref{fig:cxy} shows the real part of optical anomalous Hall conductivity at different values of $M/b$. In the insulating phase at zero frequency the anomalous Hall conductivity goes to a nonzero value. %although there is   emergent time reversal symmetry in IR. 
Furthermore, the optical anomalous Hall conductivity has a smooth change at $\omega=\Delta$ in the insulating phase.

As already explained in section \ref{sec:holoWSM}, the order parameter of the quantum phase transition 
is the DC anomalous Hall conductivity. In the topological phase, 
the DC conductivities can be analytically obtained 
$
\sigma_\text{AHE}=8\alpha A_z(0), \sigma_{xx}=\sigma_{yy}= 0\,.
$
In the gapped phase, there is no simple analytical formula for $\sigma_\text{AHE}$. The DC anomalous Hall conductivity can only be calculated numerically by taking $\omega\to 0$ limit of $\text{Re}[\sigma_\text{AH}(\omega)]$, which is different from the case for topological trivial semimetal in section \ref{sec:holoWSM}. 

\begin{figure}[h!]
\begin{center}
\includegraphics[height=6.7cm, width=0.65\textwidth]{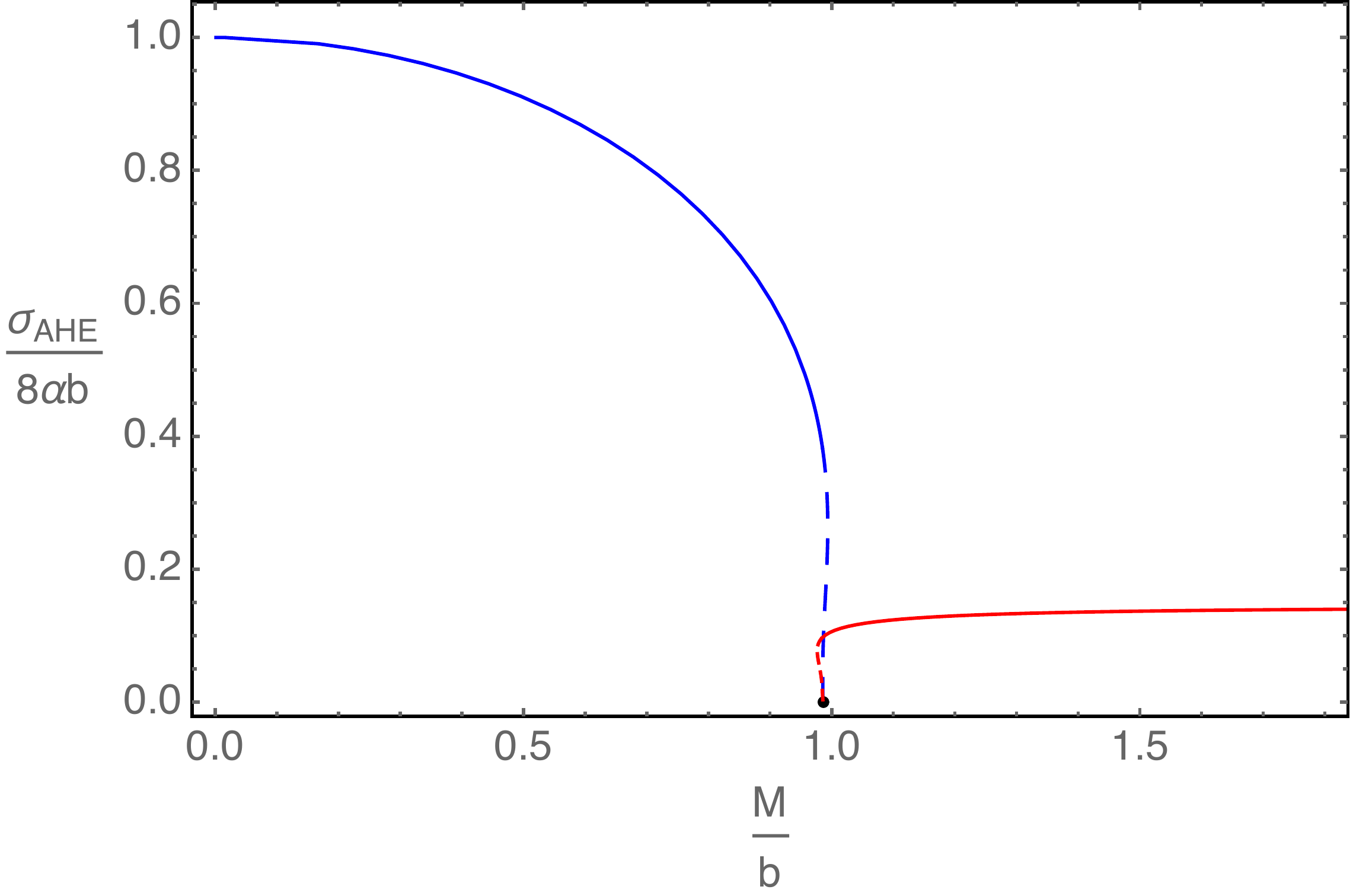}
\hspace{0.1in}
\includegraphics[height=6.6cm, width=0.3\textwidth]{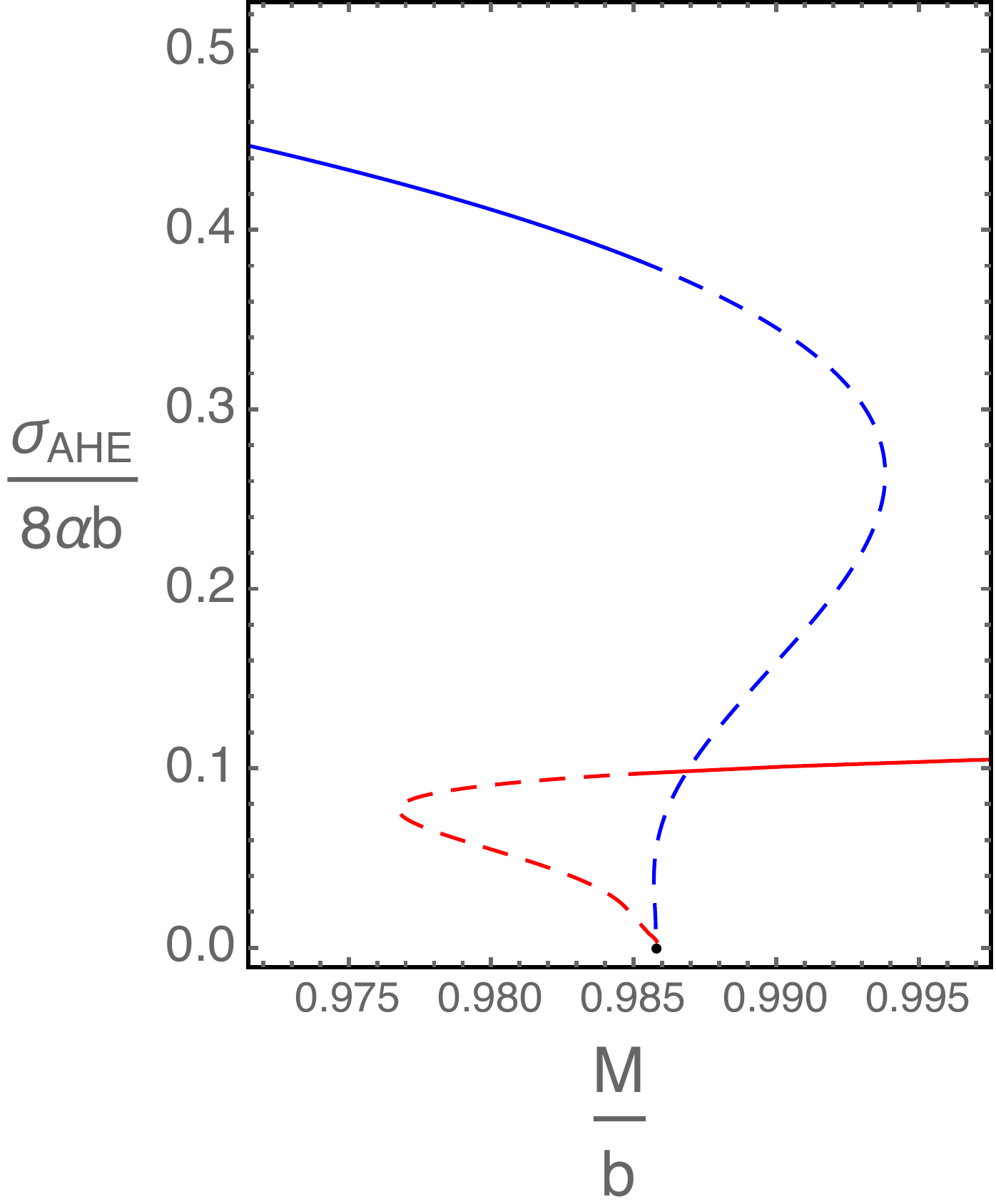}
\end{center}
\vspace{-0.6cm}
\caption{\small Both plots are for $\sigma_\text{AHE}$ at zero frequency and zero temperature from holography. The right plot is a zoomed in version of the left plot close to the quantum  phase transition point. The solid and dashed lines are for stable and unstable phases separately.  
Figure from \cite{Liu:2018spp}.}
\label{fig:ahe-wsmci}
\end{figure}

Figure \ref{fig:ahe-wsmci} shows the DC $\sigma_\text{AHE}$ depending on $M/b$. 
When $M/b$ increases the anomalous Hall conductivity decreases and jumps at to another nonzero value. After the phase transition, $\sigma_\text{AHE}$ looks insensitive to $M/b$. 
The discontinuity in $\sigma_\text{AHE}$ indicates that the holographic quantum phase transition is indeed of first order. After the phase transition, in the diagonal components of the conductivities there exists a continuous gapless spectrum above a hard gap, whereas the DC $\sigma_\text{AHE}$ is nonvanishing. These properties are exactly of a topological Chern insulator. Thus the holographic model introduced in this section describes a first order quantum phase transition from a strongly interacting topological Weyl semimetal to a topological Chern insulator. 

%%%%%%%%%%%%%%%%%%%%%%%%%%%%%
%%%%%%%%%%%%%%%%%%%%%%%%%%%%%

From the field theoretical approach the phase diagrams for interacting Weyl semimetals have been studied in \cite{burkov1,cm-1,Roy:2016amv, Roy:2016rqw}. In \cite{Roy:2016rqw} it was found that for sufficiently strong interactions the Weyl semimetal can go through a first order quantum phase transition to a normal band insulator. The holographic model shows that the strongly interacting Weyl semimetal can also go through a first order quantum phase transition to a Chern insulator.  Thus it reveals the intriguing phase structure for strongly correlated topological Weyl semimetal. The holographic model provides a novel framework to further explore the physics of strongly interacting  topological states of matter.  

%%%%%%%%%%%%%%%%%%%
%%%%%%%%%%%%%%%%%%%%%%%%%%%%%
\section{Holographic nodal line semimetals}
\label{sec:nlsm}
%%%%%%%%%%%%%%%%%%%%%%%%%%%%%
%%%%%%%%%%%%%%%%%%%
In addition to Weyl semimetals, there are several other examples of topological states of matter,  
including topological nodal line semimetal (NLSM),  topological insulators, anomalous Hall states, topological superconductors and so on. In this section, we will focus on the physics of nodal line semimetals from holography. 

In a NLSM \cite{burkov0} the shape of Fermi surface is a one dimensional circle in stead of nodal points in Weyl semimetals %under certain symmetries, e.g. mirror reflection symmetry 
(see \cite{rev1} for a review). When it is topologically nontrivial, the system cannot be gapped by small perturbations unless going through a topological phase transition to another state. 

%%%%%%%%%%%%%%%%%%%
 \subsection{Quantum field theoretical model} 
%%%%%%%%%%%%%%%%%%%

In Weyl semimetal, two Weyl points in the momentum space are separated at $b$ while in nodal line semimetal there is a one dimensional circle of nodal line. Their low energy effective theories are also different.  The Weyl semimetal is described by a Dirac field coupled to a time reversal symmetry breaking field, i.e. the axial gauge field $A_z$. Whereas for NLSM, it is described by a Dirac field coupled to a two form effective field $b_{\mu\nu}$ which breaks both time reversal and charge conjugate symmetry,
\be\label{eq:1}
\mathcal{L}=-i\bar{\psi}\big(\gamma^\mu\partial_\mu-m-\gamma^{\mu\nu} b_{\mu\nu}\big)\psi
\ee
where $\gamma^{\mu\nu}=\frac{i}{2}[\gamma^\mu, \gamma^\nu]\,$ and $b_{\mu\nu}=-b_{\nu\mu}$ is an antisymmetric two form field. Note that we take the $(-,+,+,+)$ signature. 
Without loss of generality, a source of a two form $b_{xy}$ is turned on. Then this system has the energy spectrum 
\be
E_\pm=\pm \sqrt{k_z^2+\Big(2b_{xy}\pm\sqrt{m^2+k_x^2+k_y^2}\,\Big)^2}\,.
\ee 
For $m^2< 4b_{xy}^2$, the system is a topologically nontrivial nodal line semimetal with a circle nodal line of radius $\sqrt{4 b_{xy}^2-m^2}$.\footnote{Note that the other components of $b_{\mu\nu}$ could also deform the nodal points to nodal line, e.g. nonzero $b_{tz}$ would generate an accidental nodal line semimetal.} %In this parameter regime the system cannot be gapped by small perturbations as for an ordinary Dirac field, i.e. a small perturbation in $m$ cannot gap this system. 
For $m^2 > 4b_{xy}^2$,  the system becomes an insulator and $m^2= 4b_{xy}^2$ is the quantum  phase transition point. In the nodal line semimetal phase, close to the nodal line, the dispersion is linear in $\sqrt{k_x^2+k_y^2}-\sqrt{4 b_{xy}^2-m^2}$ with velocity $\sqrt{1-\frac{m^2}{4b_{xy}^2}}$ at $k_z=0$ and linear in $k_z$ with velocity $1$ when $\sqrt{k_x^2+k_y^2}=\sqrt{4 b_{xy}^2-m^2}$.

\begin{figure}[h!]
\begin{center}
\includegraphics[width=0.4\textwidth]{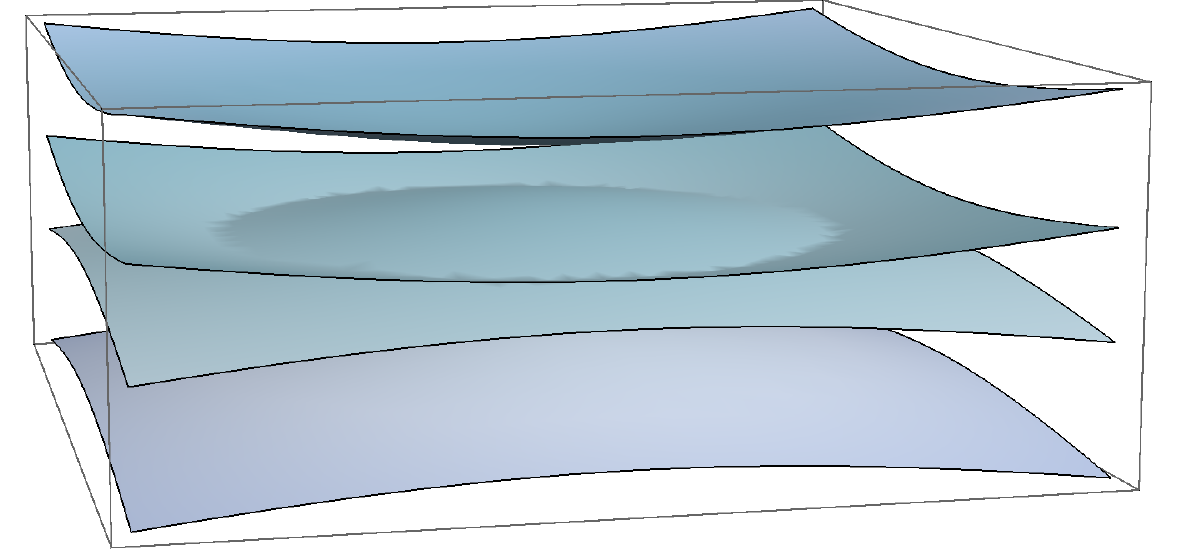}
\includegraphics[width=0.4\textwidth]{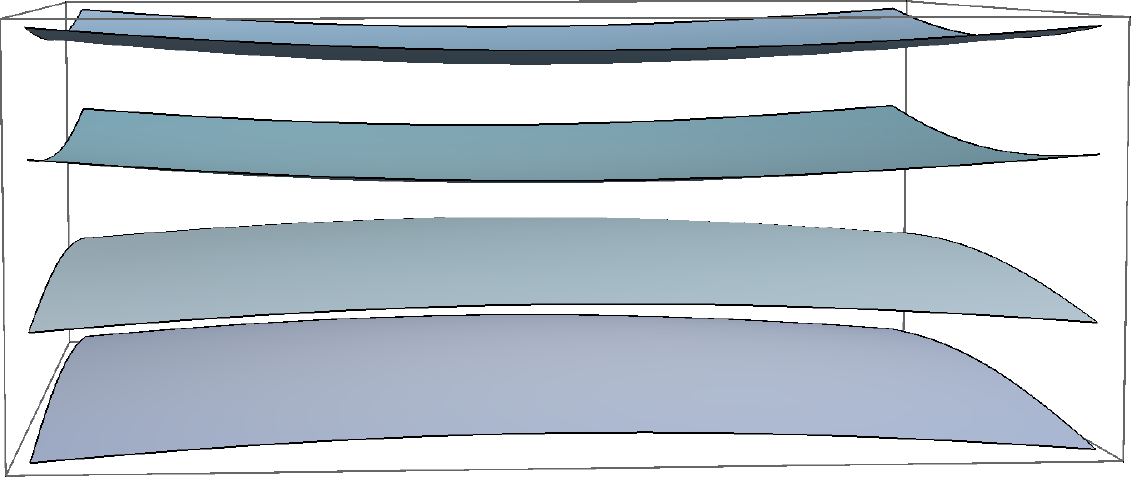}
\end{center}
\vspace{-0.3cm}
\caption{\small The energy spectrum as a function of $k_x, k_y$ for $k_z=0$. {\it Left}:  a nodal line appears at the band crossing when $m^2< 4b_{xy}^2$. {\it Right}: for $m^2 > 4b_{xy}^2$ the system is gapped. Figure from \cite{Liu:2018bye}.}
\label{fig:phase}
\end{figure}

With a nontrivial background two form field $b_{\mu\nu}$,  the electric current $J^\mu=\bar{\psi} \gamma^\mu \psi$ is still conserved while the axial current $J_5^\mu=\bar{\psi} \gamma^\mu \gamma^5\psi$ is no longer conserved. The following are the conservation equations
\bea\label{conservation}
\partial_\mu J^\mu&=&0\,,~~~\partial_\mu J_5^\mu=-2m\bar{\psi}\gamma^5\psi-2b_{\mu\nu}\bar{\psi}\gamma^{\mu\nu}\gamma^5\psi\,,
\eea where anomaly terms have been ignored.

%%%%%%%%%%%%%%%%%%%
\subsection{Holographic model}
%%%%%%%%%%%%%%%%%%%
The NLSM model (\ref{eq:1}) has shown that a coupling term $\bar{\psi}\gamma^{\mu\nu}\psi$ plays the role of deforming the single nodal point to a nodal loop. 
In holography we introduce on the gravity side a massive 2-form field $B_{ab}$ which is dual to an antisymmetric tensor  operator and is expected to play a similar role as the $\bar{\psi}\gamma^{\mu\nu}\psi$ operator. Similar to the holographic Weyl semimetal model, we also introduce an axially charged complex scalar field $\Phi$ %which is dual to the operator $\bar{\psi}\psi$ 
whose boundary value explicitly breaks the axial symmetry to generate the gap. The holographic NLSM can be realized from the following action \cite{Liu:2018bye} 
\bea
S&=&\int d^5x\sqrt{-g}\bigg[\frac{1}{2\kappa^2}\bigg(R+\frac{12}{L^2}\bigg)-\frac{1}{4}\mathcal{F}^2-\frac{1}{4}F^2+\frac{\alpha}{3}\epsilon^{abcde}A_a \bigg(3\mathcal{F}_{bc}\mathcal{F}_{de}+F_{bc}F_{de}\bigg)
\nn\\
&&-(D_a \Phi)^*(D^a\Phi)-V_1(\Phi)-\frac{1}{3\eta}\big(\mathcal{D}_{[a}B_{bc]}\big)^*\big(\mathcal{D}^{[a}B^{bc]}\big)
-V_2(B_{ab})-\lambda|\Phi|^2B_{ab}^*B^{ab}\bigg]\nn
\eea
where parts that do not involve $B_{ab}$ are the same as the holographic Weyl semimetal (\ref{eq:holomodel}) in section \ref{sec:holoWSM}.  
$B_{ab}$ has to be axially charged since its dual operator's source explicitly breaks the axial symmetry. The potential terms in the action are
\be
V_1=m_1^2 |\Phi|^2+\frac{\lambda_1}{2} |\Phi|^4\,,~~~~%\\
V_2=m_2^2 B^*_{ab}B^{ab}\,,
\ee 
where $m_{1,2}$ are mass of $\Phi$ and $B_{ab}$. 
%Note that we introduced a mass term for $B_{ab}$ because it does not correspond to a conserved operator. 
Without loss of generality, $B_{xy}$ component will be turned on in the following. 
%Similar to the models in the previous sections, the $\lambda$ term which denotes the interaction between the operators $\bar{\psi}\psi$ and the antisymmetric operator, is crucial to the existence of the topological structure and the topological phase transition. 
In the following we set $q_1=q_2=1, \lambda=\eta=1, \lambda_1=0.1$ for simplicity.  
%Note that the same potential for the scalar field is introduced as the one in the holographic WSM in section \ref{sec:holoWSM}.  

%From the field theoretical model (\ref{eq:1}), the operators $\bar{\psi}\psi$ and $\bar{\psi}\gamma^5\psi$ correspond to the real and imaginary parts of the complex scalar field in the bulk separately. 

Since the operators $\bar{\psi}\gamma^{\mu\nu}\psi$ and $\bar{\psi}\gamma^{\mu\nu}\gamma^5\psi$ are not independent, which indicates that there should be a self-duality in the real and imaginary part of the complex dual field $B_{ab}$, our strategy here is instead to consider a two form antisymmetric operator different from $\bar{\psi}\gamma^{\mu\nu}\psi$ %which nevertheless is antisymmetric in the two indices 
and does not have the property of self duality. Note that some holographic QCD models \cite{Arutyunov:1998xt,Alvares:2011wb} considered the self-duality effect of the two form field.

We will focus again on the zero temperature physics and take  the following ansatz 
\bea
ds^2&=&u(-dt^2+dz^2)+\frac{dr^2}{u}+f(dx^2+dy^2)\,,~~
\Phi=\phi(r)\,,~~
B_{xy}=B(r)\,.
\eea

Near the UV boundary $r\to \infty$, the expansions for the two matter fields $\phi(r)$ and $B(r)$ are
\be
\phi= \frac{M}{r}+\cdots\,,~~~~~ B= b r +\cdots\,,
\ee where $M$ and $b$ are the sources associated to the dual operators. %We will work at $b=1$, i.e. all dimensionful parameters are compared to $b$. 
At zero temperature, it turns out there are again three different kinds of near horizon geometries. Similar to the holographic Weyl semimetal, adding some irrelevant deformations, the  near horizon geometries flow to an AdS$_5$ in the UV with some values of $M/b$.

\noindent {\it Topological phase.} The first kind of near horizon geometry is
\bea
u&=&\frac{1}{8}(11+3\sqrt{13}) r^2\Big(1+\delta u\, r^{\alpha_1} \Big)\,,~~f= \sqrt{\frac{2\sqrt{13}}{3}-2}\, b_0 r^\alpha \Big(1+\delta f\, r^{\alpha_1} \Big)\,,\nn\\
\phi &=& \phi_0 r^{\beta}\,,~~~B=b_0 r^\alpha \Big(1+\delta b\, r^{\alpha_1} \Big)\,,\nn
\eea
where $(\alpha, \beta, \alpha_1)=(0.183, 0.290
, 1.273)$, $(\delta f, \delta b)=(-2.616, -0.302)\delta u$.  We can further set $b_0$ set to $1$. At leading order there is a Lifshitz symmetry for the solution 
\be\label{eq:lif}
(t,z,r^{-1})\to c(t,z,r^{-1})\,,~~~(x,y)\to c^{\alpha/2} (x,y)\,,
\ee
which can set $\delta u =\pm 1$ where $\delta u=-1$ flows the geometry to AdS$_5$. Thus we have a unique free parameter $\phi_0$ in the system.
 
It turns out we only get solutions with $M/b< 1.717$ in the UV. As $\phi_0$ grows from $0$, $M/b$ also grows from the value $0$ and becomes closer and closer to the critical value $1.717$.  From the property of holographic fermion spectral functions one concludes the dual phase is a topological nodal line semimetal.

\noindent{\it Critical point.} The second kind of near horizon geometry including irrelevant deformations is 
\bea
u&=& u_c r^2 (1+\delta u\, r^{\beta_1})\,,~~f= f_c r^{\alpha_c} (1+\delta f\, r^{\beta_1})\,,\nn\\
\phi &=& \phi_c  (1+\delta \phi\, r^{\beta_1})\,,~~~
B= b_c r^{\alpha_c} (1+\delta b\, r^{\beta_1})\,,\nn
\eea
with  
$
(u_c, f_c, \alpha_c, \phi_c)\simeq(3.076, 0.828 b_c, 0.292, 0.894)\,,
$ and 
$
\beta_1=1.272\,,
(\delta u, \delta f, \delta b)=(1.177,\\ -2.771,-0.409)\delta\phi\,.$

We can set $b_c$ to be $1$. At the leading order there exists a same type of Lifshitz symmetry (\ref{eq:lif}) with a different scaling exponent $\alpha_c$ instead of $\alpha$. 
This Lifshitz symmetry can set  $\delta\phi$ to be $-1$ to flow to $AdS_5$ in UV. Therefore there is only one single such solution. We get the solution with the critical value $M/b\simeq 1.717$.

\noindent{\it Trivial phase.} The third kinds of near horizon geometry is
\bea
u&=&\big(1+\frac{3}{8\lambda_1}\big)r^2\,,~~~f= r^2\,,\nn\\
\phi&=&\sqrt{\frac{3}{\lambda_1}}+\phi_1 r^{\frac{2\sqrt{160\lambda_1^2+84\lambda_1+9}}{3+8\lambda_1}-2}\,,~~~B= b_1 r^{2\sqrt{2}\sqrt{\frac{3\lambda+\lambda_1}{3+8\lambda_1}}}\,.\nn
\eea 
The $\phi_1$- and $b_1$-terms above are the irrelevant deformations that flow the geometry to 
asymptotic $AdS_5$ solutions. In this case we only get solutions with $M/b>1.717$. %For both the critical and trivial solutions, the scalar field $\phi$ is a finite constant at the horizon and the system is only partially gapped. 
 
\begin{figure}[h!]
\begin{center}
\includegraphics[width=0.43\textwidth]{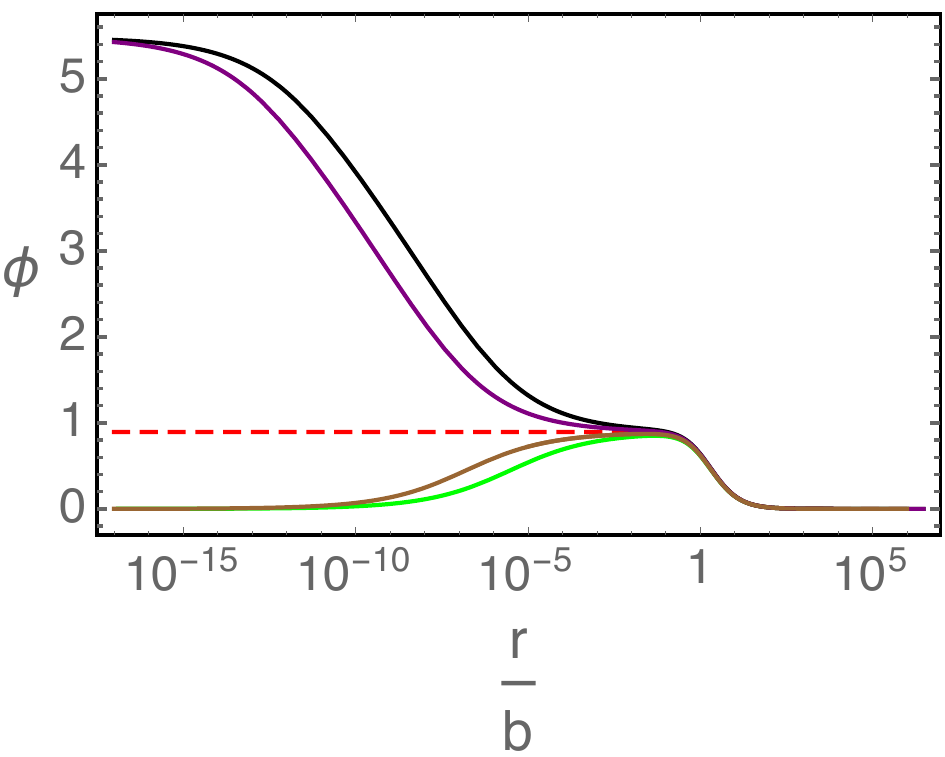}
\includegraphics[width=0.44\textwidth]{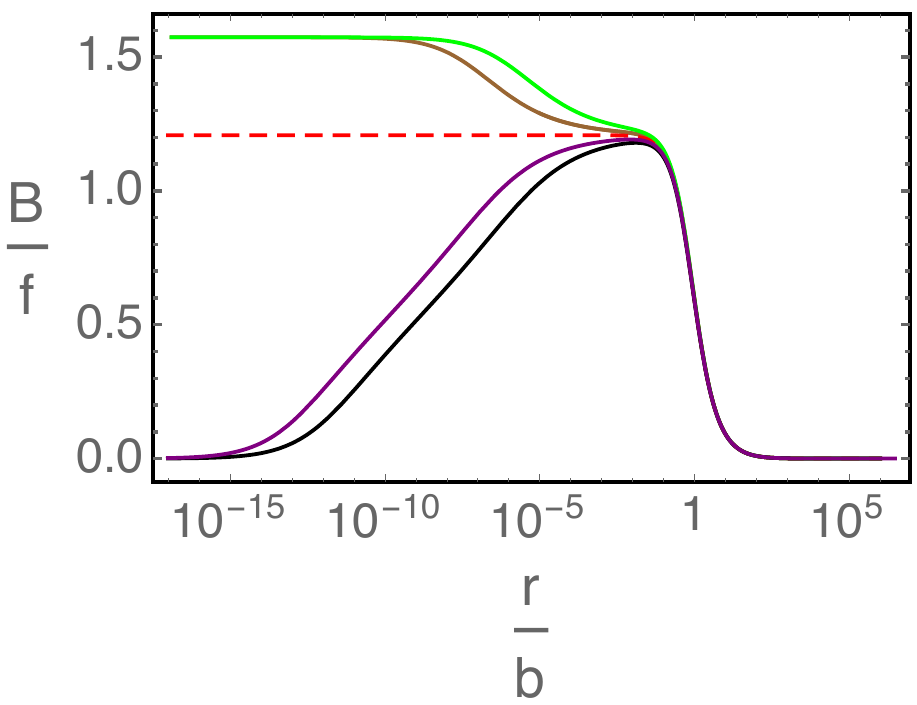}
\end{center}
\vspace{-0.48cm}
\caption{\small The bulk profile for the scalar field $\phi$ and the two form field $B/f$ for $M/b=1.682$ (green), $1.702$ (brown), $1.717$ (red), $1.733$ (purple), $1.750$ (black). Figure from \cite{Liu:2018bye}.}
\label{fig:bg}
\end{figure}

Figure \ref{fig:bg} shows the bulk profiles of matter fields $\phi$ and $B/f$ at different $M/b$. %which correspond to different near horizon geometries. 
Close to the critical $M/b$ the IR solution flows quickly to the one for  critical solution.  The free energy of the system can be numerically studied and we found that when the phase transition occurs, the system is very continuous though the bulk IR solutions are discontinuous at the horizon. In holography this is a quite common feature for continuous quantum phase transitions. 

%%%%%%%%%%%%%%%%%%%%%
\subsection{A generic framework for topological states from hologrphy} 
%%%%%%%%%%%%%%%%%%%%
This holographic nodal line semimetal has the same mathematical structure as the holographic 
Weyl semimetal in section \ref{sec:holoWSM}. \cite{Liu:2018bye} proposed a general framework in holography to describe the strongly coupled gapless topological states. The  bulk topological structure arises as follows. 
\vspace{-.2cm}
\begin{itemize}
\item 
In the holographic system, there are at least two interacting matter fields. One of them dual to the operator which plays the role of mass effect, and the other dual to an operator which  deforms the topology of the Fermi surface. For illustration these two fields are labeled as $\phi$ and $A$. The interaction between $\phi$ and $A$ in deep IR generates interesting  topological structure of the solution space.
\item 
At zero temperature usually there exist three different kinds of solutions at the horizon. Two of them are the solution that at leading order $A$ (or $\phi$) is nonvanishing with $r^{-\delta_-^{A,\phi}}$ while at subleading order $\phi$ (or $A$) is sourced by $A$ (or $\phi$). There also exists a critical solution where both $\phi$ and $A$ are subleading and sourcing each other. Because these two fields cannot be of leading order at the same time with $r^{-\delta_-^{A,\phi}}$ in IR, the semimetal phase cannot be gapped by small perturbations and is therefore topologically nontrivial.
\end{itemize}

The existence of a universal topological structure in the bulk  suggests that in principle from holography we could obtain a large class of topologically nontrivial strongly coupled gapless systems. 
%We expect that a general holographic topological semimetal state shares the properties above.  

%%%%%%%%%%%%%%%%%%%%%%%%%%%%%%%%%%%%%%%%%%%
\subsection{Fermionic probe on the holographic nodal line semimetal}
%%%%%%%%%%%%%%%%%%%%%%%%%%%%%%%%%%%%%%%%%%%
Although in NLSMs there is no sharp order parameter like anomalous Hall conductivity for Weyl semimetals, we could show that indeed there exists a circle of nodal loop in the dual fermionic spectral functions by probing fermions in the bulk. 

Similar to the discussion for holographic Weyl semimetal in section \ref{secti}, we utilize two spinors in the bulk to describe a Dirac operator in the dual field theory. 
In the bulk the coupling terms between the spinors and the  scalar field are the same as the ones in the Weyl semimetal. There is one most natural way to couple the two bulk spinors to the  two form field $B_{ab}$. We use the following action for the probe fermion 
\bea
S&=&S_1+S_2+S_\text{int}\,,\\
S_1&=&\int d^5x \sqrt{-g} i\bar{\Psi}_1\Big(\Gamma^a D_a -m_f\Big)\Psi_1\,,\nonumber \\
S_2&=&\int d^5x \sqrt{-g} i\bar{\Psi}_2\Big(\Gamma^a D_a +m_f \Big)\Psi_2\,,\nonumber \\
S_\text{int}&=&-\int d^5x \sqrt{-g}\Big( i\Phi\bar{\Psi}_1 \Psi_2+i \Phi^*\bar{\Psi}_2 \Psi_1+
\mathcal{L}_B\Big)\,,
\eea  
and 
\be\label{sb}
\mathcal{L}_B=
-i(\eta_2 B_{ab}\bar{\Psi}_1 \Gamma^{ab}\gamma^5\Psi_2-\eta_{2}^*B_{ab}^*\bar{\Psi}_2\Gamma^{ab}\gamma^5\Psi_1)\,.\ee 
Note that in the bulk the Lorentz invariance in the tangent space has been explicitly broken. 

\begin{figure}[h!]
\begin{center}
\includegraphics[width=0.4\textwidth]{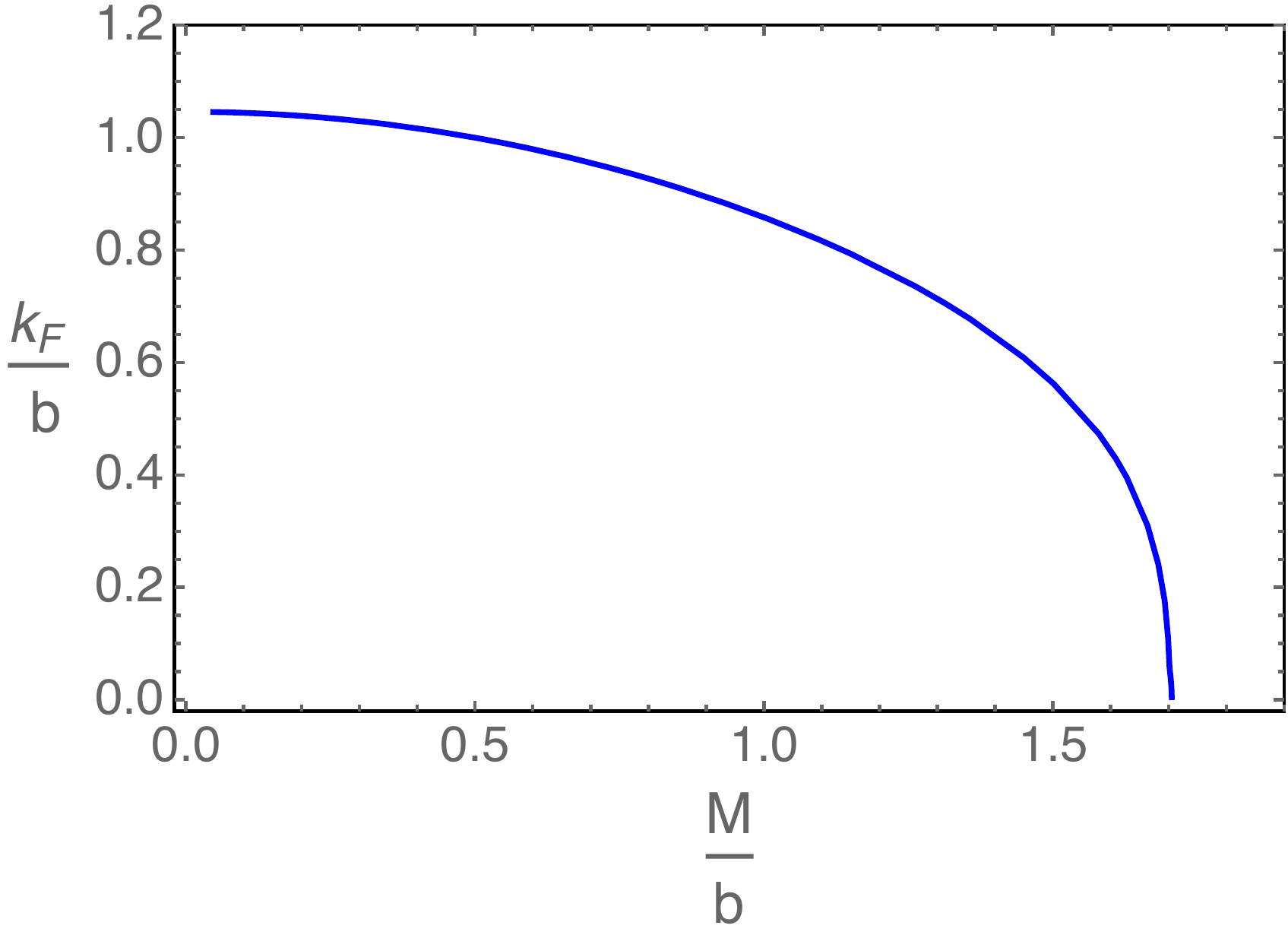}
\includegraphics[width=0.43\textwidth]{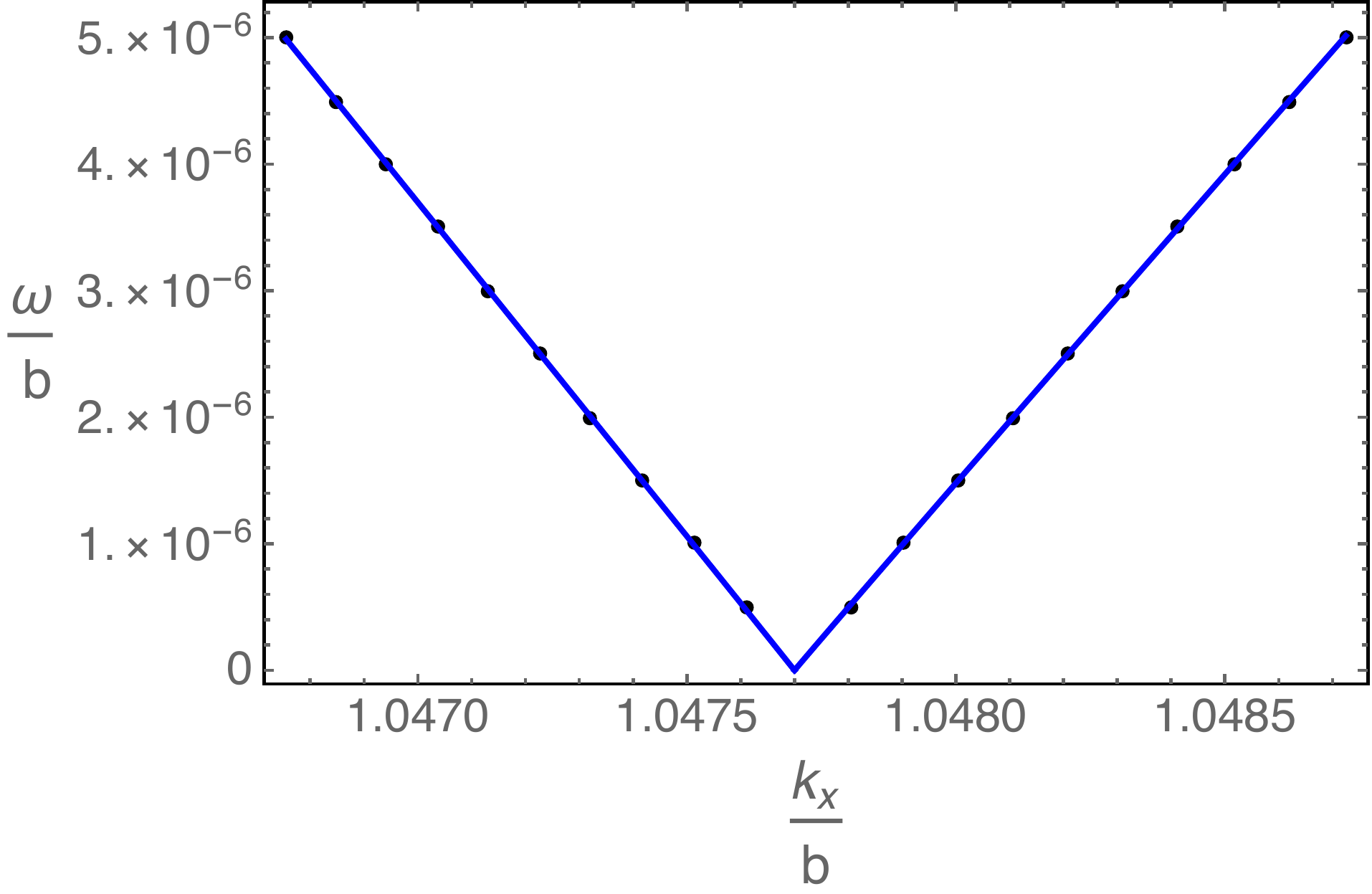}
\end{center}
\vspace{-0.4cm}
\caption{\small {\em Left:} The nodal line momentum $k_F=\sqrt{k_x^2+k_y^2}$ and $k_z=0$. In both the critical and trivial phases, no Fermi surface exists at finite $k$ whereas the pole is at $k=\omega=0$ which is consistent with the fact that only partial degrees of freedom are gapped. {\em Right:} The dispersion relation associated to the nodal momenta in the left plot at $M/b\simeq 0.0013$. The best fit are for $k_x<k_F$, $\omega\simeq 0.005 (1.0477-k_x)^{0.998}$; while for $k_x>k_F$, $\omega\simeq 0.005 (k_x-1.0477)^{0.994}$. 
Figure from \cite{Liu:2018djq}.}
\label{fig:nlsm-fs}
\end{figure}

The system has a rotation symmetry in the $k_x$-$k_y$ plane and only depends on $k_{x-y}=\sqrt{k_x^2+k_y^2}$. Without loss of generality we set $k_y=0$. From the holographic dictionary, we can compute the retarded Green's function $G$. Then we could get its four eigenvalues and the spectral function. In the following we summarize the properties for the Green's function in the holographic nodal line semimetal.
\vspace{-0.3cm}
\begin{itemize}
\item In all the three phases, for nonzero $k_z$, the retarded Green's function at zero frequency is real. % and does not have an imaginary part.  

\item In the trivial phase, %due to the special near horizon geometry 
the retarded Green's function is real for all values of $k_x, k_y, k_z\neq 0$. The pole is located at $k_x=k_y=k_z=0$ and this is consistent with the explanation that this trivial semimetal phase is only partially gapped. 

\item For the critical point, among the four eigenvalues of the Green's function, two of them have peaks in the imaginary part at $k_x=k_y=0$ and the other two are still small for all $k_x,k_y$.

\item Figure \ref{fig:spec} shows the spectral function $G^{-1}(0, k_x)$ for a finite regime of $k_x$ at $k_z=\omega=0$, $M/b\simeq 0.0013$ and $m_f=-1/4$. All the Green's function's four eigenvalues are real. They have the form $(g_1,-g_1,g_2,-g_2)$ with both $g_1$ and $g_2$ are positive and  $g_1\geq g_2$. The two eigenstates with eigenvalues $g_1,-g_1$ are labeled as ``bands I'' and the other two with $g_2,-g_2$ as ``bands II''.  In figure \ref{fig:spec} different colors are used to distinguish different bands. 
Furthermore, since $-G^{-1}(0, k)$ can be treated as a topological Hamiltonian \cite{wang-prx, interaction1}, the spectral density plot should qualitatively agree with the plot for eigenvalues in the $\omega$-$k_x$ plane.

\begin{figure}[h!]
\begin{center}
\includegraphics[width=0.6\textwidth]{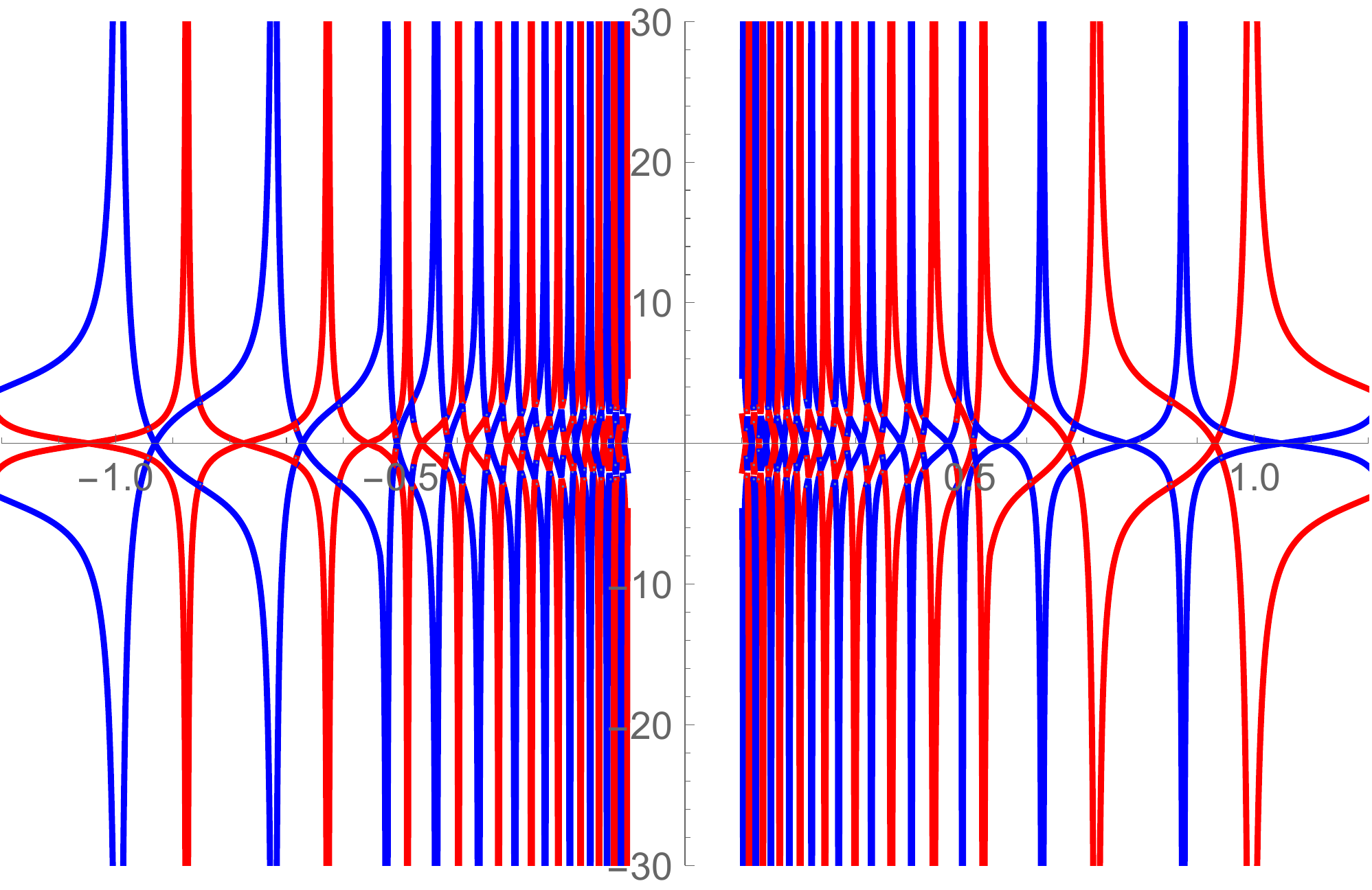}
\end{center}
\vspace{-0.4cm}
\caption{\small Eigenvalues of $-G^{-1}(0,k_x)$ for $M/b\simeq 0.0013$. %representing the qualitative behavior of the bands, which should agree qualitatively with the spectral density plot in the $\omega$-$k_x$ plane. 
Red colour is for bands I and blue colour is for bands II. 
%The two bands with red colour are referred to as bands I and the two bands with blue colour as bands II. %The distance between adjacent poles are becoming larger as $k_x$ increases. 
Figure from \cite{Liu:2018djq}.}
\label{fig:spec}
\end{figure}

\item From figure \ref{fig:spec} we can see that  between each two adjacent poles bands I and II always and only intersect once in the upper frequency plane. %Different from the weakly coupled NLSM system where the four bands are divided into two gapless bands and two gapped bands which are always gapped, now the two gapped bands in the NLSM phase would become gapless at a larger $k_x$ and exchange the role with the other two bands. 
Between each two adjacent band crossing points there is always one pole and one zero of the Green's function.

\item In the strongly coupled nodal line semimetal phase from holography there are multiple and discrete Fermi surfaces in the $k_x$-$k_y$ plane at $k_F^{i}=\sqrt{k_x^2+k_y^2}$ and $k_z=0$, $\omega\to 0$.  The dual system has more complicated topological structure.  
At each nodal line momentum, there is a sharp peak (a pole at $\omega=0$) in the imaginary part of two eigenvalues of the Green's function whereas the imaginary part of the other two is very small, which means that they are gapped. 

\item When $k_x$ increases, the distance between adjacent poles becomes larger. At small $k_x$ the poles are very close to each other. We have not plotted this regime in figure \ref{fig:spec} because the nodal loops are too dense to reveal all the poles and a much heavier numerics is required.

\item When $M/b$ increases, each nodal line momentum decreases and goes to zero at the  transition point.  In figure \ref{fig:nlsm-fs}, the left plot shows the behavior of one $k_F^{i}$ depending on $M/b$ and the right plot shows the dispersion in the $k_x$ direction at $M/b\simeq 0.0013$. Note that the dispersions in both the $k_z$ and $k_x$ directions are almost linear for each branch of nodal lines.

\end{itemize}

%The spectral function behavior confirms that the topological phase corresponds to a NLSM with multiple nodal lines while the critical and trivial solutions correspond to trivial semimetals. In the NLSM phase due to strong coupling effect there are more complicated topological structures.  %Moreover, in the holographic NLSM phase as the subleading order of $\phi$ grows bigger the radius of the nodal line circle becomes smaller. This confirms that small perturbations would not gap the system while only make the Fermi nodal circle bigger or smaller.

%%%%%%%%%%%%%%%%%%%%%%%%%%%%%%%%%%%%%%%%%%%
\subsection{Topological invariants}
%%%%%%%%%%%%%%%%%%%%%%%%%%%%%%%%%%%%%%%%%%%

In nodal line semimetals there are two kinds of topological invariants  \cite{rev1}. The first one is the Berry phase around a one dimensional closed line which links with the nodal loop in the momentum space.  
This one is related to the stability of the nodal loop under small perturbations in the system.
The second one is the Berry flux around a sphere enclosing the whole nodal loop. This topological invariant is to describe  whether the critical point is topological or not, which will not be discussed here.

From  holography the strongly interacting NLSM phase has multiple while discrete nodal lines in the $k_x$-$k_y$ plane and $k_z=0$. Since the circle that links with two or more nodal lines at the same time can be continuously deformed to two or more separate circles each enclosing only one nodal line inside, we can focus on the Berry phase associated with each nodal line.  From the Green's function at $\omega=0$ we found that these poles are from two different sets of bands (band I and  band II) which indicates that along the $k_x$ axis the two gapped bands and two gapless bands exchange their roles alternatively. We could calculate the Berry phase numerically by choosing discrete points along the circle and found that 
there is a nontrivial Berry phase $\pi$ associated with poles from bands I  and for poles from bands II the Berry phase is undetermined. For the zeros of the Green's functions we have a trivial Berry phase of zero \cite{Liu:2018djq}.

%%%%%%%%%%%%%%%%%%%%%%%%%%%
\section{Alternative approaches} 
\label{sec:alt}
%%%%%%%%%%%%%%%%%%%%%%%%%%%
An approach to strongly coupled model of Weyl fermions based on holography that differs somewhat from the one reviewed in this article was presented in \cite{Gursoy:2012ie}. 
There the idea is to study fermions which are strongly coupled in a holographic theory. This is similar to the study of probe fermions in holographic backgrounds \cite{Liu:2009dm,Cubrovic:2009ye}. The fermions are treated as probes which might or might not reflects the underlying physics of the dual field theory. It has been used for example to study the electric conductivity in \cite{Jacobs:2014nia}.  It would be interesting to explore the other characteristic features of Weyl semimetals including surface states, anomalous Hall effect etc in this approach and to go beyond the probe limit to include the backreaction of the probe fermions to the gravitational background.

It is known that Weyl semimetals can also be generated from Dirac systems by applying a rotating
electric field \cite{Zhang:floquet,Huebner:floquet}. More precisely the Dirac fermions split into left- and right-handed
Weyl fermions under the application of a fast rotating electric field.
The question if this also happens in holography has been investigated in \cite{Hashimoto:2016ize}. The construction
is based on probe D7-branes in an AdS$_5$ $\times$ S$^5$ background. The background serves also
as energy reservoir and allows the formation of a non-equilibrium steady state. This allows to 
compute the Hall conductivity as function of the applied frequency of the driving electric field but 
also beyond the regime of linear response. In this top-down model the field content of the dual theory is clear, which in principle provides more constraints and insights  into the physics of the boundary field theory. However, this model works in the probe brane limit while possible backreaction is not clear. It would be also interesting to explore more physics of the dual system from this approach.

%%%%%%%%%%%%%%%%%%%%%%%%%%%
\section{Summary and outlook on further research}
\label{sec:sumout}
%%%%%%%%%%%%%%%%%%%%%%%%%%%

We have reviewed the holographic construction of  models capable of reproducing key features of the physics of topological semimetals such as Weyl and nodal line semimetals. Amongst them the quantum phase transition to a topologically trivial state, the anomalous Hall conductivity, surface states, topological invariants, 
a new understanding of the axial Hall conductivity. Some of the new results derived from that model is the appearance
of anomalous Hall viscosity in the quantum critical region at finite temperature of the phase transition. 
A short summary is given in table \ref{tab:two} for Weyl semimetals and in table \ref{tab:three} in the case of nodal line semimetals. 

In these tables, we list different features of the weakly coupled semimetal and strongly coupled semimetal from holography for comparison, including the symmetries, transports/features, edge states, topological invariants and material realisation. In the nodal line semimetals, there is no sharp transport signature like anomalous Hall conductivity to distinguish the topological phase and trivial phase. The Fermi surfaces of the system show interesting features with nodal loop in the weakly coupled case and multiple loops in the strongly coupled case. The question marks are the items which are not clear yet. 

\begin{table}[h]
\begin{center}
\label{tab:sum-hwsm}
\begin{tabular}{|c|c|c|}
\hline
%\backslashbox{properties}{coupling}
 & Weakly coupled WSM & Holographic WSM \\
\hline\hline
Symmetries& time reversal or inversion  & time reversal  \\
&symmetry breaking&symmetry breaking\\
\hline
Transports& anomalous Hall conductivity &AHE\,, odd viscosity \\
\hline
Edge states & Fermi arc& surface current\\
\hline
Topological invariants & $\pm$ 1 & $\pm$ 1 \\
\hline
Material& TaAs\,, TaP\,, etc. &  WP$_2$?\\
\hline
\end{tabular}
\end{center}
\vspace{-0.4cm}
\caption{\small The summary of holographic Weyl semimetal.  \label{tab:two}}
\end{table}

\begin{table}[h]
\begin{center}
\label{tab:sum-hnlsm}
\begin{tabular}{|c|c|c|}
\hline
%\backslashbox{properties}{coupling}
&Weakly coupled NLSM & Holographic NLSM \\
\hline\hline
Symmetry& symmetry protected by mirror  & symmetry protected by\\
&refection symmetry\,, inversion symmetry& inversion symmetry\\
\hline
features&nodal loop & multiple nodal loops  \\
\hline
Edge states & no & ?\\
\hline
Topological invariants & $\pi$ & one set is $\pi$, \\
&&another undetermined\\
\hline
Material& PbTaSe$_2$, ZrTe, etc. & ?\\
\hline
\end{tabular}
\end{center}
\vspace{-0.4cm}
\caption{\small The summary of holographic nodal line semimetal. \label{tab:three}}
\end{table}

So far only a small subset of the parameter space of these models has been explored.  There are many open questions. 
An incomplete list is as follows. 
\begin{itemize}
\item  It would be interesting to include chemical potentials for vector and axial symmetries and study the chiral magnetic effect in these models. This would tell us about the CME in a strongly interacting Weyl semimetal. Meanwhile, it would be interesting to study negative magnetoresistivity in this model. 

\item In the quantum critical region a new anomaly related transport coefficient, anomaluos Hall viscosity appears.
It would be interesting to develop the full hydrodynamics of the quantum critical region. 

\item The Weyl cones in Weyl semimetals can be tilted and so-called type II Weyl semimatals can appear if the tilt 
exceeds the ``lightcone'' defined by the Fermi velocity. Can one also construct holographic models of type II Weyl semimetals? 

\item The holographic Weyl semimetal in section \ref{sec:holoWSM} describes a holographic dual for Weyl semimetal with two Weyl nodes. It would be interesting to consider the holographic dual for multiple Weyl nodes. 

\item The quantum phase transition in the holographic WSM/Chern insulator model is of first order. It would be interesting to study if it is still first order for more general holographic phase transitions between Weyl semimetal and insulating phases.
 
\item  It would be interesting to explore the disorder effects or other momentum dissipation effects, finite temperature physics, transport physics etc. in the holographic WSM/Chern insulator model. 

\item  The holographic insulating phase is a Chern insulator. It would be interesting to explore the topological invariants, to explore effects of surface states, to realise the phase transition to a normal insulator and so on.  

\item It would be interesting to construct the smoking gun transport in the holographic nodal line semimetals. 

\item In the holographic nodal line semimetal, it would be interesting to consider the holographic model with a self-dual two form field.

\item It would be interesting to study the behavior of nonlocal quantities, e.g. entanglement entropy, Renyi entropy, complexity etc., across the topological phase transition, to characterize the changes of dynamical degrees of freedom during the transition. 

\end{itemize}

These studies should be helpful in building holographic models for 
more complicated topological states of mater towards a classification of strongly interacting topological matter. We hope to explore some of these questions further in the future.

%%%%%%%%%%%%%%%%%%%
\subsection*{Acknowledgments}
%%%%%%%%%%%%%%%%%%%
We would like to thank  Daniel Arean, Matteo Baggioli, Rong-Gen Cai, Alberto Cortijo, Chen Fang, Carlos Hoyos, Amadeo Jimenez,  Eugenio Meg{\'\i}as, Elias Kiritsis,  Koenraad Schalm, Francisco Pena-Benitez, Maria Vozmediano, Zhong Wang, Jan Zaanen, Fuchun Zhang for useful discussions and C. Copetti, J. Fernandez-Pend{\'a}s, X.~Ji,  X.~M.~Wu, J.~K.~Zhao for enjoyable collaboration. 

This work is supported by the National Key R\&D Program of China (Grant No. 2018FYA0305800) and by the Thousand Young Talents Program of China. The work of Y.L. was also supported by the National Natural Science Foundation of China grant No.11875083. The work of Y.W.S. has also been partly supported by starting grants from University of Chinese Academy of Sciences and Chinese Academy of Sciences, and by the Key Research Program of 
Chinese Academy of Sciences (Grant No. XDPB08-1), the Strategic Priority Research Program of Chinese Academy of Sciences, 
Grant No. XDB28000000. The work of K.L. is supported by the grants SEV-2016-0597, FPA2015-65480-P and PGC2018-095976-B-C21 from MCIU/AEI/FEDER, UE.

\end{document}

%% file: conductivity-differentparameters1.tex
% GNUPLOT: LaTeX picture with Postscript
\begingroup
  \makeatletter
  \providecommand\color[2][]{%
    \GenericError{(gnuplot) \space\space\space\@spaces}{%
      Package color not loaded in conjunction with
      terminal option `colourtext'%
    }{See the gnuplot documentation for explanation.%
    }{Either use 'blacktext' in gnuplot or load the package
      color.sty in LaTeX.}%
    \renewcommand\color[2][]{}%
  }%
  \providecommand\includegraphics[2][]{%
    \GenericError{(gnuplot) \space\space\space\@spaces}{%
      Package graphicx or graphics not loaded%
    }{See the gnuplot documentation for explanation.%
    }{The gnuplot epslatex terminal needs graphicx.sty or graphics.sty.}%
    \renewcommand\includegraphics[2][]{}%
  }%
  \providecommand\rotatebox[2]{#2}%
  \@ifundefined{ifGPcolor}{%
    \newif\ifGPcolor
    \GPcolortrue
  }{}%
  \@ifundefined{ifGPblacktext}{%
    \newif\ifGPblacktext
    \GPblacktexttrue
  }{}%
  % define a \g@addto@macro without @ in the name:
  \let\gplgaddtomacro\g@addto@macro
  % define empty templates for all commands taking text:
  \gdef\gplbacktext{}%
  \gdef\gplfronttext{}%
  \makeatother
  \ifGPblacktext
    % no textcolor at all
    \def\colorrgb#1{}%
    \def\colorgray#1{}%
  \else
    % gray or color?
    \ifGPcolor
      \def\colorrgb#1{\color[rgb]{#1}}%
      \def\colorgray#1{\color[gray]{#1}}%
      \expandafter\def\csname LTw\endcsname{\color{white}}%
      \expandafter\def\csname LTb\endcsname{\color{black}}%
      \expandafter\def\csname LTa\endcsname{\color{black}}%
      \expandafter\def\csname LT0\endcsname{\color[rgb]{1,0,0}}%
      \expandafter\def\csname LT1\endcsname{\color[rgb]{0,1,0}}%
      \expandafter\def\csname LT2\endcsname{\color[rgb]{0,0,1}}%
      \expandafter\def\csname LT3\endcsname{\color[rgb]{1,0,1}}%
      \expandafter\def\csname LT4\endcsname{\color[rgb]{0,1,1}}%
      \expandafter\def\csname LT5\endcsname{\color[rgb]{1,1,0}}%
      \expandafter\def\csname LT6\endcsname{\color[rgb]{0,0,0}}%
      \expandafter\def\csname LT7\endcsname{\color[rgb]{1,0.3,0}}%
      \expandafter\def\csname LT8\endcsname{\color[rgb]{0.5,0.5,0.5}}%
    \else
      % gray
      \def\colorrgb#1{\color{black}}%
      \def\colorgray#1{\color[gray]{#1}}%
      \expandafter\def\csname LTw\endcsname{\color{white}}%
      \expandafter\def\csname LTb\endcsname{\color{black}}%
      \expandafter\def\csname LTa\endcsname{\color{black}}%
      \expandafter\def\csname LT0\endcsname{\color{black}}%
      \expandafter\def\csname LT1\endcsname{\color{black}}%
      \expandafter\def\csname LT2\endcsname{\color{black}}%
      \expandafter\def\csname LT3\endcsname{\color{black}}%
      \expandafter\def\csname LT4\endcsname{\color{black}}%
      \expandafter\def\csname LT5\endcsname{\color{black}}%
      \expandafter\def\csname LT6\endcsname{\color{black}}%
      \expandafter\def\csname LT7\endcsname{\color{black}}%
      \expandafter\def\csname LT8\endcsname{\color{black}}%
    \fi
  \fi
    \setlength{\unitlength}{0.0500bp}%
    \ifx\gptboxheight\undefined%
      \newlength{\gptboxheight}%
      \newlength{\gptboxwidth}%
      \newsavebox{\gptboxtext}%
    \fi%
    \setlength{\fboxrule}{0.5pt}%
    \setlength{\fboxsep}{1pt}%
\begin{picture}(7200.00,5040.00)%
    \gplgaddtomacro\gplbacktext{%
      \csname LTb\endcsname%
      \put(814,1071){\makebox(0,0)[r]{\strut{}$0$}}%
      \put(814,1804){\makebox(0,0)[r]{\strut{}$0.2$}}%
      \put(814,2538){\makebox(0,0)[r]{\strut{}$0.4$}}%
      \put(814,3271){\makebox(0,0)[r]{\strut{}$0.6$}}%
      \put(814,4005){\makebox(0,0)[r]{\strut{}$0.8$}}%
      \put(814,4738){\makebox(0,0)[r]{\strut{}$1$}}%
      \put(946,484){\makebox(0,0){\strut{}$0$}}%
      \put(1847,484){\makebox(0,0){\strut{}$0.2$}}%
      \put(2748,484){\makebox(0,0){\strut{}$0.4$}}%
      \put(3649,484){\makebox(0,0){\strut{}$0.6$}}%
      \put(4550,484){\makebox(0,0){\strut{}$0.8$}}%
      \put(5451,484){\makebox(0,0){\strut{}$1$}}%
      \put(6352,484){\makebox(0,0){\strut{}$1.2$}}%
    }%
    \gplgaddtomacro\gplfronttext{%
      \csname LTb\endcsname%
      \put(176,2739){\makebox(0,0){\Large$\frac{\sigma_\text{AHE}}{8 \alpha b}$}}%
      \put(3874,154){\makebox(0,0){\Large$\frac{M}{b}$}}%
    }%
    \gplbacktext
    \put(0,0){\includegraphics{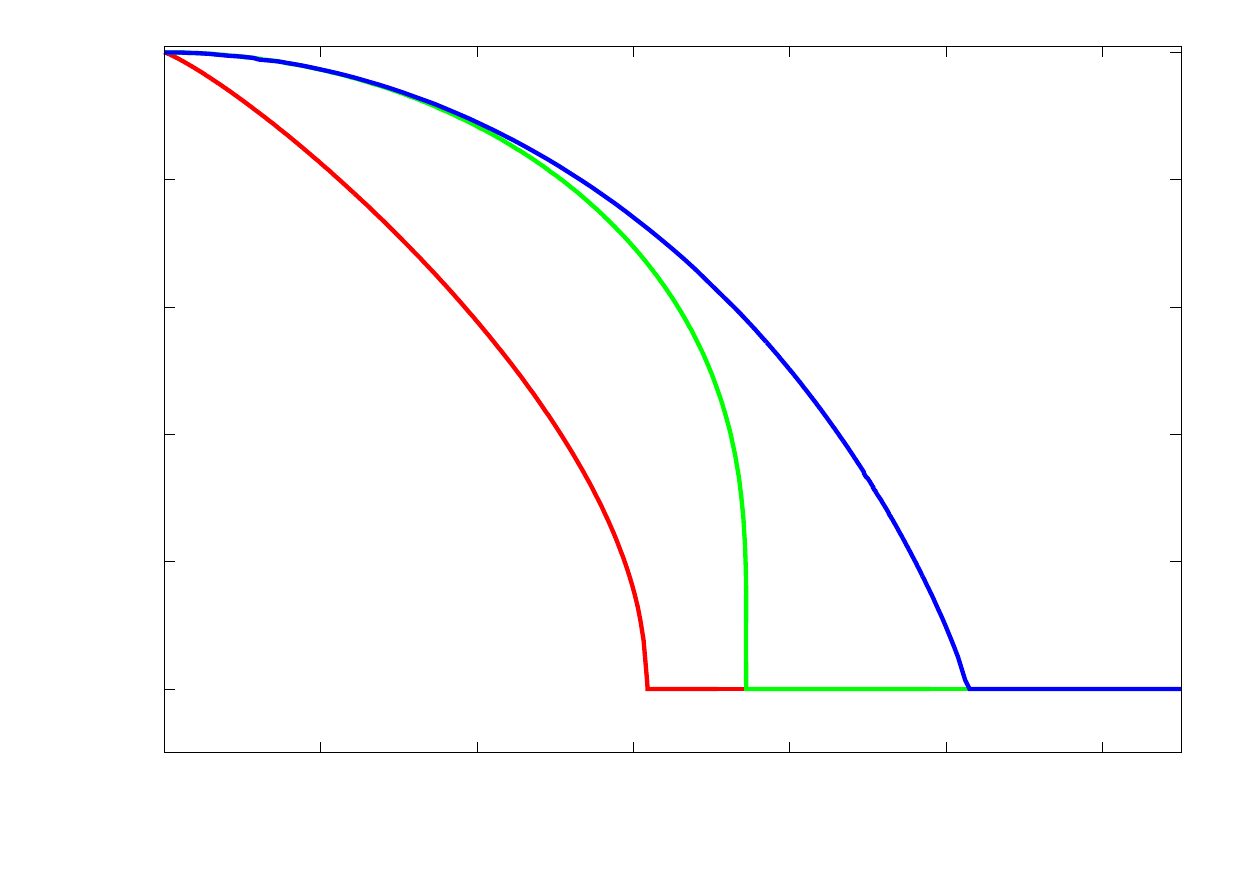}}%
    \gplfronttext
  \end{picture}%
\endgroup

%% file: lambda-map1.tex
% GNUPLOT: LaTeX picture with Postscript
\begingroup
  \makeatletter
  \providecommand\color[2][]{%
    \GenericError{(gnuplot) \space\space\space\@spaces}{%
      Package color not loaded in conjunction with
      terminal option `colourtext'%
    }{See the gnuplot documentation for explanation.%
    }{Either use 'blacktext' in gnuplot or load the package
      color.sty in LaTeX.}%
    \renewcommand\color[2][]{}%
  }%
  \providecommand\includegraphics[2][]{%
    \GenericError{(gnuplot) \space\space\space\@spaces}{%
      Package graphicx or graphics not loaded%
    }{See the gnuplot documentation for explanation.%
    }{The gnuplot epslatex terminal needs graphicx.sty or graphics.sty.}%
    \renewcommand\includegraphics[2][]{}%
  }%
  \providecommand\rotatebox[2]{#2}%
  \@ifundefined{ifGPcolor}{%
    \newif\ifGPcolor
    \GPcolortrue
  }{}%
  \@ifundefined{ifGPblacktext}{%
    \newif\ifGPblacktext
    \GPblacktexttrue
  }{}%
  % define a \g@addto@macro without @ in the name:
  \let\gplgaddtomacro\g@addto@macro
  % define empty templates for all commands taking text:
  \gdef\gplbacktext{}%
  \gdef\gplfronttext{}%
  \makeatother
  \ifGPblacktext
    % no textcolor at all
    \def\colorrgb#1{}%
    \def\colorgray#1{}%
  \else
    % gray or color?
    \ifGPcolor
      \def\colorrgb#1{\color[rgb]{#1}}%
      \def\colorgray#1{\color[gray]{#1}}%
      \expandafter\def\csname LTw\endcsname{\color{white}}%
      \expandafter\def\csname LTb\endcsname{\color{black}}%
      \expandafter\def\csname LTa\endcsname{\color{black}}%
      \expandafter\def\csname LT0\endcsname{\color[rgb]{1,0,0}}%
      \expandafter\def\csname LT1\endcsname{\color[rgb]{0,1,0}}%
      \expandafter\def\csname LT2\endcsname{\color[rgb]{0,0,1}}%
      \expandafter\def\csname LT3\endcsname{\color[rgb]{1,0,1}}%
      \expandafter\def\csname LT4\endcsname{\color[rgb]{0,1,1}}%
      \expandafter\def\csname LT5\endcsname{\color[rgb]{1,1,0}}%
      \expandafter\def\csname LT6\endcsname{\color[rgb]{0,0,0}}%
      \expandafter\def\csname LT7\endcsname{\color[rgb]{1,0.3,0}}%
      \expandafter\def\csname LT8\endcsname{\color[rgb]{0.5,0.5,0.5}}%
    \else
      % gray
      \def\colorrgb#1{\color{black}}%
      \def\colorgray#1{\color[gray]{#1}}%
      \expandafter\def\csname LTw\endcsname{\color{white}}%
      \expandafter\def\csname LTb\endcsname{\color{black}}%
      \expandafter\def\csname LTa\endcsname{\color{black}}%
      \expandafter\def\csname LT0\endcsname{\color{black}}%
      \expandafter\def\csname LT1\endcsname{\color{black}}%
      \expandafter\def\csname LT2\endcsname{\color{black}}%
      \expandafter\def\csname LT3\endcsname{\color{black}}%
      \expandafter\def\csname LT4\endcsname{\color{black}}%
      \expandafter\def\csname LT5\endcsname{\color{black}}%
      \expandafter\def\csname LT6\endcsname{\color{black}}%
      \expandafter\def\csname LT7\endcsname{\color{black}}%
      \expandafter\def\csname LT8\endcsname{\color{black}}%
    \fi
  \fi
    \setlength{\unitlength}{0.0500bp}%
    \ifx\gptboxheight\undefined%
      \newlength{\gptboxheight}%
      \newlength{\gptboxwidth}%
      \newsavebox{\gptboxtext}%
    \fi%
    \setlength{\fboxrule}{0.5pt}%
    \setlength{\fboxsep}{1pt}%
\begin{picture}(7200.00,5040.00)%
    \gplgaddtomacro\gplbacktext{%
      \csname LTb\endcsname%
      \put(946,704){\makebox(0,0)[r]{\strut{}$0$}}%
      \put(946,1518){\makebox(0,0)[r]{\strut{}$2$}}%
      \put(946,2332){\makebox(0,0)[r]{\strut{}$4$}}%
      \put(946,3147){\makebox(0,0)[r]{\strut{}$6$}}%
      \put(946,3961){\makebox(0,0)[r]{\strut{}$8$}}%
      \put(946,4775){\makebox(0,0)[r]{\strut{}$10$}}%
      \put(1078,484){\makebox(0,0){\strut{}$0$}}%
      \put(2223,484){\makebox(0,0){\strut{}$2$}}%
      \put(3368,484){\makebox(0,0){\strut{}$4$}}%
      \put(4513,484){\makebox(0,0){\strut{}$6$}}%
      \put(5658,484){\makebox(0,0){\strut{}$8$}}%
      \put(6803,484){\makebox(0,0){\strut{}$10$}}%
    }%
    \gplgaddtomacro\gplfronttext{%
      \csname LTb\endcsname%
      \put(176,2739){\makebox(0,0){\Large 
      $\left(\frac{M}{b}\right)_{\rm c}$}}%
      \put(3940,154){\makebox(0,0){\Large 
      $\lambda$}}%
    }%
    \gplbacktext
    \put(0,0){\includegraphics{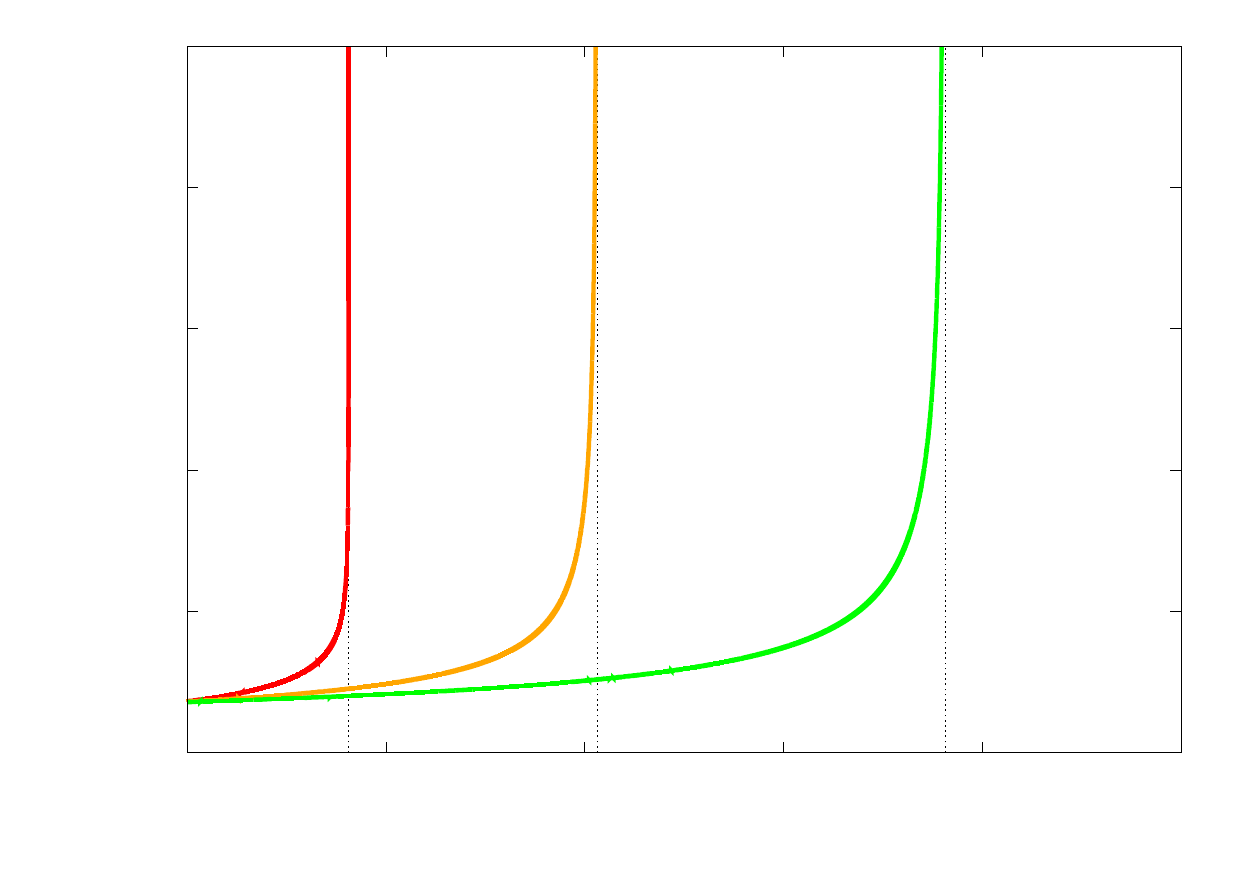}}%
    \gplfronttext
  \end{picture}%
\endgroup